\documentclass[11pt,a4paper]{article}
\usepackage{jheppub}
\usepackage{placeins}
\usepackage{braket}
\usepackage{array}
\newcolumntype{P}[1]{>{\centering\arraybackslash}p{#1}}
\usepackage{amsfonts}
\usepackage{mathtools}
\usepackage{mathrsfs}
\usepackage{amssymb}
\usepackage{float}
\usepackage{tikz}
\usepackage{comment}
\usepackage[normalem]{ulem}
\usetikzlibrary{decorations.markings}
\usetikzlibrary{decorations.markings,arrows,calc}
\usepackage{dsfont}
\newcommand{\id}{\mathds{1}}
\allowdisplaybreaks
\newcommand{\cA}{\mathcal A}
\newcommand{\cB}{\mathcal B}

\newcommand{\eps}{\epsilon}

\newcommand{\be}{\begin{equation}}
\newcommand{\ee}{\end{equation}}
\newcommand{\heff}{\hbar_{\rm eff}}
\newcommand{\cP}{\mathcal{P}}
\newcommand{\cM}{\mathcal{M}}
\newcommand{\cC}{\mathcal C}

\title{Generalised Gibbs Ensembles from the ODE/IM Correspondence: the Lee--Yang Model}

\author[a,b,c]{Sujay K. Ashok,}
\author[d]{Max Downing,}
\author[e]{Faisal Karimi,}
\author[f,b]{Adarsh Sudhakar,}
\author[c]{and Roberto Tateo}

\affiliation[a]{The Institute of Mathematical Sciences,
IV Cross Road, C.I.T. Campus, Taramani, Chennai 600113, India}
\affiliation[b]{Homi Bhabha National Institute,
Training School Complex, Anushakti Nagar, Mumbai 400094, India}
\affiliation[c]{Dipartimento di Fisica, Universit\`a di Torino
and INFN, Sezione di Torino, Via Pietro Giuria 1, I-10125 Torino, Italy}
\affiliation[d]{SISSA and INFN Sezione di Trieste, Via Bonomea 265, 34136 Trieste, Italy}
\affiliation[e]{Department of Mathematics, King's College London,
Strand, London WC2R 2LS, United Kingdom}
\affiliation[f]{Harish-Chandra Research Institute,
Chhatnag Road Jhunsi,
Allahabad 211019, India
}

\emailAdd{sashok@imsc.res.in}
\emailAdd{mdowning@sissa.it}
\emailAdd{faisal.karimi@kcl.ac.uk}
\emailAdd{adarshsudhakar@hri.res.in}
\emailAdd{roberto.tateo@unito.it}

\abstract{
We introduce a systematic scheme for constructing the ODE/IM counterpart
of generalised Gibbs ensembles, and apply it to the chiral sector of the Lee--Yang conformal field theory. We encode the chemical potentials in a deformation of the cubic oscillator, fixed order by order by requiring that the large-energy WKB period reproduces exactly the driving term of the corresponding thermodynamic Bethe ansatz equation; the subleading WKB periods then yield, as predictions, the deformed eigenvalues of the local conserved charges. We work out the spin-$5$ ensemble in detail for two states of the model: the ground state, in the module of the primary $\phi$, dual to the deformed cubic oscillator, and the primary of the identity module, dual to the same oscillator supplemented by an apparent singularity,
a double pole whose position is fixed by a trivial-monodromy condition.
In both cases, the Stokes multiplier obeys the Lee--Yang $Y$-system, and the deformed eigenvalues match, to the orders checked, an independent conformal-field-theory computation. The Borel resummation of the resulting series for the ground-state energy reproduces the numerical TBA solution over many orders of magnitude in the chemical potential. The scheme relies only on general features of the ODE/IM correspondence and applies to generalised Gibbs
ensembles in other integrable models.
}

\begin{document}

\maketitle

\section{Introduction}
\label{sec:intro}

The ODE/IM correspondence relates the spectral data of ordinary differential equations to the functional relations and integral equations of two-dimensional integrable models~\cite{Dorey:1998pt,Bazhanov:1998wj,Dorey:1999uk,Dorey:2007zx,Dorey:2025odeim}.
Massive extensions of the correspondence, in which the ODE is replaced by the linear problem of a classical integrable equation, have been developed in a number of related settings
\cite{Gaiotto:2009hg,Lukyanov:2010rn,Dorey:2012bx,Ito:2018eon,
Fioravanti:2021gbd,Aramini:2022wxi},
and higher-order and pseudo-differential versions, associated with
higher-rank integrable structures and $W$-algebras, have been formulated in a variety of ways
\cite{Dorey:1999pv,Bazhanov:2001xm,DoreyDunningMasoero:2006,
FeiginFrenkel:2007,MasoeroRaimondo:2018,Ito:2021hdy, Ashok:2024zmw, Ashok:2024ygp,  ItoZhu:2024,Kudrna:2025bzg}.
A particularly simple connection between the TBA and spectral problems, involving Airy functions, was also found in~\cite{Fendley:1999gb}.

In its simplest form, the Stokes multipliers of the Schr\"odinger equation
\be
\left[-\frac{d^2}{dx^2}+x^{2M}-E\right]\psi(x)=0
\label{eq:intro-ode}
\ee
satisfy the same functional relations as the $T$-functions
(cf.~\cite{Kuniba:2010ir}) of the thermodynamic Bethe ansatz (TBA) for the
conformal limit of a family of perturbed minimal
models~\cite{Dorey:1999uk}. Their large-$E$ asymptotics, computed using WKB methods, reproduce the driving term of the TBA, and the higher-order WKB periods
give the eigenvalues of the local conserved charges of the quantum KdV
hierarchy. For $M=3/2$, the equation is the cubic oscillator and the model is the Lee--Yang minimal model $\cM_{2,5}$~\cite{Dorey:1998pt}.

In this paper we ask how the correspondence extends to generalised Gibbs ensembles (GGEs), in which chemical potentials are switched on for the
higher local charges~\cite{Rigol:2006py,Essler:2014mya,
Maloney:2018yrz,Dymarsky:2018lhf}. We focus throughout on the Lee--Yang model, whose local charges $I_s$ have spin $s = 1\pmod6$ and $s=5 \pmod 6$. Switching on chemical potentials for a finite set $\mathcal S$ of such spins,
$s>1$, amounts in the massless TBA to modifying the driving term,
\be
e^{\theta}\;\longrightarrow\;e^{\theta}+\sum_{s\in\mathcal S}
\kappa_{{\rm TBA},s}\,e^{s\theta},
\label{eq:intro-driving-general}
\ee
while the kernel, and hence the $Y$-system, are unchanged. Such ensembles
have recently been studied from the modular point of
view~\cite{Downing:2023lnp,Downing:2021mfw,Downing:2023lop,Downing:2024nfb,Downing:2025huv,Ashok:2026cyi},
and it is natural to ask which ODE reproduces them.

The answer is not obvious. Although a nonlinear redefinition of the
spectral parameter can formally generate higher powers in the large-$E$
expansion, it merely reparametrises the undeformed cubic spectral problem
and produces a constrained tower of terms. It does not provide the
independent source sectors required by the GGE. We therefore seek a
genuine deformation of the potential whose WKB period reproduces the
prescribed driving terms.

Our main result is a systematic scheme that produces such a deformation,
fixed by a single requirement. The ground state, in the module of the
primary $\phi$, is described by the deformed cubic oscillator
\be
\left[-\frac{d^2}{dx^2}+x^3+Q(E,\boldsymbol{\kappa})\,x-P(E,\boldsymbol{\kappa})\right]\psi(x)=0,
\label{eq:intro-ODE-main}
\ee
where $\boldsymbol{\kappa}=\{\kappa_s\}_{s\in\mathcal S}$ collects one ODE
coupling for each switched-on spin, and $P$ and $Q$ are polynomials in
$E$ whose coefficients are power series in the couplings, fixed by the
requirement
\be
\oint_\gamma S(x;E,\boldsymbol{\kappa})\,dx=e^{\theta}+\sum_{s\in\mathcal S}
\kappa_{{\rm TBA},s}(\boldsymbol{\kappa})\,e^{s\theta}
+O\!\left(e^{-\theta}\right),\qquad e^{\theta}=D(\boldsymbol{\kappa})\,E^{5/6}.
\label{eq:defining-problem0}
\ee
Here $S$ is the formal WKB solution of the associated Riccati equation,
$\gamma$ is the cycle encircling the complex pair of turning points,
continued from $\boldsymbol{\kappa}=0$, and $D(\boldsymbol{\kappa})$ is a
normalisation fixed by the construction. The ODE couplings are not the
physical chemical potentials: the latter are nonlinear functions
$\kappa_{{\rm TBA},s}(\boldsymbol{\kappa})$, determined together with $P$
and $Q$. Condition \eqref{eq:defining-problem0} states that, at large
$E$, the period contains exactly the sources of the TBA driving term and
no other positive spin. Since every term in the period carries a
definite spin, it determines $P$, $Q$ and the
$\kappa_{{\rm TBA},s}(\boldsymbol{\kappa})$ order by order in the
couplings, and the coefficients of the remaining terms $e^{-s\theta}$
are then predictions: the deformed eigenvalues of the local charges
$I_s$.

We work out in full detail the simplest case, a single chemical
potential for the spin-$5$ charge, $\mathcal S=\{5\}$,
\be
e^{\theta}\;\longrightarrow\;e^{\theta}+\kappa_{\rm TBA}\,e^{5\theta}.
\label{eq:intro-driving}
\ee
The classical deformation of the potential is then found exactly, in
closed form, and the first quantum correction is obtained from the same
elliptic periods; beyond that, the recursion is purely algebraic and we
carry it to high order. The resulting Stokes multiplier obeys the
Lee--Yang $Y$-system, and the deformed eigenvalues of $I_1$, $I_5$ and
$I_7$ agree, to the orders available, with an entirely independent
construction of the deformed charges directly in conformal field
theory~\cite{Downing:2024nfb,Karimi:2026kdv}. The agreement includes the
contributions of non-KdV components of the deformed charges, whose spin
is forbidden for the KdV charges themselves. The series for the deformed
energy is divergent but Borel summable, and its resummation reproduces
the numerical solution of the TBA over many orders of magnitude in
$\kappa_{\rm TBA}$.

We then extend the analysis to the primary state of the identity module.
At $\kappa=\kappa_{\rm TBA}=0$ it is described by the cubic oscillator
with the additional term $2/x^2$, a singular point around which all
solutions are single-valued. With the deformation, this trivial-monodromy
property forces the pole to move with the coupling, to a position $z_0$
fixed by $3z_0^2+Q=0$, and the WKB periods acquire one new basic period
associated with the pole. With this modification the construction goes
through as for the ground state, and the deformed eigenvalues again match
the conformal field theory results.

This paper is organised as follows. Section~\ref{sec:lee-yang} reviews
various aspects of the Lee--Yang model and its GGE. This involves the deformed TBA and its asymptotics, the associated $Y$-system, and the deformed KdV charges
constructed in conformal field theory, whose eigenvalues provide the
predictions that we test in this work. Section~\ref{sec:derivation} introduces the
deformed cubic oscillator and the defining condition, and solves it for
the spin-$5$ ensemble at the classical and first quantum levels.
Section~\ref{sec:functional} shows that the deformed spectral problem is
covariant under the Sibuya rotation, derives the $Y$-system on the ODE
side, and relates the Stokes multiplier to the WKB period.
Section~\ref{sec:energy-wkb} extracts from the WKB period the deformed 
eigenvalues of the conserved charges and compares them with those obtained from the
conformal field theory construction. Section~\ref{sec:resummation}
resums the series for the ground-state energy and compares it with the
numerical solution of the TBA. Section~\ref{sec:excited} analyses the deformed spectral problem associated to the primary of the identity module.
Section~\ref{sec:multicoupling} generalises the construction to an
arbitrary set of chemical potentials, and Section~\ref{sec:conclusions}
summarises the results and open directions. Technical details on the
elliptic periods and the WKB reductions are collected in the 
Appendices.
\section{The Lee--Yang model and its spin-\texorpdfstring{$5$}{5} GGE}
\label{sec:lee-yang}
The Lee--Yang conformal field theory is the Virasoro minimal model
$\cM_{2,5}$, with central charge $c=-22/5$. It has two irreducible
representations, associated with the identity and with a primary field
$\phi$ of weight $h_\phi=-1/5$. The latter is the conformal ground state,
and the effective central charge is $c_{\rm eff}=c-24h_\phi=2/5$. The
massive integrable perturbation of the model contains a single particle,
with a two-body scattering  amplitude~\cite{Cardy:1989fw,Smirnov:1989fa}
\be
S(\theta)=\frac{\sinh\theta+i\sin(2\pi/3)}{\sinh\theta-i\sin(2\pi/3)}\,.
\label{eq:Smatrix}
\ee
In the conformal limit, the ground-state TBA reads
\cite{Zamolodchikov:1989cf,Klassen:1990dx}
\be
\epsilon(\theta)=e^{\theta}-\int_{-\infty}^{\infty}d\theta'\,
\Phi(\theta-\theta')\,\log\left(1+e^{-\epsilon(\theta')}\right),
\label{eq:undeformedTBA}
\ee
with the kernel
\be
\Phi(\theta)=-\frac{i}{2\pi}\frac{d}{d\theta}\log S(\theta)
=-\frac{2\sqrt3}{\pi}\,\frac{\cosh\theta}{1+2\cosh 2\theta}~.
\label{eq:kernel}
\ee

\subsection{The GGE}
The model has an infinite set of commuting local charges $\hat I_s$. The
quantum KdV hierarchy provides one for every odd spin, but in
$\cM_{2,5}$ those of spin $3,9,15,\dots$ vanish
identically~\cite{Bazhanov:1994ft,Maloney:2018hdg}, leaving one charge for each
$s=1 \pmod6$ and $s=5 \pmod6$; $\hat I_1$ is proportional to the Hamiltonian. We
consider the generalised Gibbs ensemble in which, besides the
Hamiltonian, the charge $\hat I_5$ enters the density matrix with its own
chemical potential. In the massless TBA this amounts to replacing the
driving term by $e^\theta+\kappa_{\rm TBA}e^{5\theta}$, leaving the
kernel unchanged:
\be
\epsilon(\theta;\kappa_{\rm TBA})=e^\theta+\kappa_{\rm TBA}e^{5\theta}
-\int_{-\infty}^{\infty}d\theta'\,\Phi(\theta-\theta')\,
L(\theta';\kappa_{\rm TBA}),
\label{eq:GGETBA}
\ee
with $L(\theta;\kappa_{\rm TBA}):=\log\bigl(1+e^{-\epsilon(\theta;\kappa_{\rm TBA})}\bigr)$,
used throughout.  At $\kappa_{\rm TBA}=0$ the equation reduces to
\eqref{eq:undeformedTBA}, and we write
$\epsilon(\theta)\equiv\epsilon(\theta;0)$.

This equation describes the ground state, the primary
$\phi$ of weight $h=-\tfrac15$; we denote the corresponding eigenvalues
by $I_s^{(\phi)}$, $\hat I_s|\phi\rangle=I_s^{(\phi)}|\phi\rangle$, and
their deformations by $I_s^{(\phi)}(\kappa_{\rm TBA})$. For real
$\kappa_{\rm TBA}\ge0$ the driving term suppresses $L$ at large $\theta$,
and the equation has a regular solution. Since for any
$\kappa_{\rm TBA}>0$ the large-$\theta$ asymptotics is dominated by the
spin-$5$ term, the limit $\kappa_{\rm TBA}\to0$ is not uniform in
$\theta$, and any expansion around $\kappa_{\rm TBA}=0$ is to be
understood as a formal (asymptotic) power series. The deformed
ground-state energy is
\begin{equation}
I_1^{(\phi)}(\kappa_{\rm TBA})
=
-\frac{1}{(2\pi)^2}\,
\int_{-\infty}^{\infty} d\theta\,
e^{\theta}
L(\theta;\kappa_{\rm TBA}).
\label{eq:GGEenergy}
\end{equation}
Using the large-$\theta$ behaviour of the kernel,
$\Phi(\theta-\theta')\simeq-\frac{\sqrt3}{\pi}\,e^{-(\theta-\theta')}$ for
$\theta\to+\infty$, together with \eqref{eq:GGEenergy}, one finds
\be
\epsilon(\theta;\kappa_{\rm TBA})=e^{\theta}+\kappa_{\rm TBA}e^{5\theta}
-4\sqrt3\,\pi\,I_1^{(\phi)}(\kappa_{\rm TBA})\,e^{-\theta}+O\!\left(e^{-5\theta}\right),
\label{eq:Z-asymptotic-energy}
\ee
so that, for every $\kappa_{\rm TBA}$, the $\kappa_{\rm TBA}$-expansion of
$I_1^{(\phi)}(\kappa_{\rm TBA})$ is the $\kappa_{\rm TBA}$-expansion of
the coefficient of $e^{-\theta}$.

This coefficient admits two readings, related by the mirror
transformation familiar from the TBA for AdS/CFT. The mirror transform is a double Wick rotation
which, through analytic
continuation in the rapidity, exchanges the energy and momentum of a theory with those of its mirror
theory~\cite{Arutyunov:2007tc,Arutyunov:2009ur}.
Firstly, in the
thermodynamic channel, \eqref{eq:GGETBA} describes a Gibbs ensemble for
the effective Hamiltonian $\hat I_1+\kappa_{\rm TBA}\hat I_5$, with
one-particle energy $e^{\theta}+\kappa_{\rm TBA}e^{5\theta}$, and the
coefficient of $e^{-\theta}$ is proportional to its free-energy density,
at inverse temperature set to one by a shift of $\theta$. Alternatively, in the energy
channel it is the ground-state energy of the theory whose mirror is this
ensemble. At $\kappa_{\rm TBA}=0$ the dispersion is relativistic, the
theory is its own mirror, and one recovers the Casimir energy of the
CFT. For $\kappa_{\rm TBA}\neq0$ the term $e^{5\theta}$ breaks boost
invariance, and the theory in the energy channel is no longer the CFT
with Hamiltonian $\hat I_1+\kappa_{\rm TBA}\hat I_5$: its Hamiltonian is the deformed charge ($\hat I_1$ corrected by an
infinite series of local fields of increasing weight), constructed
directly in conformal field theory~\cite{Downing:2024nfb,Karimi:2026kdv}, and
$I_1^{(\phi)}(\kappa_{\rm TBA})$ is its ground-state eigenvalue. This is
the sense in which $I_1^{(\phi)}(\kappa_{\rm TBA})$, and the analogous
coefficients below, are \emph{deformed} eigenvalues. The situation
parallels the massive case, where the relativistic driving term
$e^{\theta}+e^{-\theta}$ corresponds in the energy channel to a relevant
bulk perturbation of the CFT; here however, the deformation is irrelevant,
consistent with the asymptotic nature of the expansion around
$\kappa_{\rm TBA}=0$.

The same argument applies to the higher order terms. Using the full
large-$\theta$ expansion of the kernel,
$\Phi(\theta)=-\frac{\sqrt3}{\pi}\sum_{n\ge0}\big(e^{-(6n+1)\theta}-e^{-(6n+5)\theta}\big)$,
one finds
\be
\epsilon(\theta;\kappa_{\rm TBA})=e^{\theta}+\kappa_{\rm TBA}e^{5\theta}
+\frac{\sqrt3}{\pi}\sum_{s}\epsilon_s\,e^{-s\theta}
\int_{-\infty}^{\infty}d\theta'\,e^{s\theta'}L(\theta';\kappa_{\rm TBA}),
\label{eq:Z-asymptotic-general}
\ee
where the sum runs over the positive integers $s\equiv\pm 1\pmod6$, with
$\epsilon_s=+1$ for $s\equiv1\pmod6$ and $\epsilon_s=-1$ for
$s\equiv5\pmod6$; the term $s=1$ is \eqref{eq:Z-asymptotic-energy}. No
term with $s\equiv3\pmod6$ is generated, consistently with the absence
of local charges at those spins. The terms
$e^{-5\theta},e^{-7\theta},\dots$ carry in the same way the
$\kappa_{\rm TBA}$-deformed eigenvalues of $\hat I_5,\hat I_7,\dots$:
\begin{equation}
I_s^{(\phi)}(\kappa_{\rm TBA})
=
\frac{\sqrt{3}}{\pi N_s}\,
\epsilon_s
\int_{-\infty}^{\infty} d\theta\,
e^{s\theta}
L(\theta;\kappa_{\rm TBA}),
\label{eq:Is-moment}
\end{equation}
where the state-independent constants $N_s$ are fixed by the
large-$\theta$ expansion of the undeformed pseudoenergy, equivalently of
$\log T$ in~\cite{Bazhanov:1996aq}, once the driving term is normalised
to $e^{\theta}$:
\be
N_1=-4\sqrt3\,\pi,\qquad
N_5=-\frac{1729}{6000}\,\frac{\Gamma(\tfrac13)^6\Gamma(\tfrac16)^6}{\pi^3},
\qquad
N_7=-\frac{124729}{16800}\,\frac{\sqrt3\,\Gamma(\tfrac13)^6\Gamma(\tfrac16)^6}{\pi^2}.
\label{eq:Ns-values}
\ee
The constant $N_1=-4\sqrt3\,\pi$ of \eqref{eq:Z-asymptotic-energy} is
therefore specific to $s=1$: each charge comes with its own $N_s$.

The same analysis applies to excited states. In particular, the primary
of the identity module, $h=0$, is the ultraviolet limit of the first
excited state of the scaling Lee--Yang model. Its TBA equation is
obtained by analytic continuation of \eqref{eq:GGETBA}, and contains, in
addition to the driving term, source terms associated with a pair of
complex singularities of the pseudoenergy, whose positions are fixed
self-consistently~\cite{Dorey:1996re,Bazhanov:1996aq}. The chemical
potential enters, as for the ground state, through the term
$\kappa_{\rm TBA}e^{5\theta}$ in the driving term~\cite{Karimi:2026kdv},
and the asymptotic expansion of the pseudoenergy yields in the same way
the deformed eigenvalues $I_s^{(\id)}(\kappa_{\rm TBA})$, with the same
constants $N_s$. We return to this state in Section~\ref{sec:excited}.
\subsection{The \texorpdfstring{$Y$}{Y}-system and the allowed sources}
Let $Y(\theta;\kappa_{\rm TBA})=e^{\epsilon(\theta;\kappa_{\rm TBA})}$, and
consider the shift operator
\be
(\mathcal D f)(\theta)=f(\theta+\tfrac{i\pi}{3})+f(\theta-\tfrac{i\pi}{3})-f(\theta),
\qquad
\mathcal D\,e^{s\theta}=\Bigl(2\cos\frac{s\pi}{3}-1\Bigr)e^{s\theta}.
\label{eq:shift-operator}
\ee
The kernel of \eqref{eq:GGETBA} satisfies $\mathcal D\Phi=-\delta$, where $\delta$ is the Dirac delta function, while both terms in the driving term are zero modes of $\mathcal D$, since
$2\cos\frac{\pi}{3}=2\cos\frac{5\pi}{3}=1$. Applying $\mathcal D$ to
\eqref{eq:GGETBA} therefore gives $\mathcal D\epsilon=L$, which, upon exponentiating, is the
well-known Lee--Yang $Y$-system
\be
Y\left(\theta+\tfrac{i\pi}{3};\kappa_{\rm TBA}\right)Y\left(\theta-\tfrac{i\pi}{3};\kappa_{\rm TBA}\right)
=1+Y(\theta;\kappa_{\rm TBA}).
\label{eq:GGEYsystem}
\ee
The derivation shows why the chemical potential leaves the $Y$-system
unchanged: the functional relation only sees the kernel, and the driving
term drops out. The information it carries survives only in the
asymptotics of $Y$ at large $|\theta|$, which, together with analyticity
in the strip $|{\rm Im}\,\theta|\le\pi/3$, selects one solution among the
many admitted by \eqref{eq:GGEYsystem}; this is the general mechanism of
the $Y$-system/TBA
framework~\cite{Ravanini:1992fi,Kuniba:2010ir,CastroAlvaredo:2022ys}.

The same argument tells us which sources can be added. By
\eqref{eq:shift-operator}, a driving term $e^{s\theta}$ is a zero mode of
$\mathcal D$, and hence compatible with \eqref{eq:GGEYsystem}, if and
only if
\be
2\cos\frac{s\pi}{3}=1,\qquad\text{i.e.}\qquad s\equiv\pm1\pmod6 .
\label{eq:spin-selection-TBA}
\ee
This is precisely the set of spins of the local charges of the model,
and not by accident. The one-particle eigenvalue of a conserved charge
of spin $s$ is proportional to $e^{s\theta}$, and consistency with the
fusion $A\times A\to A$ of the Lee--Yang scattering theory \cite{Cardy:1989fw}, in which the
bound state at rapidity $\theta$ is formed by two particles at
$\theta\pm\tfrac{i\pi}{3}$, requires the eigenvalues to add up,
$e^{s(\theta+i\pi/3)}+e^{s(\theta-i\pi/3)}=e^{s\theta}$, which is again
\eqref{eq:spin-selection-TBA}. The excluded spins $s\equiv3\pmod6$ are
those of the charges that vanish in $\cM_{2,5}$. Thus, for any finite
set $\mathcal S$ of allowed higher spins, the same TBA equation admits
the driving term $e^\theta+\sum_{s\in\mathcal S}\kappa_s^{\rm TBA}e^{s\theta}$,
with the coefficient of the highest spin positive so that $L$ is
suppressed at large $\theta$. The spin-$5$ equation \eqref{eq:GGETBA} is
the one-parameter slice used for the detailed CFT and numerical
comparisons below. We shall
see in Section~\ref{sec:rotation} that the same
selection rule reappears on the ODE side, as the condition for the
deformed spectral problem to be covariant under the Sibuya rotation.

\subsection{The perturbative expansion and the deformed KdV charges}
\label{sec:deformed-kdv}

Let us describe the perturbative regime in which the ODE/IM
correspondence developed in the following sections will be tested. In~\cite{Downing:2024nfb,Ashok:2026cyi,Karimi:2026kdv} the expansion
around $\kappa_{\rm TBA}=0$ was studied from the CFT point of view, the
basic question being which operators of the CFT have eigenvalues
reproduced by the moments of the deformed TBA introduced above. The
answer is a family of deformed charges, defined as formal series
\begin{equation}
    \hat{\mathcal I}_s
    =
    \hat I_s
    +
    \sum_{m=1}^{\infty}
    \frac{\gamma_{\rm CFT}^m}{2^{m-1}(m-1)!}\,
    \hat J_{s,s+4m},
    \label{eq:deformed-charge-expansion}
\end{equation}
where hats denote operators, $\hat I_s$ is the KdV charge of spin $s$,
and $\hat J_{s,s+4m}$ is the zero-mode of a quasi-primary field of
weight $s+4m+1$, whose construction follows from the analysis of the
modular transformation of the
GGE~\cite{Downing:2024nfb,Ashok:2026cyi,Karimi:2026kdv}. The deformation
parameter $\gamma_{\rm CFT}$ is proportional to the TBA chemical
potential,
\begin{equation}
    \gamma_{\rm CFT}=\frac{1729\pi^2}{2025\sqrt{3}}
    \left[\frac{\Gamma(\tfrac16)}{\Gamma(\tfrac23)}\right]^6
    \kappa_{\rm TBA}.
    \label{eq:gamma-kappa-TBA}
\end{equation}
On a common eigenstate the eigenvalues, denoted by the same symbols
without hats, inherit the expansion
\eqref{eq:deformed-charge-expansion}, and order by order in
$\kappa_{\rm TBA}$ the moments of the TBA are identified with them. When
both Lee--Yang primaries are involved we label the eigenvalues by the
state, $\hat I_s|\phi\rangle=I_s^{(\phi)}|\phi\rangle$ and
$\hat I_s|\id\rangle=I_s^{(\id)}|\id \rangle$; unless
stated otherwise, $I_s$ without a label refers to the ground state, the
primary $\phi$ of weight $h=-\tfrac15$.

An important feature of this construction is that the individual
components need not commute with one another, even though the complete
deformed charges form a commuting hierarchy. For the deformed
Hamiltonian,
\begin{equation}
    \hat{\mathcal I}_1
    =
    \hat I_1
    +\gamma_{\rm CFT}\, \hat J_{1,5}
    +\frac{\gamma_{\rm CFT}^2}{2}\,\hat J_{1,9}
    +\frac{\gamma_{\rm CFT}^3}{8}\,\hat J_{1,13}
    +O(\gamma_{\rm CFT}^4),
    \label{eq:deformed-hamiltonian}
\end{equation}
the first order component is simply the KdV charge,
$\hat J_{1,5}=\hat I_5$, while the components of spin $9$ and $13$ are
not KdV charges and do not commute individually with $\hat I_5$,
\begin{equation}
    [\hat J_{1,9},\hat I_5]\neq0,
    \qquad
    [\hat J_{1,13},\hat I_5]\neq0,
\end{equation}
although the complete deformed charges do,
$[\hat{\mathcal I}_s,\hat{\mathcal I}_r]=0$. The appearance of
$\hat J_{1,9}$ is particularly striking. Its spin, $9\equiv3\pmod6$, is
one at which the KdV charge of the Lee--Yang model vanishes identically,
and at which, by the selection rule~\eqref{eq:spin-selection-TBA}, no
source can be added to the TBA. The deformed hierarchy nevertheless
contains a non-vanishing spin-$9$ component, at second order in
$\gamma_{\rm CFT}$. Any extension of the ODE/IM correspondence to the GGE
must therefore, first of all, accommodate such non-KdV contributions.
The GGE TBA does so, and its moments have been shown to agree with the
CFT construction~\cite{Downing:2024nfb,Ashok:2026cyi,Karimi:2026kdv}.

This is consistent with the selection rule. In the TBA, $J_{1,9}$ never
appears as a separate term $e^{-9\theta}$ in the large-$\theta$
expansion of the pseudoenergy, where the kernel only generates the
allowed exponentials $e^{-(6n\pm1)\theta}$
of~\eqref{eq:Z-asymptotic-general}; it enters instead through the
$\kappa_{\rm TBA}^2$ coefficient of the energy term $e^{-\theta}$, that
is, through the expansion of the moment~\eqref{eq:GGEenergy}. More
generally, every component $J_{s,s+4m}$ of a deformed charge is carried
by the single, allowed term $e^{-s\theta}$, whatever the spin $s+4m$.

The CFT predictions that we shall test are collected in
Table~\ref{tab:cft-eigenvalues}. On the ground state the component
eigenvalues are those of~\cite{Karimi:2026kdv}, so that
\begin{equation}
    I_1^{(\phi)}(\kappa_{\rm TBA})
    \sim
    -\frac1{60}
    +\frac{89}{756000}\,\gamma_{\rm CFT}
    -\frac{1729}{21600000}\,\gamma_{\rm CFT}^2
    +\frac{1077983}{2592000000}\,\gamma_{\rm CFT}^3
    +O(\gamma_{\rm CFT}^4).
    \label{eq:I1-cft-ground}
\end{equation}
On the identity primary, $h=0$, we have evaluated the zero-modes of the
same fields, obtaining
\begin{equation}
    I_1^{(\id)}(\kappa_{\rm TBA})
    \sim
    \frac{11}{60}
    +\frac{341}{756000}\,\gamma_{\rm CFT}
    -\frac{2821}{21600000}\,\gamma_{\rm CFT}^2
    +\frac{1475507}{2592000000}\,\gamma_{\rm CFT}^3
    +O(\gamma_{\rm CFT}^4).
    \label{eq:I1-cft-identity}
\end{equation}
\begin{table}[t]
\centering
\renewcommand{\arraystretch}{1.6}
\begin{tabular}{c|c|c}
 & $\phi$, \ $h=-\tfrac15$ & $\id $, \ $h=0$ \\ \hline
$I_1$ & $-\dfrac{1}{60}$ & $\dfrac{11}{60}$ \\
$J_{1,5}=I_5$ & $\dfrac{89}{756000}$ & $\dfrac{341}{756000}$ \\
$J_{1,9}$ & $-\dfrac{1729}{10800000}$ & $-\dfrac{2821}{10800000}$ \\
$J_{1,13}$ & $\dfrac{1077983}{324000000}$ & $\dfrac{1475507}{324000000}$
\end{tabular}
\caption{Eigenvalues of the components of the deformed Hamiltonian
\eqref{eq:deformed-hamiltonian} on the two Lee--Yang primaries. The
values on $\phi$ are from~\cite{Karimi:2026kdv}; those on $\id$
are obtained by evaluating the zero-modes of the same fields on the
identity primary.}
\label{tab:cft-eigenvalues}
\end{table}
In Sections~\ref{sec:karimi-comparison} and~\ref{sec:excited-results}
we show that the eigenvalues obtained from the ODE/IM construction
reproduce \eqref{eq:I1-cft-ground} and \eqref{eq:I1-cft-identity}
exactly, together with the analogous expansions of $I_5$ and $I_7$.

\section{An ODE for the GGE}
\label{sec:derivation}
In this section, we propose an ODE dual to the generalised Gibbs ensembles associated with the ground state of the Lee--Yang model, whose WKB periods yield the eigenvalues of the deformed KdV charges. We begin with a review of the cubic oscillator and then discuss the deformed spectral problem relevant to the GGE. 

\subsection{The cubic oscillator and the Lee--Yang ground state}
\label{sec:cubic}

The ODE/IM correspondence for the ground state of the Lee--Yang model,
with central charge $c=-\frac{22}{5}$ and conformal dimension
$h=-\frac15$, is built on the cubic oscillator~\cite{Dorey:1998pt}
\be
\left[-\frac{d^2}{dx^2}+x^3-E\right]\psi(x)=0 .
\label{eq:LY-ODE}
\ee
The cubic growth of the potential divides the complex $x$-plane into five
Stokes sectors, with neighbouring sectors related by the discrete rotation
$x\to e^{2\pi i/5}x$.  The relevant WKB cycle $\gamma$ joins the two
complex turning points
$x_\pm=E^{1/3}e^{\pm2\pi i/3}$ of \eqref{eq:LY-ODE}.  Rescaling
$x=E^{1/3}\xi$ gives
\be
\oint_\gamma dx\,\sqrt{x^3-E}
=E^{5/6}\oint_\gamma d\xi\,\sqrt{\xi^3-1} = \cA \,  E^{5/6} \,, \quad  \cA = \frac{\sqrt3\,\Gamma(\tfrac13)\Gamma(\tfrac16)}{5\sqrt\pi}
\label{eq:WKB-scaling}
\ee
so that the natural spectral variable is $E^{5/6}$. 

Let $S:=\psi'/\psi$ be the logarithmic derivative of the solution
subdominant along the positive real axis, expanded semiclassically.
The Stokes multiplier relevant here is the lateral connection coefficient
$T$ between the subdominant solutions in two non-adjacent Stokes sectors.
Sibuya~\cite{Sibuya:1984fe} already studied its functional relation.
Together with its Lee--Yang $Y$-system interpretation, this is the simplest
example of the ODE/IM correspondence~\cite{Dorey:1998pt,Bazhanov:1998wj,
Dorey:1999uk,Dorey:2007zx,ItoShu:2025,Dorey:2025odeim}. In the present conventions, the large-$E$
asymptotics of $T(E)$ is governed by the WKB period
along $\gamma$,
\be
\log T(E)\simeq\oint_\gamma S(E,x) \,dx = \cA\,E^{5/6} + O(E^{-5/6}),
\qquad E\to+\infty.
\label{eq:undeformed-defining}
\ee
For the Lee-Yang model, the ODE/IM correspondence  gives the identification 
$T(E) = Y(\theta) = e^{\eps(\theta)}$, with (cf.  (\ref{eq:undeformedTBA}) and  (\ref{eq:undeformed-defining})) 
\be
e^{\theta}=\cA\,E^{5/6}
\label{eq:E-theta-map}
\ee
No other positive power of $E^{5/6}$ appears at any order, while
the negative-power tail in equation \eqref{eq:undeformed-defining} encodes the ground-state eigenvalues of the local
charges $I_s$.  Section~\ref{sec:stokes-period} derives
\eqref{eq:undeformed-defining} in detail for the deformed problem introduced
below, and it reduces to the present case in the $\kappa=0$ limit.
\subsection{The deformed spectral problem and its semiclassical expansion}
\label{sec:family}
Once we have identified the spectral parameter in the ODE, the natural generalisation that encodes the switching on of the spin-$5$ chemical potential replaces \eqref{eq:undeformed-defining} by
\be
\oint_\gamma S\,dx=e^{\theta}+\kappa_{\rm TBA}(\kappa)\,e^{5\theta}+O\!\left(e^{-\theta}\right).
\label{deformedperiod}
\ee
The $e^{5\theta}$ term gives one further term in the large-$E$ expansion of the period,
matching the one extra term \eqref{eq:intro-driving} in the
driving term of the TBA. 
Here $\gamma$ is the cycle of Section~\ref{sec:cubic}, continuously
deformed away from $\kappa=0$ along with the two complex turning points
it joins. Section~\ref{sec:stokes-period} shows that, to this order,
the Stokes multiplier of the deformed problem is given by the WKB
period along $\gamma$, and Section~\ref{sec:ode-ysystem} shows that it
satisfies the same functional relation as the Lee--Yang $Y$-function.

In this section, we shall adopt a more hands-on approach to the problem. We look for the deformed ODE whose Stokes multiplier has precisely the asymptotic expansion written out in \eqref{deformedperiod}. Guided by the  preceding discussions, we deform
the cubic oscillator by a single coupling $\kappa$,
\be
\begin{aligned}
&\left[-\frac{d^2}{dx^2}+V(x;E,\kappa)\right]\psi(x)=0,\\
&V(x;E,\kappa)=x^3+Q(E,\kappa)\,x-P(E,\kappa),
\qquad P(E,0)=E,\ \ Q(E,0)=0,
\end{aligned}
\label{eq:deformed-ODE}
\ee
with $P,Q$ formal power series in $\kappa$ with polynomial coefficients
in $E$; the leading large-$x$ behaviour is kept cubic, and an $x^2$
term has been removed by a translation of $x$.\footnote{For the vacuum
module the translation would move the singularity $2/x^2$ and
additional structures are allowed.  We continue to consider only the
conformal ground state in this section and shall revisit this problem in a later section.}  The coefficients of $P$ and $Q$ are
the unknowns of the construction, to be fixed order by order in $\kappa$. 

Before we proceed with the analysis, let us comment upon why we restrict $P$ and $Q$ to have polynomial coefficients in $E$. At first glance, this might seem odd, given that the fractional power $E^{5/6}$ already appears in the undeformed case, in \eqref{eq:undeformed-defining}. However, that exponent is the growth rate of $\log T(E)$ in the asymptotic expansion at the irregular singular point $E=\infty$, not a statement about multivaluedness in the finite $E$-plane. What makes $T(E)$ single-valued enough to possess such an expansion at all is that $\psi(x;E)$ solves a linear ODE whose coefficients $P,Q$ are polynomials in $E$ at each order in their $\kappa$-expansions, so that $\psi(x;E)$ is an entire function of $E$ for each fixed $x$ at each order in $\kappa$. A branch cut in $P,Q$ would instead make $V(x;E,\kappa)$, and hence the solutions themselves, genuinely multivalued in $E$, so that $T(E)$ would not be single-valued to begin with. A pole in $E$ is equally fatal for a different reason: it would make $V$ singular at some specific, spurious value of $E$ with no counterpart in the physical spectrum, corrupting the analyticity in $E$ on which the later identification with the TBA driving term depends.

To organise this order-by-order construction, we use the standard WKB
expansion of the ODE. Writing $S=\psi'/\psi$ turns
\eqref{eq:deformed-ODE} into the Riccati equation $S'+S^2=V$; to
expand it order by order we introduce a bookkeeping parameter $\hbar$
through $S\to\hbar S$,\footnote{Equivalently, $[-\hbar^2\partial_x^2+V]\psi=0$
with $S=\hbar\psi'/\psi$.} giving $\hbar S'+S^2=V$, and expand
$S=\sum_{n\ge0}\hbar^nS_n$. Collecting the terms order by order in $\hbar$ leads to a triangular system of equations:
\be
S_0^2=V,\qquad S_0'+2S_0S_1=0,\qquad S_1'+2S_0S_2+S_1^2=0,\qquad\dots,
\ee
so that $S_0=\sqrt V$ and, at every subsequent order,
\be 
S_n=-\frac{1}{2S_0}\big[S_{n-1}'+\sum_{i=1}^{n-1}S_iS_{n-i}\big] ~.
\ee
Each $S_n$ is a specific rational function of $V$ and its first $n$
derivatives, with the same branch of $\sqrt V$ throughout. Explicitly,
\be
S_1=-\frac{V'}{4V},\qquad
S_2=\frac{V''}{8V^{3/2}}-\frac{5(V')^2}{32V^{5/2}}.
\label{eq:S0S1S2}
\ee
Only the even-order $S_{2K}$ contribute to a period integral along a closed
cycle; the odd-order terms are total derivatives and their period integrals vanish. We then set $\hbar=1$, since
\eqref{eq:deformed-ODE} itself contains no such parameter, and call 
\be
\Pi(E,\kappa)=\oint_\gamma S\,dx=\sum_{K\ge0}\Pi_{2K}(E,\kappa)
\label{eq:total-period-general}
\ee
the Lee--Yang period. The first term in the expansion, $\Pi_0=\oint_\gamma\sqrt V\,dx$, we denote the
classical, or Bohr--Sommerfeld, period while the level-$K$ periods $\Pi_{2K} = \oint_\gamma S_{2K}\,dx$, $K\ge1$ will be called
the quantum periods; the whole series is the semiclassical, or WKB,
period.

Although $\hbar$ has been set to $1$, the hierarchy still organises into a
genuine expansion at large $E$.  Rescaling $x=E^{1/3}\xi$ turns
\eqref{eq:deformed-ODE} into
\be
\left[-\heff^2\frac{d^2}{d\xi^2}+v(\xi)\right]\psi=0,
\qquad
v(\xi)=\xi^3+a\xi-b,
\qquad
a=\frac{Q}{E^{2/3}},\quad b=\frac PE,
\label{eq:hbar-effective}
\ee
with $\heff=E^{-5/6}$.

The same $\hbar$ that organised the formal
Riccati expansion becomes, after this rescaling, the genuine small
parameter $\heff$ of the physical problem, and at fixed $(a,b)$ the
level-$K$ period $\Pi_{2K}$ is suppressed by $\heff^{2K}$ relative to
$\Pi_0$. This is what makes the hierarchy an asymptotic expansion at
large $E$, rather than just a bookkeeping device.

We first find the leading behaviour required of $Q$, by asking how
the classical period $\Pi_0$ must change, to first order in
$\kappa$, in order to supply the new term $\kappa_{\rm TBA}e^{5\theta}$ in \eqref{deformedperiod}.  Since $Q(E,0)=0$, the leading $\kappa$-dependent
correction to $\Pi_0$ comes from differentiating with respect to $Q$
at $Q=0$,
\be
\delta\Pi_0=\oint_\gamma\frac{x\,dx}{2\sqrt{x^3-E}}\,\delta Q~. 
\ee
Rescaling $x=E^{1/3}\xi$ as in \eqref{eq:WKB-scaling} shows the
integral itself scales as $E^{1/6}$. For $\delta\Pi_0$ to reproduce the new source
$\kappa_{\rm TBA}e^{5\theta}\sim\kappa E^{25/6}$, we therefore need
$\delta Q\sim E^{25/6-1/6}=E^4$ at order $\kappa^1$: the deformation
$Qx$ must begin as $\kappa E^4x$. This fixes the combination: 
\be
z:=\kappa E^{10/3}
\label{eq:z-def}
\ee
as the one that the leading correction depends on, since
$a=Q/E^{2/3}\sim\kappa E^{10/3}=z$; we shall find that $z$ organises
the whole construction, order by order in $\kappa$, at every WKB
level.
\subsubsection{A counting argument}
So far this has fixed only the first term of $Q$. We now show that the same single variable $z$
organises the entire construction: every term generated at every WKB level, to all orders in
$\kappa$, depends on $E$ and $\kappa$ only through $z$. The argument is a weight-counting
(dimensional-analysis) one, and it is worth stating before carrying it out: assign to $x,E,\kappa$
a formal scaling weight, chosen so that $z$ itself carries weight $0$; track how this weight
propagates through the Riccati recursion that builds each $S_n$; then check the result against the
known, independent scaling $\Pi_{2K}\sim E^{5/6}\heff^{2K}$. Demanding consistency between the two scalings will force
$a$ and $b$ to depend on $\kappa$ and $E$ through $z$ alone, level by level. 

Concretely, assign weights under $x\to\lambda^2x$, $E\to\lambda^6E$, $\kappa\to\lambda^{-20}\kappa$,
and write $[\,\cdot\,]$ for the resulting weight of any quantity, so that $[x]=2$, $[E]=6$,
$[\kappa]=-20$. These particular powers are fixed uniquely by the classical, $\kappa=0$ relation
$V_0=x^3-E$, which demands $[x^3]=[E]$, together with $[z]=[\kappa]+\frac{10}{3}[E]=0$.
This scaling assignment gives $[V_0]=6$. If the classical potential in the deformed case, $V_0=x^3+Q_0x-P_0$, is to scale uniformly with weight $6$, then one must also impose $[Q_0]=4$ and $[P_0]=6$.

Since $[V_0]=6$, and each $x$-derivative lowers the weight by $[d/dx]=-2$, we find
that $[V_0^{(j)}]=6-2j$. Equation~\eqref{eq:S0S1S2} then gives the weight of the first two WKB
densities directly: $[S_0]=\tfrac12[V_0]=3$, and
\be
[S_2]=[V_0'']-\tfrac32[V_0]=2-9=-7=[S_0]-10.
\ee
The same pattern continues at every order: writing the Riccati recursion as
\be 
S_n=-\frac{1}{2S_0}\bigl[S_{n-1}'+\sum_{i=1}^{n-1}S_iS_{n-i}\bigr]~,
\ee
and assuming inductively
that $[S_i]=3-5i$ for all $i<n$, every term on the right has the same weight,
\be
[S_{n-1}']=[S_{n-1}]-2=6-5n,\qquad [S_iS_{n-i}]=(3-5i)+(3-5(n-i))=6-5n\ \ \text{for every }i,
\ee
so that $[S_n]=(6-5n)-[S_0]=3-5n$, closing the induction. Restricting to even order,
\be
[S_{2K}]=3-10K,
\ee
and, since integrating over $x$ adds $[\oint dx]=2$,
\be
[\Pi_{2K}]=5-10K.
\ee
Now compare this to what we already know independently: $\Pi_{2K}\sim
E^{5/6}\heff^{2K}\times(\text{function of }a,b)$, and since $[E]=6$, the prefactor
$E^{5/6}\heff^{2K}$ itself carries weight exactly $5-10K$, regardless of the specific values of $a,b$, matching
$[\Pi_{2K}]$ precisely. This is the key consistency check: it means the
function of $(a,b)$ multiplying the prefactor must itself be weight-neutral. But $z$ is, by
construction, the unique weight-$0$ combination of $\kappa,E$, so a weight-neutral function of
$a,b$ can only depend on $\kappa,E$ through $z$. This is the precise
sense in which $z$ organises the whole construction: at every WKB level separately, $a$ and $b$
are functions of $z$ alone, not of $\kappa$ and $E$ independently.

Having established that $z$ is the right variable at every
level, the same weight-counting tool can be reused to determine which specific monomials $\kappa^nE^m$ can appear in $P$, and which monomials $\kappa^nE^mx$ in $Q$, at a given WKB level
$K$. This is the constraint that will actually let us solve for $P,Q$
order by order in the next two subsections.

A monomial \(\kappa^nE^m\) of \(P\) contributes to \(b=P/E\) of \eqref{eq:hbar-effective} as \(\kappa^nE^{m-1}=z^nE^{m-1-10n/3}.\)
If this term appears as a quantum correction to \(b\) at WKB level \(K\), then the remaining \(E\)-dependence must be precisely the characteristic suppression \(\heff^{2K}\) of that level. Since \(\heff^{2K}=E^{-5K/3}\), and
we require
\be
z^nE^{m-1-10n/3}=z^n\heff^{2K},
\ee
we find
\be\label{m eqn}
m-1-\frac{10n}{3}=-\frac{5K}{3}
\qquad\Longleftrightarrow\qquad
6m-20n=6-10K.
\ee 
The same argument for a monomial $\kappa^nE^mx$ of $Q$, contributing
to $a=Q/E^{2/3}$, gives
\be
6m+2-20n=6-10K.
\ee
At linear order ($n=1$, $K=0$) only the second equation has an integer
solution, $m=4$: this is the seed just found, and the corresponding
equation for $P$ has no polynomial solution, which is
the ODE-side reflection of the selection rule proved in the next
subsection. 
\subsubsection{Appearance of higher spins} 
It is convenient to repackage this counting in a single integer. A term
at WKB level $K$ that is homogeneous of degree $n$ in $z$ (carrying
$\heff^{2K}$ from its level and $z^n=\kappa^nE^{10n/3}$ from its own
$z$-dependence) scales, relative to the undeformed period
$\Pi_0\sim E^{5/6}$, as
\be
E^{5/6}\,\heff^{2K}\,E^{10n/3}=\left(E^{5/6}\right)^{r},
\qquad
r=1+4n-2K.
\label{eq:spin-formula}
\ee
We call $r$ the spin of the term: since $e^\theta=\cA E^{5/6}$,
a term of spin $r$ is the
term that would multiply $e^{r\theta}$ in the TBA driving term, so $r$ labels which physical coupling the term belongs to. Combining the spin with the two weight equations \eqref{m eqn}
above,
\be
m=\frac{5r+1}{6}\quad\hbox{for a monomial of }P,
\qquad
m=\frac{5r-1}{6}\quad\hbox{for a monomial of }Q,
\label{eq:monomial-spin-map}
\ee
so that a positive spin $r\equiv1\pmod6$ can be matched by exactly one
monomial of $P$ (for the given $n,K$), a positive spin $r\equiv5\pmod6$
by exactly one monomial of $Q$. For $r\equiv3\pmod6$ neither
exponent is an integer, so no monomial of either polynomial can ever
carry that spin. This last fact will reappear repeatedly: whenever the
construction below produces a term of spin $r\equiv3\pmod6$, it cannot
be absorbed into $P$ or $Q$ and must instead cancel on its own, which
is exactly the Lee--Yang selection rule, seen here from the ODE side.
The form of $P,Q$ thus derived has a further consequence, exploited in
Section~\ref{sec:rotation} to derive the $Y$-system of the deformed
spectral problem: it makes them covariant under the five-fold rotation
underlying the model's functional relations.
\subsection{The defining condition}
\label{sec:defining-condition}
The counting above lets us state the construction precisely, as a single condition on the WKB period: the spin-$5$ spectral problem is defined by
\be
\begin{aligned}
&\left[-\frac{d^2}{dx^2}+x^3+Q(E,\kappa)\,x-P(E,\kappa)\right]\psi=0,\\[4pt]
&\oint_\gamma S(x;E,\kappa)\,dx=e^{\theta}+\kappa_{\rm TBA}(\kappa)e^{5\theta}
+O\!\left(e^{-\theta}\right),\qquad e^{\theta}=D(\kappa)E^{5/6}.
\end{aligned}
\label{eq:defining-problem}
\ee
Here $\kappa$ is the ODE coupling, $\gamma$ is the cycle of
Section~\ref{sec:cubic}, continuously deformed away from $\kappa=0$,
$\kappa_{\rm TBA}(\kappa)$ is
the physical chemical potential, and $D(\kappa)$ is the deformed
generalisation of the period $\cA$ of \eqref{eq:E-theta-map}, to
which it reduces at $\kappa=0$. The two source sectors $r=1$ and
$r=5$ are not cancelled, but instead define $D(\kappa)$ and the
(generally nonlinear) map $\kappa\mapsto\kappa_{\rm TBA}(\kappa)$,
order by order in $\kappa$, with
the reparametrisation convention that the corresponding free potential
monomials are set to zero. Every other positive spin, by
\eqref{eq:monomial-spin-map}, is cancelled by a unique coefficient of
$P$ or $Q$; the negative-spin tail is left free and constitutes the
prediction of the construction. We solve this order by order below,
starting from the classical level $K=0$.

\subsection{The classical solution}
\label{sec:one-spin-classical}
\label{sec:spins}
\label{sec:classical}

Write $P=P_0(E,\kappa)+O(\heff^2)$, $Q=Q_0(E,\kappa)+O(\heff^2)$ for the
classical, $K=0$, part of the deformation. The leading WKB order of
\eqref{eq:defining-problem} involves only the classical period of the
deformed potential,
\be
\Pi_0(E,\kappa)=\oint_\gamma dx\,\sqrt{x^3+Q_0(E,\kappa)\,x-P_0(E,\kappa)}\,.
\label{eq:classical-period}
\ee
Rescaling $x=E^{1/3}\xi$ and introducing
\be
a=\frac{Q_0(E,\kappa)}{E^{2/3}},\qquad b=\frac{P_0(E,\kappa)}{E},
\label{eq:ab-def}
\ee
one has $\Pi_0=E^{5/6}\cP(a,b)$ with
\be
\cP(a,b)=\oint_\gamma d\xi\,\sqrt{\xi^3+a\xi-b}\,,
\label{eq:elliptic-period}
\ee
the period of the elliptic curve $y^2=\xi^3+a\xi-b$. Rescaling further
by $\xi=b^{1/3}s$ shows that the only non-trivial dependence is through
$u=a^3/b^2$,
\be
\cP(a,b)=b^{5/6}\,\cP\!\left(u^{1/3},1\right),\qquad u=\frac{a^3}{b^2}.
\label{eq:u-def}
\ee
Expanding $\cP(u^{1/3},1)$ around $u=0$ is a standard computation for
the periods of an elliptic curve (Appendix~\ref{app:periods} gives it
in full, via the Beta-function representation of
$\oint_\gamma\xi^n(\xi^3-1)^{1/2-n}d\xi$): only two of the three residue classes of
$n\bmod3$ survive, resumming into the two independent Picard--Fuchs
solutions. The result is written in terms of the hypergeometric functions:
\be
\begin{aligned}
&\cP(a,b)=b^{5/6}\left[\cA R_0(u)+\frac{\cB}{2}u^{1/3}R_1(u)\right],\\
\text{with}\quad &R_0(u)={}_2F_1\!\left(-\tfrac{5}{12},\tfrac{1}{12};\tfrac23;-\tfrac{4u}{27}\right),
\qquad
R_1(u)={}_2F_1\!\left(-\tfrac{1}{12},\tfrac{5}{12};\tfrac43;-\tfrac{4u}{27}\right),
\end{aligned}
\label{eq:R0R1}
\ee
where the two constants appearing as coefficients are the basic periods of the undeformed curve (see also Appendix~\ref{app:periods}): 
\be
\cA=\oint_\gamma\sqrt{\xi^3-1}\,d\xi= \frac{\sqrt3\,\Gamma(\tfrac13)\Gamma(\tfrac16)}{5\sqrt\pi},\qquad
\cB=\oint_\gamma\frac{\xi\,d\xi}{\sqrt{\xi^3-1}}=\frac{8\pi^{3/2}}{\Gamma(\tfrac13)\Gamma(\tfrac16)}~.
\label{eq:AB}
\ee
The discriminant locus of the elliptic curve $4a^3+27b^2=0$ corresponds
to $u=-27/4$, or equivalently to unit hypergeometric argument $-4u/27=1$,
which is the other singular point of the hypergeometric equation besides $u=0$.
In the rest of this section it is convenient to measure the period in
units of its undeformed value, $\cP\to\cP/\mathcal A$, so that $\cP(0,1)=1$ and
the second class in \eqref{eq:R0R1} carries the ratio
$\mathcal B/2\mathcal A$; the factor $\mathcal A$ is restored in
Section~\ref{sec:classical-alone}.

Recall from \eqref{eq:z-def} that $z=\kappa E^{10/3}$, and that
$e^\theta=D(\kappa)E^{5/6}$: the classical period $\Pi_0=E^{5/6}\cP(a,b)$
is the leading, largest term of the full WKB period at large $E$, and
matching it to the two terms of \eqref{eq:defining-problem} requires
\be
\cP(a,b)=1+c_1z,
\label{eq:period-linear-condition}
\ee
for some constant $c_1$, which encodes the classical values of the
normalisations $D(\kappa)$ and $\kappa_{\rm TBA}(\kappa)$: no power of $z$ beyond the first may appear in the evaluation of the period integral appearing on the left. This is one equation, but $a$ and $b$ (equivalently $Q_0(E,\kappa)$ and
$P_0(E,\kappa)$) are two unknown functions of $\kappa$, so
\eqref{eq:period-linear-condition} alone does not determine them; an
extra input is needed to turn it into two equations.
That input is that $P_0,Q_0$ are polynomials in $E$ at every order in
$\kappa$: an order-$\kappa^n$ term has spin $r=1+4n$ by
\eqref{eq:spin-formula}, and by \eqref{eq:monomial-spin-map} only
$r\equiv1\pmod6$ ($n\equiv0\pmod3$) can be a term of $P_0$ and only
$r\equiv5\pmod6$ ($n\equiv1\pmod3$) a term of $Q_0$; the third class,
$n\equiv2\pmod3$, i.e.\ $r\equiv3\pmod6$, contributes nothing, since its
integrals along the cycle $\gamma$ vanish identically. This fact recurs
at every WKB level, and is proved once, for all levels together, in
Appendix~\ref{app:periods}. Consequently
\be
\frac{P_0(E,\kappa)}{E}=F_0(z^3),\qquad
\frac{Q_0(E,\kappa)}{\kappa E^4}=G_0(z^3),
\label{eq:FG-structure}
\ee
with $F_0(0)=G_0(0)=1$: that is, $b=F_0(z^3)$ is a series in the
single variable $z^3$, and, since $Q_0=\kappa E^4G_0(z^3)$ gives
$a=Q_0/E^{2/3}=zG_0(z^3)$, so is $a/z$; consequently
$u=a^3/b^2$ is a series in $z^3$ as well. So $a(z)$ and $b(z)$, which
in principle could have been arbitrary functions of $z$, are pinned
down to this restricted two-function form.

This is exactly what is needed to turn \eqref{eq:period-linear-condition}
into two equations. Substituting $b=b(z^3)$, $u=u(z^3)$ into
\eqref{eq:R0R1},
\be
\cP(a,b)=\underbrace{b^{5/6}R_0(u)}_{\text{a series in }z^0,z^3,z^6,\dots}
\ +\ \underbrace{b^{5/6}u^{1/3}R_1(u)}_{z\ \times\ (\text{a series in }z^0,z^3,z^6,\dots)},
\ee
so the left-hand side of \eqref{eq:period-linear-condition} splits into
two independent series, one running over $z^{3m}$ and the other over
$z^{3m+1}$. The right-hand side, $1+c_1z$, has exactly one term in each
family, $1$ and $c_1z$, and none at all in $z^2,z^5,\dots$ (consistent
with the vanishing theorem of Appendix~\ref{app:periods}) or beyond $z^3,z^4$
respectively. Equating the two series term by term is therefore
equivalent to the two separate conditions
\be
b^{5/6}R_0(u)=1,\qquad b^{5/6}u^{1/3}R_1(u)=c_1z,
\label{eq:flat-conditions}
\ee
one equation for each of the two independent hypergeometric series,
which is how a single scalar condition on $\cP(a,b)$ determines the two
unknown functions $a(z)$ and $b(z)$.

The first equation gives $b$
directly as a function of $u$. In the second, $c_1$ is not free: $\kappa$
is already normalised by the linear seed $\kappa E^4$ in $Q$, i.e.\
$a=z+O(z^4)$ and hence $u^{1/3}=z+O(z^4)$, so comparing the two sides at
leading order fixes $c_1=\cB/2\cA$, and dividing the second equation by
the first then gives $z$ as a function of $u$ with no free constant.
Thus, the solution of the classical deformation is given by 
\be
\frac{P_0(E,\kappa)}{E} =R_0(u)^{-6/5},\qquad
\frac{Q_0(E,\kappa)}{\kappa E^4}=\frac{R_0(u)^{1/5}}{R_1(u)}~,\\
\ee
with
\be
z:=\kappa E^{10/3} =u^{1/3}\frac{R_1(u)}{R_0(u)} .
\label{eq:P0Q0-exact}
\ee
The last relation determines $u$ as a function of $z=\kappa E^{10/3}$,
and the first two then give $P_0$ and $Q_0$ in closed form: thus, the
classical deformation, to all orders in $\kappa$, is the inverse of an
elementary hypergeometric period map.

Expanding around $z=0$, with $u=z^3+\frac{5}{432}z^6-\frac{17}{435456}z^9
+\dots$ (the series inversion is given in Appendix~\ref{app:periods}),
\begin{align}
\frac{P_0}{E}&=1-\frac{z^3}{108}+\frac{z^6}{4320}
-\frac{281}{22044960}z^9+\frac{121523}{130947062400}z^{12}+O(z^{15}),
\label{eq:P0-expansion}\\
\frac{Q_0}{\kappa E^4}&=1-\frac{z^3}{432}+\frac{19}{204120}z^6
-\frac{131}{22044960}z^9+\frac{198071}{425577952800}z^{12}+O(z^{15}),
\label{eq:Q0-expansion}
\end{align}
that is,
\be
P_0=E-\frac{1}{108}\kappa^3E^{11}+O(\kappa^6),\qquad
Q_0=\kappa E^4-\frac{1}{432}\kappa^4E^{14}+O(\kappa^7).
\label{eq:P0Q0-kappa}
\ee
All coefficients are rational, and there is no classical term at order
$\kappa^2$, as expected from \eqref{eq:FG-structure}: the first
correction at that order appears only at the first quantum level,
which we compute next.
\subsection{The first quantum level: a first run}
\label{sec:quantum}
We solve the defining condition one WKB level at a time, exactly as at
the classical level: at level $K=1$ (order $\heff^2$) the unknowns are
the two functions $F_1,G_1$ collecting the monomials of $P,Q$ allowed
at that level, and they are fixed by requiring that the period, now
including the first quantum correction, still reproduces only
$e^\theta+\kappa_{\rm TBA}(\kappa)e^{5\theta}$. We first derive the
general formula for the first quantum period.
\subsubsection{The first quantum period}
\label{sec:first-quantum}
Recall $S_0,S_1,S_2$ from \eqref{eq:S0S1S2} and $\Pi_0,\Pi_2$ from
\eqref{eq:total-period-general}, valid for the potential $V$ of
\eqref{eq:deformed-ODE}; how corrections to $V$ at successive orders
in $\heff^2$ combine with $S_2$ itself is made precise in
Section~\ref{sec:hbar2-level}. For now, we keep $V$ implicit, and at second order we have
\be 
\Pi_2=\oint_\gamma S_2\,dx~.
\ee 
Since $d(V'/V^{3/2})=[V''/V^{3/2}-\frac32(V')^2/V^{5/2}]dx$ integrates to zero
on the closed cycle, the two terms of $S_2$ are proportional, and we obtain 
\be
\Pi_2=\left(\frac18-\frac{5}{32}\cdot\frac23\right)\oint_\gamma dx\,\frac{V''}{V^{3/2}}
=\frac{1}{48}\oint_\gamma dx\,\frac{V''}{V^{3/2}} .
\label{eq:Pi2-reduced}
\ee
\subsubsection{Order \texorpdfstring{$\kappa^2$}{kappa squared}}
\label{sec:order-k2}
To find the first quantum correction to the deformation it is enough to
insert the leading classical potential $V=x^3-E+\kappa E^4x+O(\kappa^3)$
into \eqref{eq:Pi2-reduced}. With $x=E^{1/3}\xi$, $V=E(\xi^3-1+z\xi)$,
$V''=6E^{1/3}\xi$, $dx=E^{1/3}d\xi$,
\be
\Pi_2=\frac18\,E^{-5/6}\oint_\gamma\frac{\xi\,d\xi}{(\xi^3-1+z\xi)^{3/2}},
\label{eq:Pi2-z}
\ee
and, expanding in $z$,
\be
(\xi^3-1+z\xi)^{-3/2}=(\xi^3-1)^{-3/2}
\left[1-\frac32\frac{z\xi}{\xi^3-1}+\frac{15}{8}\frac{z^2\xi^2}{(\xi^3-1)^2}+O(z^3)\right].
\label{eq:Pi2-expansion}
\ee
The $z^0$ term is the undeformed first quantum period, which encodes
$I_1$. The $z^1$ term has spin $3$; its integrand is
$\xi^2f(\xi^3)d\xi$ and its period vanishes by the same argument that
eliminated the class $n\equiv2$ in \eqref{eq:In-beta}. The $z^2$ term is
\be
\Pi_2^{(2)}=\frac{15}{64}\,z^2E^{-5/6}\oint_\gamma\frac{\xi^3\,d\xi}{(\xi^3-1)^{7/2}} ~.
\label{eq:Pi2-order2}
\ee
The coefficient of the integral, written in terms of $\kappa$, takes the form
\be 
\qquad
z^2E^{-5/6}=\kappa^2E^{35/6}=\kappa^2\left(E^{5/6}\right)^7~.
\ee 
Thus it is a source of spin $7$ at order $\kappa^2$. We note that this is consistent with the general counting argument done previously: $7=1+4\cdot2-2\cdot1$ (order $\kappa^2$, level $K=1$, by \eqref{eq:spin-formula}). Thus, without a compensating term,
the deformed problem would describe a GGE with a chemical potential $\propto\kappa^2$ for $I_7$. The integrand is
$\xi^0\cdot\xi^3f(\xi^3)d\xi$, in the class $r\equiv1\pmod6$, which is
one of the two non-vanishing classes identified in
Section~\ref{sec:one-spin-classical}: unlike the $z^1$ term above, this
period is generically nonzero, and it must therefore be cancelled by a
correction to the spectral problem. That is, we need to add a quantum correction to the potential. 

The weight counting fixes the form of this correction: at order $\kappa^2$ the only
monomial with the weight of a first-level quantum correction is
$\kappa^2E^6$ in $P$. We set
\be
P(E,\kappa)=P_0(E,\kappa)+p_6\,\kappa^2E^6+O(\kappa^3),
\qquad \delta V=-p_6\,\kappa^2E^6 .
\label{eq:P-quantum}
\ee
Such a counterterm changes the \emph{classical} period, by
$\delta\Pi_0=\oint\delta\sqrt V\,dx=\frac12\oint\delta V/\sqrt V\,dx$
with the same branch as before,
\be
\delta\Pi_0=-\frac{p_6}{2}\,\kappa^2E^6\,E^{\frac13-\frac12}
\oint_\gamma\frac{d\xi}{(\xi^3-1)^{1/2}}
=-\frac{p_6}{2}\,z^2E^{-5/6}\oint_\gamma\frac{d\xi}{(\xi^3-1)^{1/2}} .
\label{eq:deltaPi0}
\ee
This has the same $E$-dependence as \eqref{eq:Pi2-order2}, as it must: a
term of the potential suppressed by $\heff^2$ contributes to the
classical period at the same order as the classical potential
contributes to the first quantum period. The classical and quantum contributions must therefore be combined:
the cancellation condition applies to the total coefficient of
$\kappa^2E^{35/6}$. The condition $\Pi_2^{(2)}+\delta\Pi_0=0$ gives
\be
p_6=\frac{15}{32}\,
\frac{\oint_\gamma\xi^3(\xi^3-1)^{-7/2}\,d\xi}{\oint_\gamma(\xi^3-1)^{-1/2}\,d\xi}.
\label{eq:p6-ratio}
\ee
Defining $J(k,s):=\oint_\gamma\xi^k(\xi^3-1)^s d\xi$, the reduction
identities of Appendix~\ref{app:periods} give
\be
J(3,-\frac72)/J(0,-\frac12)=14/405\,,
\ee
and therefore
\be
p_6=\frac{7}{432}.
\label{eq:period-ratio}
\ee
The same ratio can be obtained by explicit evaluation of each integral in terms of the Euler Beta function (see 
Appendix~\ref{app:periods}). 
The deformed spectral problem
therefore begins as
\be
V(x;E,\kappa)=x^3+Q_0(E,\kappa)\,x-P_0(E,\kappa)-\frac{7}{432}\,\kappa^2E^6+\cdots
\label{eq:quantum-completed-potential}
\ee
The coefficient is not an independent normalisation: it is the unique
value for which the first quantum period does not introduce a spin-$7$
source. Note that the counterterm also contributes to $\Pi_2$ itself at
order $\kappa^2$, through the change of $V^{3/2}$; by
\eqref{eq:spin-formula} that contribution has spin $5$ and is part of
the renormalisation of the coefficient of $\kappa e^{5\theta}$, one of
the reparametrisations of Section~\ref{sec:defining-condition}.
\subsection{The  first quantum level: a Picard--Fuchs reduction}
\label{sec:hbar2-level}
The computation of the previous subsection can be repeated order by
order in $\kappa$: at order $\kappa^3$ the first quantum period produces
a spin-$11$ source, to be cancelled by the monomial $\kappa^3E^9x$ in $Q$,
and so on. Each step requires the expansion of $\Pi_2$ to one more order
in $z$ and one more period ratio. There is, however, a shortcut which
solves the whole first quantum level at once, and which we now explain.

The classical period of the rescaled potential $v=\xi^3+a\xi-b$ is a
function of the two moduli,
\be
\cP(a,b)=\oint_\gamma\sqrt{v}\,d\xi .
\ee
Differentiating under the integral sign with respect to the moduli
lowers the power of $\sqrt v$ in the integrand and brings down
polynomials in $\xi$:
\be
\partial_b\sqrt{v}=-\frac{1}{2\sqrt v},\qquad
\partial_a\sqrt{v}=\frac{\xi}{2\sqrt v},\qquad
\partial_a\partial_b\sqrt{v}=\frac{\xi}{4\,v^{3/2}},\qquad
\partial_b^2\sqrt v=-\frac{1}{4\,v^{3/2}},\qquad\dots
\label{eq:PF-derivatives}
\ee
Conversely, any integrand of the form $\xi^k v^{-n-1/2}$ with $k\le n$
can be written as a combination of derivatives of $\sqrt v$ with respect
to $a$ and $b$, and integrands with higher powers of $\xi$ are reduced
to these by integration by parts, exactly as in
Section~\ref{sec:order-k2}. This is the content of the Picard--Fuchs
structure of the elliptic curve: all the periods of all the differentials
$\xi^kv^{s}d\xi$ are generated by $\cP$ and its derivatives with respect
to the moduli. 

What this essentially means is that the quantum period can be written in closed form as a derivative of the classical period: 
\be
\Pi_2(E,\kappa)=\frac18E^{-5/6}\oint_\gamma\frac{\xi\,d\xi}{(\xi^3+a\xi-b)^{3/2}}
=E^{5/6}\,\heff^2\;\frac12\,\partial_a\partial_b\cP(a,b),
\label{eq:Pi2-picard-fuchs}
\ee
where the factor $E^{-5/6}=E^{5/6}\heff^2$ is the expected suppression of
the first quantum period. The first quantum period of the deformed curve
therefore inherits its
hypergeometric representation with no further integration.\footnote{The
same branch of $\sqrt v$ is used in both periods, as required by the
Riccati recursion; expanding \eqref{eq:Pi2-picard-fuchs} at order $a^2$
and comparing with \eqref{eq:Pi2-order2} reproduces the period ratio
\eqref{eq:period-ratio} as a check.} 
The same mechanism applies to the higher WKB
coefficients: $S_{2k}$ is a sum of terms with derivatives of $V$ in the
numerator and $V^{(2k-1)/2+r}$ in the denominator, and for the cubic the
derivatives are $V'$, $V''$ and the constant $V'''$, so that every
$\Pi_{2k}$ is a differential operator in $(a,b)$ acting on $\cP$
(see Appendix~\ref{app:wkb} for details), in line with the Picard--Fuchs treatment of
quantum WKB periods developed in~\cite{Ito:2021hdy}. We shall use this in
Section~\ref{sec:energy-wkb}.

The goal is to find the order-$\heff^2$ corrections to $Q,P$ by the
same method as at the classical level: write down what the defining
condition requires at this order, and solve it. We write the
deformation through the first quantum level as
\be
\frac{P}{E}=b_0(z)+\heff^2\,b_1(z)+\dots,\qquad
\frac{Q}{E^{2/3}}=a_0(z)+\heff^2\,a_1(z)+\dots,
\label{eq:ab-hbar-expansion}
\ee
where $(a_0,b_0)$ is the classical solution \eqref{eq:P0Q0-exact}
and $a_1,b_1$ collect the monomials allowed at level $\heff^2$, namely
$\kappa^{3m+2}E^{6+10m}$ in $P$ and $\kappa^{3m+3}E^{9+10m}x$ in $Q$
(Section~\ref{sec:family}). In terms of $z$ these are
\be
b_1(z)=z^2F_1(z^3),\qquad a_1(z)=z^3G_1(z^3),
\ee
with $F_1(0)=p_6$ the counterterm already found and $G_1(0)$ the
coefficient of $\kappa^3E^9x$; $F_1,G_1$ are the new unknowns.

At order $\heff^2$, the full period $\cP(a,b)$ (we drop the overall
factor $E^{5/6}$, as at the classical level) receives contributions
from two sources: the shift of its arguments from $(a_0,b_0)$ to
$(a_0+\heff^2a_1,\,b_0+\heff^2b_1)$, and the intrinsic first quantum
period $\frac12\partial_a\partial_b\cP$ of
\eqref{eq:Pi2-picard-fuchs}, evaluated on the classical curve. Expanding
to first order in $\heff^2$,
\be
\begin{aligned}
&\cP(a_0+\heff^2a_1,\,b_0+\heff^2b_1)+\heff^2\cdot\tfrac12\partial_a\partial_b\cP(a_0,b_0)\\
&\quad=\cP(a_0,b_0)+\heff^2\Big[a_1\partial_a\cP+b_1\partial_b\cP+\tfrac12\partial_a\partial_b\cP\Big]_{(a_0,b_0)}+O(\heff^4).
\end{aligned}
\label{eq:hbar2-expansion}
\ee
This is the same interplay between a classical and a quantum period
that fixed $p_6$ in Section~\ref{sec:order-k2}, now kept to all orders
in $z$: a shift of the classical solution changes the classical
period, and this change must cancel the intrinsic quantum period. The
defining condition of Section~\ref{sec:defining-condition} requires
that, at order $\heff^2$, no positive power of $z$ survive: since
$\cP(a_0,b_0)=1+c_1z$ already saturates the classical order, the
bracket in \eqref{eq:hbar2-expansion} must reduce to its own $z^0$ part,
\be
\left.a_1\,\partial_a\cP+b_1\,\partial_b\cP+\frac12\,\partial_a\partial_b\cP\,\right|_{(a_0(z),b_0(z))}
=\left.\frac12\,\partial_a\partial_b\cP\,\right|_{(0,1)} .
\label{eq:hbar2-constraint}
\ee
Everything in this equation is known except $a_1$ and $b_1$: the
derivatives of $\cP$ are explicit hypergeometric expressions, and
$(a_0,b_0)$ is the classical inverse period map. The equation is linear
in the unknowns, because the shift of the moduli enters only to first
order. It is one equation for two functions, but, as at the classical
level, it separates.
\subsubsection{Separation of the two classes}
Split the period into its two hypergeometric classes,
\be
\cP=A+B,\qquad A=b^{5/6}R_0(u),\qquad B=c_1\,a\,b^{1/6}R_1(u),\qquad u=\frac{a^3}{b^2},
\ee
where $c_1=\cB/2\cA$ is the coefficient of the linear term in
$\cP(a_0,b_0)=1+c_1z$; the undeformed value $\cP(0,1)=1$ is already
fixed by this same normalisation, \eqref{eq:period-linear-condition}.
Exactly as at the classical level (Section~\ref{sec:one-spin-classical}),
$A$ is a series in $z^{3m}$ and $B$ in $z^{3m+1}$, with the third,
grade-$1$ class (spin $\equiv3\pmod6$) vanishing identically; the two
terms of \eqref{eq:hbar2-constraint} can therefore be matched
independently, giving
\be
a_1\,\partial_aA+b_1\,\partial_bA=-\tfrac12\,\partial_a\partial_bA,
\qquad
a_1\,\partial_aB+b_1\,\partial_bB=\tfrac{c_1}{12}-\tfrac12\,\partial_a\partial_bB,
\label{eq:hbar2-split}
\ee
one equation from each class. Every term of the second equation is
proportional to $c_1$, which drops out, leaving two linear equations
for the two unknowns $a_1,b_1$.

Differentiating $A,B$ with respect to $a,b$ is elementary
(Appendix~\ref{app:periods} gives the six derivatives in full); the
system \eqref{eq:hbar2-split} is then linear and algebraic, and we obtain
\be
a_1=\frac{r_A\,\partial_bB-r_B\,\partial_bA}{D},\qquad
b_1=\frac{r_B\,\partial_aA-r_A\,\partial_aB}{D},\qquad
D=\partial_aA\,\partial_bB-\partial_bA\,\partial_aB,
\label{eq:hbar2-solution}
\ee
with $r_A=-\frac12\partial_a\partial_bA$, $r_B=\frac{c_1}{12}-\frac12
\partial_a\partial_bB$, everything evaluated on the classical solution
$a_0=zR_0^{1/5}/R_1$, $b_0=R_0^{-6/5}$, $z^3=uR_1^3/R_0^3$. The first
quantum level of the deformation is thereby expressed in closed form
through $R_0$, $R_1$ and their first two derivatives at the inverse
period map. No new transcendental function appears: the same two
hypergeometric functions that describe the classical deformation also
describe its first quantum correction.
\subsubsection{Leading coefficients and the series expansion}
The first terms can be read off by hand and provide a check. At order
$z^2$ only the grade-$2$ equation contributes, with $\partial_bA\to\frac56$
and $\partial_a\partial_bA\to3a^2\cdot(-\frac76)R_0'(0)=-\frac72\cdot\frac{5}{648}a^2$,
so that 
\be
F_1(0)=p_6=\frac{7}{432},
\ee
which reproduces \eqref{eq:period-ratio}, sign included, from the single
coefficient $\frac{5}{648}$ of $R_0$ and without any period ratio. At
order $z^3$ the grade-$0$ equation gives, after expanding
$b_0^{-5/6}=R_0=1+\frac5{648}u+\dots$ and $R_1=1+\frac{5}{1296}u+\dots$,
\be
G_1(0)=-\frac{F_1(0)}{6}+\frac12\left[\frac{15}{2}\cdot\frac{5}{1296}
-\frac16\cdot\frac{5}{1296}-\frac16\cdot\frac{5}{648}\right]
=\frac{7}{648}\,,
\label{eq:q9-result}
\ee
that is, the counterterm that cancels the spin-$11$ source at order
$\kappa^3$ is
\be
Q(E,\kappa)=Q_0(E,\kappa)+\frac{7}{648}\,\kappa^3E^9+\cdots .
\label{eq:Q-quantum}
\ee
Expanding
\eqref{eq:hbar2-solution} to higher orders, we obtain the series expansions:  
\begin{align}
F_1(z^3)&=\frac{7}{432}-\frac{553}{233280}\,z^3+\frac{3277}{10077696}\,z^6
-\frac{6606293}{149653785600}\,z^9+O(z^{12}),
\label{eq:F1-series}\\
G_1(z^3)&=\frac{7}{648}-\frac{2107}{1399680}\,z^3+\frac{1033}{5038848}\,z^6
-\frac{6231127}{224480678400}\,z^9+O(z^{12}),
\label{eq:G1-series}
\end{align}
with rational coefficients throughout, as expected from the structure
of the computation. All the ingredients are ratios of Gamma functions
with rational arguments, evaluated on a series with rational
coefficients. Through the first quantum level, the solution of the deformed spectral problem is given by: 
\begin{align}
P(E,\kappa)&=E\left[F_0(z^3)+\kappa^2E^5F_1(z^3)+O(\heff^4)\right]
\nonumber\\
&=E-\frac{\kappa^3E^{11}}{108}+\dots+\frac{7}{432}\kappa^2E^6-\frac{553}{233280}\kappa^5E^{16}+\dots,
\label{eq:P-through-hbar2}\\
Q(E,\kappa)&=\kappa E^4\left[G_0(z^3)+\kappa^2E^5G_1(z^3)\right]+E^{2/3}O(\heff^4)
\nonumber\\
&=\kappa E^4-\frac{\kappa^4E^{14}}{432}+\dots+\frac{7}{648}\kappa^3E^9-\frac{2107}{1399680}\kappa^6E^{19}+\dots,
\label{eq:Q-through-hbar2}
\end{align}
with $F_0=P_0/E$, $G_0=Q_0/(\kappa E^4)$ as in
\eqref{eq:P0-expansion}--\eqref{eq:Q0-expansion}. In words: the
classical deformation is the inverse period map of the curve, and the
first quantum correction is obtained from the same map by
differentiation and the solution of a $2\times2$ linear system. Nothing
prevents the same procedure from being carried out at the next levels,
where there are again two unknown functions per level and the source terms
are higher derivatives of $\cP$. 

\subsubsection{Summary}
\label{sec:classical-alone}

It is worth collecting the classical and first-quantum results in one
place, in the same language, to see directly that the construction
does what it was designed to do: reproduce
$e^\theta+\kappa_{\rm TBA}(\kappa)e^{5\theta}$ and nothing else at
positive powers of $E^{5/6}$. Restoring the factor $\cA$ removed in
Section~\ref{sec:classical}, the classical period is
\be
\Pi_0=E^{5/6}\Big[\cA\,b_0^{5/6}R_0(u)+\tfrac12\cB\,b_0^{5/6}u^{1/3}R_1(u)\Big]
=E^{5/6}\big(\cA+\tfrac12\cB\,z\big),
\ee
the last equality being exactly the defining condition
\eqref{eq:period-linear-condition} at the classical level, with
$c_1=\cB/2\cA$. By construction, $\Pi_0$ already contains only the two
allowed powers $E^{5/6}$ and $E^{25/6}$.

The first quantum period, evaluated not on the classical potential
alone but on the corrected one, $(a,b)=(a_0+\heff^2a_1,\,b_0+\heff^2b_1)$,
is fixed by \eqref{eq:hbar2-constraint} to be exactly
\be
\heff^2\,\Pi_2\Big|_{\rm corrected}=E^{5/6}\heff^2\left[a_1\partial_a\cP+b_1\partial_b\cP+\tfrac12\partial_a\partial_b\cP\right]_{(a_0,b_0)}
=E^{5/6}\heff^2\cdot\tfrac12\,\partial_a\partial_b\cP(0,1).
\ee
Every positive power of $z$ that the naive (uncorrected) first
quantum period would have produced (at order $\kappa^2$ the source of
spin $7$ found in \eqref{eq:Pi2-order2}, and at order $\kappa^3$ the
spin-$11$ source found in Section~\ref{sec:hbar2-level}) has been
cancelled by construction, exactly as \eqref{eq:hbar2-constraint}
requires. Adding
the two,
\be
\Pi_0+\heff^2\Pi_2\Big|_{\rm corrected}
=E^{5/6}\big(\cA+\tfrac12\cB\,z\big)+\heff^2\,\tfrac{1}{24}\cB\,E^{5/6}+O(\heff^4),
\ee
so that through this order the full WKB period contains only
$e^\theta$, $\kappa_{\rm TBA}(\kappa)e^{5\theta}$, and a term of order
$\heff^2E^{5/6}\sim E^{-5/6}\sim e^{-\theta}$: exactly the
$O(e^{-\theta})$ tail that \eqref{eq:defining-problem} leaves free. 
This term carries the first quantum correction to the ground-state energy
(we shall have a more detailed discussion of this point in Section~\ref{sec:energy-wkb}). The asymptotic form required by the
defining condition is maintained at this order, by exactly the same mechanism (cancellation between a classical
and a quantum contribution) that will maintain it at every subsequent order.
\section{Functional relations and the correspondence}
\label{sec:functional}
The defining condition in \eqref{eq:defining-problem} matches the asymptotics of the
quantum period to the TBA driving term. The standard ODE/IM correspondence identifies the quantum period with a Stokes multiplier, which in turn satisfies the same
functional relation as the TBA $Y$-function. The functional relation, together with specified asymptotics and analyticity in a strip determines the function uniquely.
Hence the two objects coincide, and the local charges follow. In this section, we derive the functional relation for the deformed problem and state the additional conditions needed to identify it with the GGE.
\subsection{The Sibuya rotation acts with \texorpdfstring{$\kappa$}{kappa} fixed}
\label{sec:rotation}
The form of $P,Q$ established in Section~\ref{sec:family} makes the
deformed spectral problem covariant under a five-fold discrete
rotation, which we now use to derive its functional relations. Let
$\omega=e^{2\pi i/5}$: for
$\psi(\omega^{-1}x;\omega^2E,\kappa)$ to solve the same equation as
$\psi(x;E,\kappa)$, the chain rule requires
\be
P(\omega^2E,\kappa)=\omega^2P(E,\kappa),\qquad
Q(\omega^2E,\kappa)=\omega^3Q(E,\kappa).
\ee
Since $\omega^{5}=1$, a monomial $E^m$ in $P$ is compatible with these
relations if and only if $m\equiv1\pmod5$, and a monomial in $Q$ if and
only if $m\equiv4\pmod5$. These are precisely the powers allowed by
\eqref{eq:monomial-spin-map}: $m=1+5t$ for $r=1+6t$ in $P$, and
$m=4+5t$ for $r=5+6t$ in $Q$. The rotation therefore acts on the
spin-$5$ spectral problem with $\kappa$ held fixed: the spectral
problems compatible with a Lee--Yang GGE are exactly those compatible
with the Lee--Yang functional relations. (When only the spin-$5$
coupling is nonzero, the rotation sends $z\to e^{2\pi i/3}z$ and
leaves $z^3$ invariant, so the discrete covariance holds order by
order in $\kappa$ and at each WKB order.)

Let $y_0(x;E,\kappa)$ be the solution subdominant along the positive real
axis, normalised by its WKB asymptotics, and define
\be
y_k(x;E,\kappa)=\omega^{k/2}\,y_0\!\left(\omega^{-k}x;\,\omega^{2k}E,\,\kappa\right).
\label{eq:sibuya-rotation}
\ee
By the covariance just established, $y_k$ solves the same equation
with the same $\kappa$, and is subdominant in the $k$-th sector. The
adjacent Wronskians
$W[y_k,y_{k+1}]=y_ky_{k+1}'-y_k'y_{k+1}$ are independent of $k$; a common constant normalisation of $y_0$, together with the
phase $\omega^{k/2}$, is chosen so that
\be
W[y_k,y_{k+1}]=1 .
\label{eq:adjacent-wronskian}
\ee
Since $\omega^{5/2}=-1$ and $\omega^{5}=1$,
\be
y_{k+5}=-y_k .
\label{eq:y-periodicity}
\ee
\subsection{The \texorpdfstring{$Y$}{Y}-system}
\label{sec:ode-ysystem}
Expanding $y_{-1}$ in the basis $\{y_0,y_1\}$ gives the Stokes relation
\be
T(E,\kappa)\,y_0=y_{-1}+y_1,\qquad T(E,\kappa)=W[y_{-1},y_1],
\label{eq:stokes-T}
\ee
and rotating $k$ times, $W[y_{k-1},y_{k+1}]=T(\omega^{2k}E,\kappa)$. The
Pl\"ucker identity for four solutions,
\be
W[y_{-1},y_1]\, W[y_0,y_2]=W[y_{-1},y_0]\, W[y_1,y_2]+W[y_{-1},y_2]\, W[y_0,y_1],
\ee
has left-hand side given by $T(E)T(\omega^2E)$, the first term on the right equal to
$1$, and, by \eqref{eq:y-periodicity}, $W[y_{-1},y_2]=-W[y_4,y_2]=W[y_2,y_4]
=T(\omega E)$. Hence
\be
T(E,\kappa)\,T(\omega^2E,\kappa)=1+T(\omega E,\kappa).
\label{eq:T-functional-E}
\ee
In the rapidity variable \eqref{eq:E-theta-map}, $E\to\omega^2E$ is $\theta\to
\theta+\frac{2\pi i}{3}$ and $E\to\omega^6E=\omega E$ is $\theta\to
\theta+\frac{i\pi}{3}$. Shifting $\theta\to\theta-\frac{i\pi}{3}$ and
setting $Y^{\rm ODE}(\theta;\kappa)=T(E(\theta),\kappa)$, we obtain 
\be
Y^{\rm ODE}\!\left(\theta+\tfrac{i\pi}{3};\kappa\right)
Y^{\rm ODE}\!\left(\theta-\tfrac{i\pi}{3};\kappa\right)
=1+Y^{\rm ODE}(\theta;\kappa),
\label{eq:ODE-Ysystem}
\ee
which is precisely the Lee--Yang $Y$-system \eqref{eq:GGEYsystem}. The derivation uses only
the rotation \eqref{eq:sibuya-rotation} with $\kappa$ fixed; it holds for
any $P$, $Q$ of the allowed form, whether or not
\eqref{eq:defining-problem} has been imposed. This is the ODE counterpart
of the observation, made in Section~\ref{sec:lee-yang}, that the
$Y$-system is insensitive to the driving term.
\subsection{From the \texorpdfstring{$Y$}{Y}-system to the GGE}
\label{sec:ode-to-tba}
The functional relation alone does not single out the physical solution
\cite{Zamolodchikov:1991et,Ravanini:1992fi,Kuniba:2010ir}. Fourier
transforming its logarithm requires shifting the contours of
$\log Y(\theta\pm i\pi/3)$ to the real axis, so $Y$ must be analytic and
nonvanishing in the strip $|\mathrm{Im}\,\theta|\le\pi/3$, with
$\log Y-\nu$ bounded there, where $\nu(\theta)$ is the driving term
encoding the asymptotics. This yields the ground-state
TBA~\cite{Zamolodchikov:1989cf,Klassen:1990dx}, unique within this class.
Excited states have additional zeros of $Y$ in the
strip~\cite{Martins:1991zy,Fendley:1991bj,Bazhanov:1996aq,Dorey:1996re,Dorey:1997rb,Bajnok:2014fca},
which generate source terms at positions fixed self-consistently by
$1+Y(\theta_j)=0$, leaving $\nu$ unchanged; a complementary approach is
based on nonlinear integral equations~\cite{Fioravanti:1996rz,Lencses:2014cba}.

Three of these ingredients are already available here: the functional
relation~\eqref{eq:ODE-Ysystem}; the driving term, fixed by the defining
condition~\eqref{eq:defining-problem} to be
$\nu(\theta)=e^{\theta}+\kappa_{\rm TBA}(\kappa)\,e^{5\theta}$, or more
generally~\eqref{eq:intro-driving-general} for the ensembles of
Section~\ref{sec:multicoupling}; and the boundary value. Every allowed
monomial in $P$ and $Q$ carries a positive power of $E$, so at $E=0$ the
equation reduces to $-y''+x^3y=0$, and on the branch connected to the
undeformed ground state $Y(-\infty)=(1+\sqrt5)/2$.

What is missing is analytic control over the strip. The defining
condition fixes only the formal, order-by-order large-$E$ expansion,
after the reparametrisations of Section~\ref{sec:derivation}: at each
perturbative order the spectral coefficients are polynomials in $E$,
but this does not by itself provide uniform bounds on an unbounded
rapidity strip, nor does continuous motion of individual zeros exclude
zeros entering from infinity. The large-$E$ and small-coupling limits
must also be distinguished: at fixed positive $\kappa_{\rm TBA}$, the
real part of the leading source $\kappa_{\rm TBA}e^{5\theta}$ changes
sign inside the strip, so the deformation cannot be treated as
everywhere small.

Subject to a coefficient-wise analytic reconstruction and uniqueness
in the ground-state class, the proposed identification is therefore
\be
Y^{\rm ODE}(\theta;\kappa)
=Y(\theta;\kappa_{\rm TBA}(\kappa)),
\label{eq:ODE-equals-TBA}
\ee
which we use as a conditional correspondence, rather than a proved
finite-coupling identity: the direct period calculations below test
its energy predictions independently through third order, while a
proof of the required analytic properties remains an open part of the
construction. As a strong check on the correctness of this proposal, we resum the ground-state energy from the asymptotic series and compare it with results obtained directly from a numerical solution of the TBA in Section \ref{sec:resummation}. 
\subsection{From the Stokes multiplier to the WKB period}
\label{sec:stokes-period}
The relation
between exact WKB periods and TBA systems for polynomial potentials has been
studied from the Riemann--Hilbert perspective in~\cite{ItoMarinoShu:2018}.
Since $T$ is a Wronskian of two subdominant solutions, while the period is
built from the semiclassical wavefunction, we make the bridge between the two
explicit. It rests on three elementary facts.

\paragraph{Formal solutions and Wronskians.} Split the Riccati series by its
parity under $\sqrt V\to-\sqrt V$,
\be
S=S_{\rm odd}+S_{\rm even},\qquad
S_{\rm odd}=S_0+S_2+S_4+\dots,\qquad S_{\rm even}=S_1+S_3+\dots
\label{eq:S-parity}
\ee
(in the bookkeeping of \eqref{eq:total-period-general}, even and odd powers of $\hbar$). The
part of the Riccati equation $S'+S^2=V$ which is odd under the flip is
$S_{\rm odd}'+2S_{\rm odd}S_{\rm even}=0$, so that
\be
S_{\rm even}=-\tfrac12\,\partial_x\log S_{\rm odd}
\label{eq:Seven}
\ee
exactly, and the two formal solutions are
\be
\psi_\pm(x;a)=S_{\rm odd}(x)^{-1/2}\exp\Big(\pm\int_a^x S_{\rm odd}\,dx'\Big).
\label{eq:formal-solutions}
\ee
For any two base points,
\be
W\big[\psi_+(\cdot\,;a),\psi_-(\cdot\,;b)\big]
=-2\exp\Big(\int_a^b S_{\rm odd}\,dx'\Big).
\label{eq:Wformal}
\ee
The exponential of a WKB action between two points is therefore the
Wronskian of two formal solutions.
Along a closed cycle the higher odd-indexed terms are total derivatives,
so that the non-trivial WKB period is
$\oint S\,dx=\oint S_{\rm odd}\,dx$; this is the period
\eqref{eq:total-period-general}.  The leading odd-indexed term
$S_1=-\frac12\partial_x\log S_0$ accounts instead for the constant
monodromy of the WKB prefactor around the turning points, an
$E$-independent factor that is absorbed in the normalisation.

\paragraph{The subdominant solutions.} In its own sector, $y_k$ is the
resummation of $\psi_-$ normalised at infinity,
\begin{align}
y_k(x) &=S_{\rm odd}(x)^{-1/2}
  \exp\Big(-\!\int_{\infty_k}^x S_{\rm odd}(x')\,dx'\Big),\\
\int_{\infty_k}^x S_{\rm odd}\,dx'
&:=\lim_{\substack{X\to\infty\\ X\in S_k}}
\Big[\int_X^x S_{\rm odd}\,dx'+\tfrac25X^{5/2}+Q\,X^{1/2}\Big],
\label{eq:ynorm}
\end{align}
the subtraction removes the part of the action that diverges at
infinity, and fixes the WKB normalisation up to one constant common
to all $k$; we choose it so that the adjacent Wronskians satisfy
\be
W[y_k,y_{k+1}]=1,
\ee
independently of $E$ and $\kappa$, in agreement with
\eqref{eq:adjacent-wronskian}.

\begin{figure}[t]
\centering
\begin{tikzpicture}[scale=1.0, >=stealth]
\foreach \a in {36,108,180,252,324}{
  \draw[dashed, gray!70, line width=0.5pt] (0,0) -- (\a:3.4);
}
\node at (0:3.2)   {\large $\boldsymbol{S_0}$};
\node at (62:3.2)  {\large $\boldsymbol{S_1}$};
\node at (144:3.1) {\large $\boldsymbol{S_2}$};
\node at (216:3.1) {\large $\boldsymbol{S_{-2}}$};
\node at (298:3.2) {\large $\boldsymbol{S_{-1}}$};
\coordinate (x0) at (0:1.3);
\coordinate (xp) at (120:1.3);   
\coordinate (xm) at (240:1.3);  
\coordinate (mid) at (-0.65,0); 
\draw[line width=1.7pt] (xp) -- (xm);
\def\rx{0.34}\def\ry{1.30}
\draw[blue, line width=1.5pt,
  postaction={decorate},
  decoration={markings, mark=at position 0.30 with {\arrow{>}}}]
  ([yshift=\ry cm]mid)
  .. controls ($(mid)+(0.55*\rx,\ry)$) and ($(mid)+(\rx,0.55*\ry)$) .. ([xshift=\rx cm]mid)
  .. controls ($(mid)+(\rx,-0.55*\ry)$) and ($(mid)+(0.55*\rx,-\ry)$) .. ([yshift=-\ry cm]mid)
  .. controls ($(mid)+(-0.55*\rx,-\ry)$) and ($(mid)+(-\rx,-0.55*\ry)$) .. ([xshift=-\rx cm]mid)
  .. controls ($(mid)+(-\rx,0.55*\ry)$) and ($(mid)+(-0.55*\rx,\ry)$) .. cycle;
\node[blue] at (-1.26,0) {\Large $\gamma$};
\filldraw (x0) circle (1.3pt);
\node at (x0) [below right=0pt] {\large $x_0$};
\filldraw (xp) circle (1.3pt);
\node[above=4pt] at (xp) {\large $x_+$};
\filldraw (xm) circle (1.3pt);
\node[below=5pt] at (xm) {\large $x_-$};
\draw[red, line width=1.5pt,
  postaction={decorate},
  decoration={markings, mark=at position 0.55 with {\arrow{>}}}
] (288:3.2)
  to[out=110,in=245] (0.28,0)
  to[out=65,in=250] (72:3.2);
\node[red] at (288:3.45) [below right] {\large $\infty_{-1}$};
\node[red] at (72:3.45)  [above right] {\large $\infty_1$};
\node[red] at (0.66,0.33) {\Large $\mathcal P$};
\filldraw (0,0) circle (0.8pt);
\end{tikzpicture}
\caption{The five Stokes sectors, the turning points $x_0,x_\pm$, and
the two contours of Section~\ref{sec:stokes-period}: $\gamma$, encircling
the cut joining $x_+$ to $x_-$, and $\mathcal P$, from $\infty_{-1}$ to
$\infty_1$ through the origin, separating $x_0$ from the pair
$\{x_+,x_-\}$. The two are shown to be homologous.}
\label{fig:contours}
\end{figure}
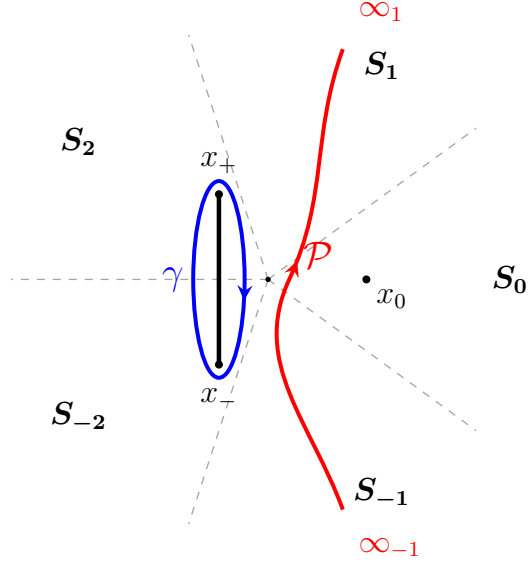

The turning points are $x_\pm=E^{1/3}e^{\pm2\pi i/3}(1+O(\kappa))$ and
$x_0=E^{1/3}(1+O(\kappa))$ (Figure~\ref{fig:contours}); we choose
continuation paths from $S_{\pm1}$ to the origin that do not cross the
Stokes line joining $x_+$ to $x_-$.
\paragraph{The Wronskian at the origin.}
Continued from their respective subdominant sectors to the origin,
$y_1$ and $y_{-1}$ arrive with opposite branches of $S_{\rm odd}$,
$S_{\rm odd}(0)=\mp i\sqrt E\,(1+\dots)$; the branch is fixed at
infinity by the decay and then continued along the chosen paths. By \eqref{eq:Wformal}, then,
\be
T(E)=\frac{W[y_{-1},y_1]}{W[y_0,y_1]}
=\exp\Big(\int_{\mathcal P}S_{\rm odd}\,dx\Big)\big[1+O(e^{-cE^{5/6}})\big],
\label{eq:TP}
\ee
where $\mathcal P$ is the path from $\infty_{-1}$ to $\infty_1$ through the
origin, with the subtractions of \eqref{eq:ynorm} at its two ends, and the
constant phase is fixed by $W[y_k,y_{k+1}]=1$ so that $T$ is real and positive
for real $E$. The exponentially small corrections are the contributions of the
other exponential, which the Stokes relation \eqref{eq:stokes-T} controls.

\paragraph{The path is the cycle.} On the curve $y^2=V(x)$ the point at infinity
is a branch point, $V$ being of odd degree, so a path starting and
ending at infinity is a closed cycle: as Figure~\ref{fig:contours}
shows, $\mathcal P$ is homologous to $\gamma$, oriented so that
$\Pi_0=\mathcal A\,E^{5/6}>0$. Hence
\be
\log T(E,\kappa)\simeq\oint_\gamma S_{\rm odd}\,dx=\sum_{K\ge0}\Pi_{2K}(E,\kappa),
\qquad E\to+\infty,
\label{eq:logTperiod}
\ee
which is the sense in which the WKB period is the asymptotic expansion of the Stokes
multiplier. At the level of the formal perturbative construction, the same argument extends order by order in $\kappa$: with the covariance of
Section~\ref{sec:rotation} the rotated solutions solve the same equation, the
$Q\,x^{1/2}$ term is part of the normalisation \eqref{eq:ynorm}, and $\gamma$ is
the cycle around the complex pair of turning points of the deformed cubic, i.e.\
the cycle of \eqref{eq:classical-period}. The homology can be checked directly at the
convergent orders: the integral of $S_{2K}$, $K\ge1$, along a path from
$\infty_{-1}$ to $\infty_1$ crossing the negative real axis between the cut and
the origin equals $\Pi_{2K}$, while along a path crossing it to the left of the
cut it vanishes. The same relation follows from the Stokes relation
\eqref{eq:stokes-T} evaluated at the origin, $T\,y_0(0)=y_{-1}(0)+y_1(0)$: the WKB
actions from $\infty_{\pm1}$ to the origin are the asymptotics of
$y_{\pm1}(0)=\omega^{\pm1/2}D_\pm(\Omega^{\pm1}E)$, where $D$ is the Dirichlet
spectral determinant; it is their sum, not their difference, that builds the
period, and $y_0(0)=D(E)$ itself, which oscillates for real $E$, does not enter
\eqref{eq:TP}, its role being taken by the $E$-independent adjacent Wronskian.

We can numerically check the proposal in \eqref{eq:logTperiod}. For the undeformed cubic potential, the periods reduce to the two basic
ones, with the identities of Appendix~\ref{app:periods}:
\begin{align}
\Pi_0&=\mathcal A\,E^{5/6},
&\Pi_2&=\tfrac{1}{24}\mathcal B\,E^{-5/6},
&\Pi_4&=0,
\nonumber\\
\Pi_6&=-\tfrac{21983}{1119744}\,\mathcal A\,E^{-25/6},
&\Pi_8&=\tfrac{26317819}{376233984}\,\mathcal B\,E^{-35/6},
&\Pi_{10}&=0,
\nonumber\\
\Pi_{12}&=-\tfrac{70877384469605}{13792092880896}\,\mathcal A\,E^{-55/6},
\label{eq:pure-cubic-periods}
\end{align}
the vanishing of
$\Pi_4$ and $\Pi_{10}$ being the selection rule for the spins $-3$ and $-9$. Integrating
the equation numerically along the three rays with the normalisation
\eqref{eq:ynorm} gives, at $E=8$, $\log T=16.50822223$, against
$16.50823110$ from $\Pi_0+\Pi_2$, $16.50822235$ adding $\Pi_6+\Pi_8$ and
$16.50822227$ adding $\Pi_{12}$; at $E=12$ the series through $\Pi_{12}$ and the
Stokes multiplier agree to $10^{-7}$, the residual accuracy of the integration.
The same integration reproduces the massless TBA, $T(E)=Y(\theta)$ with
$e^\theta=\mathcal A\,E^{5/6}$, to $10^{-7}$ over $4\le E\le24$, and for the deformed
problem at first order, $\partial_\kappa\log T|_{\kappa=0}=\tfrac12\mathcal B\,E^{25/6}$ to
$3\cdot10^{-4}$ at $E=4,8,12$.
\section{Eigenvalues of higher charges from the WKB periods}
\label{sec:energy-wkb}
Given the identification of the logarithm of the $T$-function with the WKB period, we now compute the deformed eigenvalues
of $I_1$ (the energy) and $I_5$ directly from the WKB periods by doing an asymptotic expansion for large $E$. 

The ODE-side coefficients are obtained without assuming the finite-coupling
identification of the previous subsection: they are read directly from the
negative powers of $E$ in $\Pi(E,\kappa)$, exactly as in the standard
correspondence. Their conversion into the conventionally normalised
eigenvalues $I_s$ uses only the asymptotic normalisation constants $N_s$
of Section~\ref{sec:lee-yang}; the novelty here is that several WKB
levels contribute to the same power of $E$, and that the reparametrisations have to be accounted for.

The WKB period is given by 
\be 
\Pi(E,\kappa) = \sum_{K\ge 0} \Pi_{2K}(E,\kappa)~,   
\ee 
where the WKB period at level $K$ is written in the form 
\be
\Pi_{2K}=E^{5/6}\heff^{2K}\pi_{2K}(a,b),\quad\text{with}\quad
\pi_{2K}(a,b):=\oint_\gamma s_{2K}\big[\xi^3+a\xi-b\big]\,d\xi,
\label{eq:pi2k-def}
\ee
which is the period integral of the $K$-th WKB coefficient of the rescaled potential,
with $s_{2K}$ the same WKB density as $S_{2K}$ in \eqref{eq:S0S1S2}, but  now evaluated on $v(\xi)=\xi^3+a\xi-b$ of \eqref{eq:hbar-effective}
instead of $V(x;E,\kappa)$. 

Unlike the undeformed case, $a,b$ are not fixed numbers here: they are
functions of $z=\kappa E^{10/3}$, and, since they are themselves
determined order by order in the same WKB construction, they carry their
own $\heff$-expansion,
\be
a=\sum_{K'\ge0}\heff^{2K'}a_{K'}(z),\qquad b=\sum_{K'\ge0}\heff^{2K'}b_{K'}(z),
\label{eq:ab-full-expansion}
\ee
with $a_0,b_0$ the classical solution \eqref{eq:P0Q0-exact} and $a_1,b_1$ the first counterterms of
Section~\ref{sec:quantum}. So $E$ enters $\Pi_{2K}$ in three separate places: once explicitly, in
the overall prefactor $E^{5/6}$; a second time, in $\heff^{2K}=E^{-5K/3}$, the bare WKB suppression
of the level $K$ itself; and a third time, hidden inside every $z^n=\kappa^nE^{10n/3}$ appearing in
$\pi_{2K}(a,b)$'s own expansion once \eqref{eq:ab-full-expansion} is substituted in. The power of $E$ carried by a single term is therefore fixed by three
integers at once (the bare level $K$, the order $K'$ at which $a,b$'s
own counterterms enter, and the power $n$ of $z$) through
\be
E^{5/6}\cdot\heff^{2K}\cdot\heff^{2K'}\cdot z^n
\;\sim\;E^{(5-10K-10K'+20n)/6}=E^{5r/6},
\qquad r=1+4n-2(K+K').
\label{eq:E-power-of-Kn}
\ee
Only the combination $K_{\rm tot}:=K+K'$, not $K$ and $K'$ separately, fixes the power of $E$:
a given total WKB order $K_{\rm tot}$ can be reached either by a high bare level $K$ acting on the
purely classical $a_0,b_0$ ($K'=0$), or by a lower bare level combined with one or more of $a,b$'s
own counterterms ($K'\ge1$), and every such combination with the same $K_{\rm tot}$ contributes to
the same power of $E$. 

Summing \eqref{eq:E-power-of-Kn} over every allowed spin $r$ therefore gives an asymptotic
expansion of the form
\be
\Pi(E,\kappa)\;\simeq\;\sum_{r}\,D_r(\kappa)\,E^{5r/6},
\ee
where each coefficient $D_r(\kappa)$ collects the entire ladder of contributions, over all
$K,K',n$ with $1+4n-2(K+K')=r$, at that spin. 
After imposing the defining condition, only the positive-spin sectors
$r=1$ and $r=5$ survive in the growing part of the period; all other
positive-spin contributions generated at intermediate stages are
cancelled by the counterterms.
The negative spins are not constrained this way; they are whatever the constructed
potential actually produces, and the selection rule of Section~\ref{sec:family} guarantees these take the form
$r=-1,-5,-7,-11,-13,\dots$, that is, minus the spins of the Lee--Yang KdV
charges (odd, excluding multiples of $3$), each giving the eigenvalue of
the corresponding deformed local charge $I_{-r}$. The
asymptotic expansion is therefore, relabelling $D_1=:D$,
\be
\Pi(E,\kappa)\;\simeq\;D(\kappa)\,E^{5/6}+D_5(\kappa)\,E^{25/6}
+\sum_{s=1,5,7,11,\dots}\widetilde I_s(\kappa)\,E^{-5s/6}.
\label{eq:logT-expansion}
\ee
Carrying out the expansion, we find that the coefficients have the following $\kappa$-expansion:
\be
\begin{aligned}
D(\kappa)&=\mathcal A\Big[1+\tfrac{637}{31104}\kappa-\tfrac{258822655}{483729408}\kappa^2\Big],\\
D_5(\kappa)&=\tfrac12\mathcal B\,\kappa\Big[1+\tfrac{637}{15552}\kappa\Big],\\
\widetilde I_1 &=\tfrac1{24}\mathcal B\Big[1-\tfrac{55913}{46656}\kappa+\tfrac{97372494583}{725594112}\kappa^2\Big],\\
\widetilde I_5 &=-\frac{21983}{1119744}\,\mathcal A
+\frac{38800584095}{17414258688}\,\mathcal A\,\kappa
-\frac{522470933556338585}{270826551115776}\,\mathcal A\,\kappa^2+O(\kappa^3).
\label{eq:I1I5tilde}
\end{aligned}
\ee
We have to compare the asymptotic expansion of the quantum period with that of $\log T(E)$, which coincides with the logarithm of the $Y$-function, which, in turn, can be derived from the TBA (see equation \eqref{eq:Z-asymptotic-general}): 
\be
\log Y(\theta;\kappa_{\rm TBA})=e^{\theta}+\kappa_{\rm TBA}e^{5\theta}
+\sum_{s}N_s\, I_s^{(\phi)}(\kappa_{\rm TBA})\, e^{-s\theta}\;.
\ee 
The constants $N_s$ have been listed in \eqref{eq:Ns-values}.
The leading term immediately gives us the $\kappa$-dependent map between the spectral parameter $E$ of the ODE and the rapidity term $e^{\theta}$ of the TBA:  
\be 
e^{\theta}=D(\kappa)E^{5/6}~.
\ee 
The coefficient of the spin-5 deformation in turn tells us that the TBA
coupling is related to the deformation parameter of the ODE by the relation
\be 
\kappa_{\rm TBA}(\kappa)=\frac{D_5(\kappa)}{D(\kappa)^5}~.
\ee 
Carrying this elimination to the order needed 
gives the ODE--TBA dictionary
\be
\kappa_{\rm TBA}=\frac{\mathcal B}{2\mathcal A^5}\,\kappa
\left[1-\frac{637}{10368}\,\kappa+\frac{1038131003}{967458816}\,\kappa^2
-\frac{9009615409677337}{30091839012864}\,\kappa^3+O(\kappa^4)\right].
\label{eq:kappa-dictionary}
\ee
The two couplings agree at leading order only up to the normalisation
$\mathcal B/2\mathcal A^5$ fixed by the choice of cycle and rapidity scale. Finally, equating the coefficients of the remaining terms relates the ODE-side quantities $\widetilde I_1,\widetilde I_5$ to the physical eigenvalues $I_1^{(\phi)}, I_5^{(\phi)}$:
\be
N_s\,I_s^{(\phi)}\big(\kappa_{\rm TBA}(\kappa)\big)=D(\kappa)^s\,\widetilde I_s(\kappa).
\label{eq:I1I5-relation}
\ee
\subsection{Comparison with the deformed KdV hierarchy}
\label{sec:karimi-comparison}
Evaluating \eqref{eq:I1I5-relation} with $\widetilde I_1,\widetilde
I_5$ from \eqref{eq:I1I5tilde}, and the map in 
\eqref{eq:kappa-dictionary} gives the ground-state eigenvalues as a power series expansion with coefficients being 
rational multiples of powers of $\mathcal A^5/\mathcal B$:
\begin{align}
I_1^{(\phi)}(\kappa_{\rm TBA})&=-\frac{1}{60}
+\frac{21983}{559872}\,\frac{\mathcal A^5}{\mathcal B}\,\kappa_{\rm TBA}
-\frac{25843717445}{2902376448}\,\frac{\mathcal A^{10}}{\mathcal B^2}\,\kappa_{\rm TBA}^2
+O(\kappa_{\rm TBA}^3),
\label{eq:I1-numeric}\\
I_5^{(\phi)}(\kappa_{\rm TBA})&=\frac{89}{756000}
-\frac{2989441}{111974400}\,\frac{\mathcal A^5}{\mathcal B}\,\kappa_{\rm TBA}
+\frac{3222566577503}{69657034752}\,\frac{\mathcal A^{10}}{\mathcal B^2}\,\kappa_{\rm TBA}^2
+O(\kappa_{\rm TBA}^3).
\label{eq:I5-numeric}
\end{align}
The expansions \eqref{eq:I1-numeric} and \eqref{eq:I5-numeric} are
written in powers of $(\mathcal A^5/\mathcal B)\,\kappa_{\rm TBA}$, while
the CFT prediction \eqref{eq:I1-cft-ground} is written in powers of
$\gamma_{\rm CFT}$; to compare them directly, we
need the exact relation between the two expansion variables. 
The period integrals can be computed simply in terms of Euler Beta functions, and are given by
(see Appendix~\ref{app:periods})
\be
\mathcal A=\frac{\sqrt3\,\Gamma(\tfrac13)\Gamma(\tfrac16)}{5\sqrt\pi},
\qquad
\mathcal B=\frac{8\pi^{3/2}}{\Gamma(\tfrac13)\Gamma(\tfrac16)}~.
\label{eq:AB-closed}
\ee
Both $\gamma_{\rm CFT}$ in
\eqref{eq:gamma-kappa-TBA} and $\mathcal A,\mathcal B$  are built from the
same four Gamma functions, and the ratio reduces, by the reflection formula
$\Gamma(\tfrac13)\Gamma(\tfrac23)=2\pi/\sqrt3$, to a rational number:
\be
\gamma_{\rm CFT}=\frac{216125}{648}\;\frac{\mathcal A^5}{\mathcal B}\;\kappa_{\rm TBA}.
\label{eq:gamma-A5B-exact}
\ee
The two expansions are therefore not written in the same variable by coincidence or convention,
but because both $\gamma_{\rm CFT}$ and $\mathcal A^5\kappa_{\rm TBA}/\mathcal B$ are, independently,
proportional to the one physical coupling $\kappa_{\rm TBA}$; \eqref{eq:gamma-A5B-exact} is the
exact constant of proportionality between them, fixed entirely by special-function identities, with
no free parameter left to adjust. Substituting \eqref{eq:gamma-A5B-exact} into
\eqref{eq:I1-cft-ground} and re-expanding in powers of $(\mathcal A^5/\mathcal B)\kappa_{\rm TBA}$
reproduces \eqref{eq:I1-numeric} term by term, 
confirming that the two constructions agree exactly.
\section{Resummation and matching the TBA}
\label{sec:resummation}
The construction of the previous sections determines the deformed ODE, and
hence $I_1^{(\phi)}(\kappa_{\rm TBA})$, order by order in $\kappa_{\rm TBA}$.
Since the spin-$5$ term dominates the driving term at large $\theta$ for any
$\kappa_{\rm TBA}>0$, this expansion is only asymptotic, and agreement with
the TBA coefficient by coefficient does not by itself show that the series
captures the full function. In this section we provide that test: we
compute the WKB coefficients to high order, determine the analytic
structure of their Borel transform, and show that the resummed series reproduces
the numerical solution of the deformed TBA over the full range of
couplings (explored numerically), from the vicinity of the conformal point
deep into the regime where the spin-5 charge dominates.

\begin{figure}[!htbp]
\centering
\includegraphics[width=0.75\textwidth]{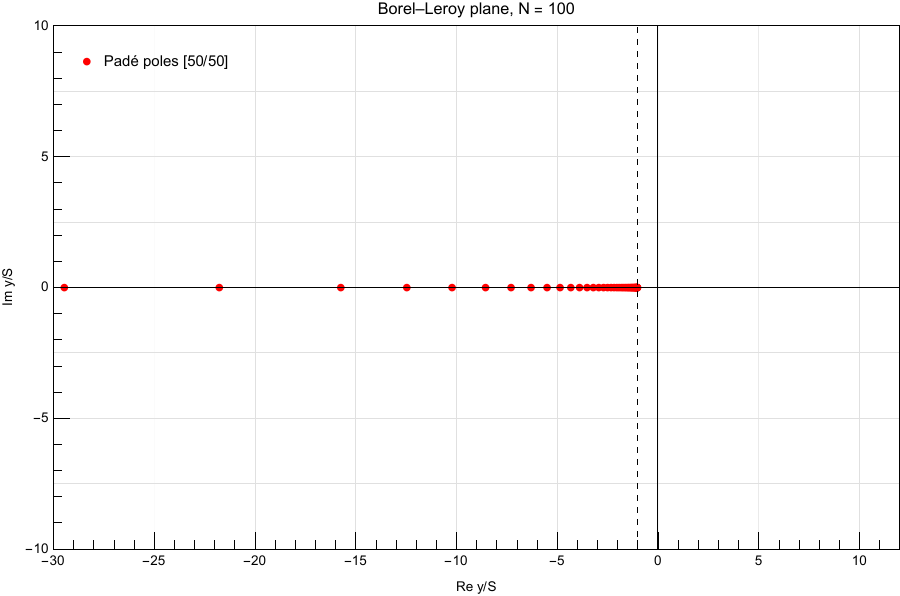}
\caption{Poles of the [50/50] Pad\'e approximant of the Borel--Leroy
transform $\mathcal B(y)$ built from $N=100$ coefficients, in units of $S$.
The poles start at $y=-S$ (dashed line), where $\mathcal B$ has a simple
pole with residue $kS$, and accumulate along the cut $(-\infty,-S]$.}
\label{fig:borel}
\end{figure}

The first few coefficients of the $\kappa_{\rm TBA}$-expansion of
$I_1^{(\phi)}$ were given in the previous section. Writing
\be
I_1^{(\phi)}(\kappa_{\rm TBA})=-\frac{1}{60}\sum_{n\ge0}f_n\,x^n,
\qquad x=\frac{\cA^5}{\cB}\,\kappa_{\rm TBA},
\label{eq:fn-def}
\ee
with $f_0=1$ and $f_1=-\tfrac{109915}{46656}$, the WKB construction can be
carried to high order, and we have computed the exact rational
coefficients $f_n$ up to $n=100$. The series is divergent, with
\be
f_n \sim k\,(-1)^n \frac{\Gamma(4n+b)}{S^n}\left[1+O\!\left(\frac{1}{n}\right)\right],
\label{eq:large-order}
\ee
and a ratio analysis of the coefficients, accelerated by Richardson
extrapolation~\cite{BenderOrszag:1978}, gives
\be
b=\frac12~,\qquad S= 5.765503277909121093\ldots~,\qquad k= 1.355771754100788651\ldots~.
\ee
The numerical value of $S$ agrees with
\be
S=\frac{4^4}{5^5}\,\frac{\cA^5}{\cB}
\label{eq:S-exact}
\ee
to $10^{-34}$, and we therefore conjecture \eqref{eq:S-exact} to be exact.
In terms of $\kappa_{\rm TBA}$,
\eqref{eq:S-exact} places the leading Borel--Leroy singularity at
$\kappa_{\rm TBA}\,t^4=-4^4/5^5$, which is precisely the critical value of
the dispersion relation $u+\kappa_{\rm TBA}u^5$ ($u=e^\theta$) at negative
coupling: its maximum, at $u_*^4=1/(5|\kappa_{\rm TBA}|)$, equals
$\tfrac45 u_*$, giving non-perturbative effects of order
$\exp\!\big[-\tfrac45(5|\kappa_{\rm TBA}|)^{-1/4}\big]$. The order four of
the Borel--Leroy transform below is thus the difference of the spins, and
the alternating sign places the singularity on the negative real axis, so
that the series is Borel summable for $\kappa_{\rm TBA}>0$.
We resum it by the Borel--Leroy transform
\be
\mathcal B(y)=\sum_{n\ge0}\frac{f_n}{\Gamma(4n+b)}\,y^n,
\qquad
\sum_{n\ge0}f_n x^n=\int_0^\infty dt\,t^{b-1}e^{-t}\,\mathcal B(x\,t^4).
\label{eq:borel-leroy}
\ee
The Pad\'e approximants of $\mathcal B(y)$, shown in
Figure~\ref{fig:borel}, exhibit a simple pole at $y=-S$, with residue
$kS$, followed by a sequence of poles accumulating along
$(-\infty,-S]$, consistent with a branch cut. No additional
singularities are resolved at this order.

Two further inputs fix the resummation. First, the pole is subtracted
exactly, $\mathcal B(y)=kS/(S+y)+\mathcal B_{\rm reg}(y)$. Second, the large-coupling behaviour of $I_1^{(\phi)}$ follows from the defining
relation \eqref{eq:defining-problem} alone. Let $I_1(c_1,\kappa_{\rm TBA})$
denote the coefficient of $e^{-\theta}$ when the leading terms of the WKB
period are $c_1e^{\theta}+\kappa_{\rm TBA}e^{5\theta}$, so that
$I_1^{(\phi)}(\kappa_{\rm TBA})=I_1(1,\kappa_{\rm TBA})$. Since the
normalisation $D$ in $e^\theta=D E^{5/6}$ is conventional, a shift
$\theta\to\theta+a$ describes the same spectral problem, with $c_1$ and
$\kappa_{\rm TBA}$ rescaled by $e^{a}$ and $e^{5a}$ and the coefficient of
$e^{-\theta}$ by $e^{-a}$. Hence $I_1$ is weighted-homogeneous,
$I_1(\lambda c_1,\lambda^5\kappa_{\rm TBA})=\lambda^{-1}I_1(c_1,\kappa_{\rm TBA})$,
and the choice $\lambda=\kappa_{\rm TBA}^{-1/5}$ gives
\be
I_1^{(\phi)}(\kappa_{\rm TBA})=\kappa_{\rm TBA}^{-1/5}\,
I_1\big(\kappa_{\rm TBA}^{-1/5},1\big)
\;\simeq\;\kappa_{\rm TBA}^{-1/5}\,I_1(0,1),
\qquad \kappa_{\rm TBA}\to\infty,
\label{eq:large-kappa}
\ee
provided the spectral problem with the spin-$5$ term alone is regular,
so that $I_1(0,1)$ is finite.
A Borel transform compatible with this large-coupling behaviour has
$\mathcal B(y)\sim y^{-1/5}$ at large $y$, with corrections in powers
of $y^{-1/5}$.
We therefore map the cut to the boundary of the unit disc with
$y=4Sw/(1-w)^2$, write $\mathcal B_{\rm reg}=(1-w)^{2/5}\,G(w)$,
approximate $G$ by a diagonal Pad\'e approximant~\cite{BenderOrszag:1978},
and evaluate the Laplace integral in \eqref{eq:borel-leroy}
numerically.\footnote{The Mathematica code that generates the high-order
WKB coefficients and carries out the resummation was developed with the
assistance of the AI models Claude Opus and GPT-6 Pro.}
The factor $(1-w)^{2/5}$ is essential: without it the resummation misses
the large-coupling behaviour \eqref{eq:large-kappa}, and a plain diagonal
Pad\'e approximant deviates from the TBA by up to $10^{-3}$, with relative
errors of order one at the largest couplings.

\begin{figure}[!htbp]
\centering
\includegraphics[width=0.8\textwidth]{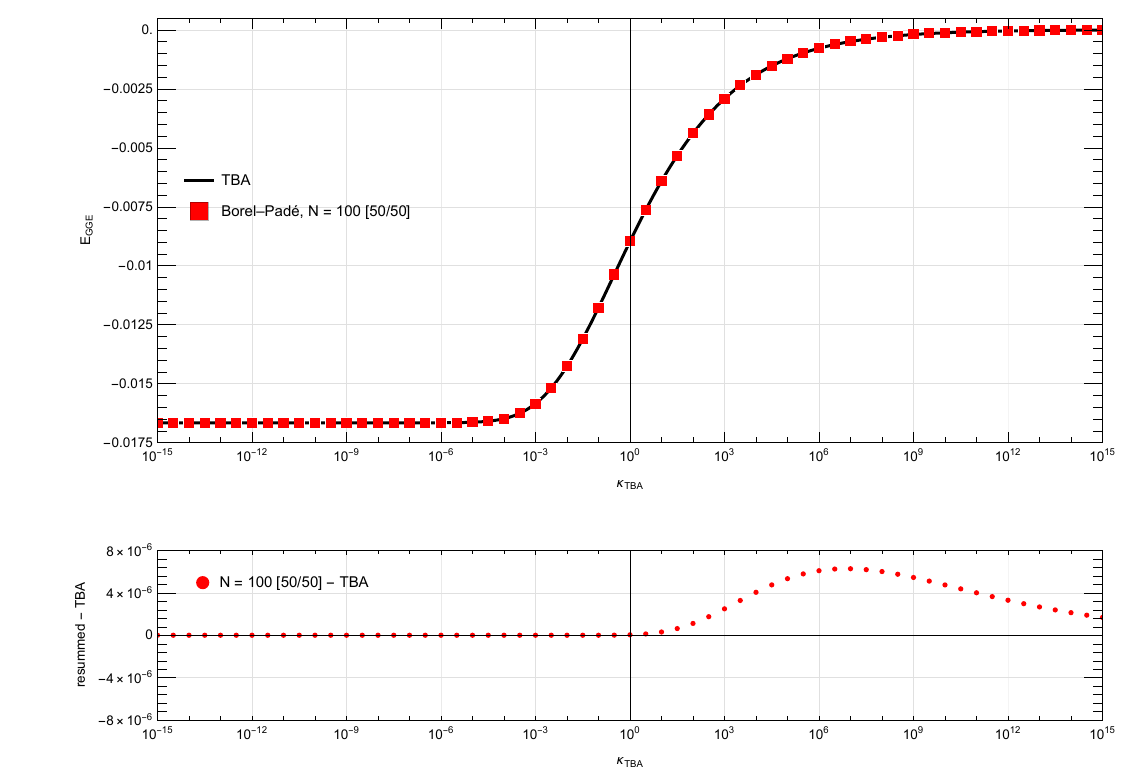}
\caption{Top: the ground-state energy $E_{GGE}=I_1^{(\phi)}$ as a function of
$\kappa_{\rm TBA}$, from the numerical solution of the deformed TBA (black
line) and from the Borel--Leroy--Pad\'e resummation of the WKB series with
$N=100$ coefficients (red squares). Bottom: their difference, which stays
below $7\times10^{-6}$ over the whole range.}
\label{fig:resummation}
\end{figure}

Figure~\ref{fig:resummation} compares the result, obtained from the [50/50]
approximant, with $I_1^{(\phi)}$ from the numerical solution of the deformed
TBA. With no input beyond its own coefficients and the exponent of the
large-coupling behaviour, the resummed series reproduces the ground-state
energy over thirty decades, $10^{-15}\le\kappa_{\rm TBA}\le10^{15}$, with a
maximum deviation of $6.3\times10^{-6}$. The residual deviation comes from
the slow convergence of the Pad\'e approximants near $w=1$, where $G(w)$ has
further non-analytic terms associated with the corrections
$\kappa_{\rm TBA}^{-2/5},\kappa_{\rm TBA}^{-3/5},\dots$ to the
large-coupling behaviour.
Before closing this section, we stress that the ground state studied here
may be a privileged case. Since the deformed TBA for $\phi$ is regular for
all $\kappa_{\rm TBA}>0$, $I_1^{(\phi)}$ is analytic on the whole positive
axis, and the only singularity in the Borel plane is the one at $y=-S$,
which is set by the dispersion relation and is therefore common to all
states. For excited states the situation can be different: independent
studies of the deformed excited-state TBA~\cite{DAST-TBA} show that
square-root branch points can appear at real positive $\kappa_{\rm TBA}$,
where two levels merge into a complex-conjugate pair. A Borel sum whose
transform is singular only on the negative axis is real and analytic for
$\kappa_{\rm TBA}>0$, and cannot reproduce such a branch point. The
corresponding Borel transform is then expected to have singularities on the
positive axis as well, and the resummation requires the inclusion of
non-perturbative contributions, along the lines of a trans-series
analysis~\cite{Aniceto:2018bis,Dorigoni:2014hea}.
\section{The identity primary: a moving double pole}
\label{sec:excited}
So far the construction has been tied to the conformal ground state, the
primary $\phi$ of weight $h_\phi=-\tfrac15$, for which the cubic
oscillator~\eqref{eq:LY-ODE} has no singularity at the origin. The
Lee--Yang model has a second irreducible module, that of the identity,
$h=0$, and the ODE/IM correspondence assigns to its primary a cubic oscillator carrying an
additional double pole, an apparent Fuchsian singularity with trivial monodromy, whose position is fixed by this
condition~\cite{Bazhanov:2003ni,Conti:2020zzk}. In this section we switch
on the spin-$5$ chemical potential on top of that term. The whole
apparatus of Sections~\ref{sec:family}--\ref{sec:hbar2-level} carries over, with two novelties: the pole must move with the
coupling, and there is one new basic period. The outcome is a closed prediction for the
deformed eigenvalues of the identity state, which we have checked against
the CFT recursion of~\cite{Karimi:2026kdv} through the orders available
there.
\subsection{The deformed spectral problem}
\label{sec:excited-ODE}
The undeformed identity state corresponds to the cubic oscillator with a double pole at the origin,
\be
\left[-\frac{d^2}{dx^2}+x^3-E+\frac{2}{x^2}\right]\psi(x)=0.
\label{eq:excited-ODE}
\ee
The strength $2=\ell(\ell+1)$ is that of angular momentum $\ell=1$, which shifts the conformal weight of the
state from $h=-\tfrac15$ at $\ell=0$ to $h=\bigl((2\ell+1)^2-9\bigr)/40=0$.
Switching on the chemical potential, we deform, exactly as in \eqref{eq:deformed-ODE}, the regular part of the potential,
\be
\begin{aligned}
&\left[-\frac{d^2}{dx^2}+V_{\rm reg}(x)+\frac{2}{(x-z_0)^2}\right]\psi(x)=0,\\
&V_{\rm reg}(x)=x^3+Q(E,\kappa)\,x-P(E,\kappa),\qquad 3z_0^2+Q(E,\kappa)=0,
\end{aligned}
\label{eq:monodromy-condition}
\ee
where $V_{\rm reg}$ is the potential without the double pole and $z_0=z_0(E,\kappa)$ is the position of the pole.

Two features are worth highlighting here. First, contrary to the undeformed problem, the pole can no longer sit at the origin.
Near a pole at $x_*$ the Frobenius exponents are the roots of $s(s-1)=2$, namely $-1$ and $2$. They differ by an
integer, so the solution with exponent $-1$ can acquire a logarithm. Expanding
$V_{\rm reg}(x)=\sum_k v_k\,(x-x_*)^k$, the recursion for its coefficients first develops a
logarithmic term at the third step, with a coefficient proportional to $v_1=V'_{\rm reg}(x_*)$. The singularity is
therefore apparent if and only if $V'_{\rm reg}(x_*)=0$, that is, $3x_*^2+Q=0$. This is the
trivial-monodromy condition imposed in \eqref{eq:monodromy-condition}. At $Q=0$ it is satisfied at the origin; for $Q\neq0$
it moves the pole to $z_0=\pm\sqrt{-Q/3}$, one of the two stationary points of the regular cubic potential.

Since the large-$x$ asymptotics are still governed by the
cubic term, the Stokes sectors at infinity are the same as for the regular
cubic. Moreover, when the pole is rotated covariantly together with the
potential, the five-fold rotation of Section~\ref{sec:rotation} and the
resulting functional relations are preserved. The Stokes multipliers
themselves do depend on the pole, as expected for an excited-state
spectral problem. This is the same mechanism by which the potentials with
apparent singularities of~\cite{Bazhanov:2003ni,Conti:2020zzk} (the
so-called ``monster'' potentials) describe genuine excited states of a
regular spectral problem.

Second, the double pole term is of order $\heff^2$ after
the rescaling $x=E^{1/3}\xi$: writing $\zeta_0=E^{-1/3}z_0$,
\be
\frac{2}{(x-z_0)^2}=\heff^2\,E\,\frac{2}{(\xi-\zeta_0)^2},
\qquad \heff^2=E^{-5/3},
\ee
so the pole term enters the rescaled potential
$v(\xi)=\xi^3+a\xi-b+\heff^2\,2/(\xi-\zeta_0)^2$ at the same order as the
first quantum correction, and contributes only from WKB level $K\ge1$
onwards. The classical period is therefore \emph{unchanged}: at $K=0$ the
identity state and the ground state share the closed-form deformation
\eqref{eq:P0Q0-exact}.

The condition $3z_0^2+Q=0$ makes
$\zeta_0=\pm\sqrt{-a/3}$ a half-integer power of the coupling,
$\zeta_0\sim\sqrt{z}$. Odd powers of $\sqrt{z}$ therefore appear in
intermediate quantities, including the counterterms and the individual
WKB periods. After imposing the defining condition, however, they
cancel from physical observables, which are analytic functions of
$z=\kappa E^{10/3}$. We have verified this cancellation explicitly,
order by order, in the deformed energy and in the eigenvalues of
$I^{(1)}_5$ and $I^{(1)}_7$, to high order in $\kappa$. 

\subsection{The defining condition and a new period}
\label{sec:excited-condition}
The defining condition is unchanged: we still require
\be
\oint_\gamma S(x;E,\kappa)\,dx
   =e^{\theta}+\kappa_{\rm TBA}(\kappa)\,e^{5\theta}+O\!\left(e^{-\theta}\right),
\qquad e^{\theta}=D(\kappa)\,E^{5/6},
\ee
with $\gamma$ the cycle around the complex pair of turning points,
continuously deformed together with the potential and the moving pole
$\zeta_0$.

The weight assignment of Section~\ref{sec:family} is untouched: the pole term is weight-neutral ($x^{-2}$ carries weight $-4$, the
$\heff^2$ restores $+4$), so the spin grading \eqref{eq:spin-formula} and
the allowed monomials \eqref{eq:monomial-spin-map} of $P$ and $Q$ are the
same as before. The counterterms are fixed by the same requirement that no unwanted
positive spin survives, together with the cancellation of the half-integer powers of $ z$ noted above;
the negative-spin tail is the prediction. In Appendix \ref{sec:id-wkb}, we show how this works at the first few orders in the WKB expansion.

Compared with the ground state analysis, two things change in practice. First, the trivial-monodromy condition
couples the orders: $\zeta_0=\sqrt{-a/3}$ must be re-expanded whenever
$a$ receives a new counterterm, so the coefficients of $P$ and $Q$ are no
longer determined one power of $\kappa$ at a time by a single algebraic
equation, but by a small triangular linear system that mixes each order
with the pole expansion. Second, the periods of the reduced integrand now
include negative powers of $\xi$. The reduction identities of
Appendix~\ref{app:periods} extend to $m<0$,
\be
J(m,s)=\frac{m+3s+4}{m+1}\,J(m+3,s),\qquad m\le-2,
\label{eq:excited-J}
\ee
(the same exact-differential identity read in the opposite direction).
Repeated use of this identity reduces the periods with $m\leq -2$
either to the usual periods $\cA,\cB$ or to the exceptional case
$m=-1$. The latter gives one genuinely new object, the period of a
meromorphic differential with a simple pole at $\xi=0$,
\be
\cC:=J\!\left(-1,\frac12\right)
=\oint_\gamma \frac{(\xi^3-1)^{1/2}}{\xi}\,d\xi .
\ee
Unlike $\cA$ and $\cB$, this period is sensitive to the winding of
$\gamma$ around the pole at $\xi=0$.
However, it appears only in the class $r\equiv 3\pmod 6$, the
class whose period vanished identically for the ground state. Here that
vanishing becomes a condition rather than an identity: the coefficient
of $\cC$ in the $r\equiv3\pmod 6$ sector must cancel. At the lowest order this coefficient is
proportional to $V'_{\rm reg}(z_0)$, which is exactly what the trivial-monodromy condition
\eqref{eq:monodromy-condition} sets to zero. This reveals a direct connection between
trivial monodromy and the Lee--Yang selection rule on the ODE side.
\subsection{Results and checks}
\label{sec:excited-results}
At $\kappa=0$ the only contribution to the coefficient of $E^{-5/6}$ in $\Pi$ is
the first quantum period $\Pi_2=E^{-5/6}\pi_2$ of the potential $v=\xi^3-1+\heff^2\,2/\xi^2$. Hence
$\widetilde I_1^{(\id)}(0)=\pi_2$, which, in turn, has one part from the cubic potential and one from the pole:
\be
\widetilde I_1^{(\id)}(0)=\frac18\oint_\gamma\frac{\xi\,d\xi}{(\xi^3-1)^{3/2}}
+\oint_\gamma\frac{d\xi}{\xi^2\sqrt{\xi^3-1}}
=\frac{\cB}{24}-\frac{\cB}{2}=-\frac{11}{24}\,\cB .
\ee
Both integrals reduce to $\cB$ by the identities of Appendix~\ref{app:periods} and \eqref{eq:excited-J}. Then the physical conserved charge is obtained by taking into account the normalisation constants (see \eqref{eq:I1I5-relation}): 
\be
I_1^{(\id)}(0)
=\frac{D(0)}{N_1}\,\widetilde I_1^{(\id)}(0)
=\frac{11}{60}\,,
\ee
which is the correct value for the identity module. We have used $N_1=-4\sqrt{3}\pi$, $D(0)=\cA$ and
$\cA\,\cB=\tfrac{8\sqrt{3}\pi}{5}$.

To go beyond leading order, recall that the expansion of the quantum period at large $E$ takes the form
\be
\Pi(E,\kappa)\;\simeq\;D(\kappa)\,E^{5/6}+D_5(\kappa)\,E^{25/6}
+\sum_{s=1,5,7,11,\dots}\widetilde I_s(\kappa)\,E^{-5s/6},
\ee
where $D$ fixes the map $e^\theta=D\,E^{5/6}$, $D_5$ fixes the TBA coupling through
$\kappa_{\rm TBA}=D_5/D^5$, and $\widetilde I_s$ are related to the eigenvalues $I_s$ by
$N_s\,I_s=D^s\,\widetilde I_s$, as before. Carrying the expansion through, as described in Appendix~\ref{sec:id-wkb}, the
coefficients for the identity state are
\be
\begin{aligned}
D(\kappa)&=\mathcal A\Big[1+\tfrac{5077}{31104}\kappa-\tfrac{448693087}{483729408}\kappa^2\Big],\\
D_5(\kappa)&=\tfrac12\mathcal B\,\kappa\Big[1+\tfrac{5941}{15552}\kappa-\tfrac{328481537}{120932352}\kappa^2\Big],\\
\widetilde I_1 &=-\tfrac{11}{24}\mathcal B\Big[1+\tfrac{11527}{46656}\kappa-\tfrac{152462829619}{7981535232}\kappa^2\Big],\\
\widetilde I_5 &=-\tfrac{84227}{1119744}\,\mathcal A\Big[1-\tfrac{8399915}{171072}\kappa+\tfrac{93550793317085}{2660511744}\kappa^2+O(\kappa^3)\Big].
\end{aligned}
\label{eq:I1I5tilde-identity}
\ee
With $x:=\cA^5\kappa_{\rm TBA}/\cB$, the ODE--TBA dictionary for the identity state is
\be
x=\frac{\kappa}{2}\Big[1-\frac{4501}{10368}\,\kappa+\frac{1944092939}{967458816}\,\kappa^2+O(\kappa^3)\Big].
\label{eq:kappa-dictionary-identity}
\ee
Inverting this series, and using the appropriate normalisation constants, we obtain the physical eigenvalues as a power series in $x$ with rational coefficients,
\begin{align}
I_1^{(\id)}&=\frac{11}{60}+\frac{84227}{559872}\,x-\frac{42166065305}{2902376448}\,x^2+O(x^3),\\
I_5^{(\id)}&=\frac{341}{756000}-\frac{4877509}{111974400}\,x+\frac{4410941121587}{69657034752}\,x^2+O(x^3).
\end{align}
Using the exact relation \eqref{eq:gamma-A5B-exact}, $\gamma_{\rm CFT}=\tfrac{216125}{648}\,x$, the expansion of
$I_1^{(\id)}$ above reproduces the CFT result \eqref{eq:I1-cft-identity} term by term through the order shown. We have checked the higher orders, as well as confirmed the results for the higher spin eigenvalues.

\subsection{Limitations and open questions}
\label{sec:excited-open}
The match with the CFT results, to the orders checked, strongly supports
our proposal for the ODE dual to the identity primary of the Lee--Yang
CFT. A few caveats, however, are worth keeping in mind.

First, the closed form available for the ground state is only partly
inherited. Since the pole term is of order $\heff^2$, the classical
deformation is the same as for the ground state. However, we have not carried
out the Picard--Fuchs reduction of Section~\ref{sec:hbar2-level} at the
first quantum level, where the puncture period $\cC$ and the
trivial-monodromy constraint enter. For this state we have computed the
first-quantum correction only as a power series in $\kappa$, order by
order, up to high order; the exact expression valid for all $\kappa$,
written in terms of the elliptic periods as for the ground state,
remains to be found.

Second, the analytic control needed to promote the formal correspondence
to a finite-coupling identity is, if anything, more delicate here. The
functional relation of Section~\ref{sec:ode-ysystem} continues to hold,
since the five-fold covariance is unaffected by a pole that rotates with
the turning points. The behaviour of the Stokes multiplier at $E=0$,
however, which fixes the limit of the $Y$-function as
$\theta\to-\infty$, is now that of the excited state, and the analysis in
the strip must accommodate the moving pole. The identification with the
excited-state TBA of~\cite{Dorey:1996re,Bazhanov:1996aq}, in which the
pseudoenergy acquires additional source terms from a pair of complex
singularities, remains conjectural at finite $\kappa$.

Third, and physically most interesting, is the resummation of the
energy of this state. A large-order analysis of the coefficients
we have computed shows the same leading Borel structure as for the
ground state, with the same $S$ and $b$ and alternating signs. This is
expected, since the leading singularity is set by the dispersion
relation and is common to all states, but it does not by itself imply
Borel summability along the positive axis: as discussed at the end of
Section~\ref{sec:resummation}, excited states can develop square-root
branch points at real positive $\kappa_{\rm TBA}$~\cite{DAST-TBA}, which
require further singularities in the Borel plane, invisible at the
order we have reached. A comparison with the numerical solution of the
deformed excited-state TBA, analogous to that of
Section~\ref{sec:resummation}, 
requires solving that equation with the GGE deformation included; this is known to be subtle already in
the undeformed theory, since the pseudoenergy develops singularities in
the physical strip~\cite{Bazhanov:1996aq}. We leave this, and the
analysis of other excited states, to future work.
\section{Generalisation to an arbitrary set of chemical potentials}
\label{sec:multicoupling}
Although we have concentrated on the spin-$5$ ensemble, nothing in the
construction is restricted to a single chemical potential.  Let
$\mathcal S$ be an arbitrary set of Lee--Yang spins,  $s\equiv\pm1\pmod 6$, denote by
\be
   \boldsymbol{\kappa}=\{\kappa_s\}_{s\in\mathcal S}
\ee
the corresponding ODE deformation parameters.  The defining asymptotic
condition is the multicoupling version of \eqref{eq:defining-problem},
with the prescribed sources given by \eqref{eq:intro-driving-general}.
The physical chemical potentials are in general nonlinear functions
$\kappa_{{\rm TBA},s}=\kappa_{{\rm TBA},s}(\boldsymbol{\kappa})$, so that
different chemical potentials mix beyond linear order.  The prescribed
positive-spin sectors are retained, all other positive-spin terms are
cancelled, and the negative-spin terms are predictions for the deformed
conserved charges.

Already at the classical level the two allowed congruence classes have a
simple universal form.  For a single source of spin $s\equiv1\pmod6$, with
\be
z_s=\kappa_s E^{5(s-1)/6} ,
\ee
one finds
\begin{equation}
 Q_0=0,\qquad
 P_0=E\left(1+\frac56 z_s\right)^{6/5}.
 \label{eq:general-class-1}
\end{equation}
For $s\equiv5\pmod6$, the same inverse period map encountered in the spin-$5$
problem applies universally:
\begin{equation}
 \frac{P_0}{E}=R_0(u)^{-6/5},\qquad
 \frac{Q_0}{\kappa_s E^{(5s-1)/6}}
 =\frac{R_0(u)^{1/5}}{R_1(u)},\qquad
 z_s=u^{1/3}\frac{R_1(u)}{R_0(u)} .
 \label{eq:general-class-5}
\end{equation}
Only the powers of $E$ depend on $s$; the period map itself is universal
within each congruence class.  These two families exhaust the allowed
Lee--Yang spins.

For several simultaneous chemical potentials and at higher WKB orders,
no new principle is required. The expansion contains arbitrary coupling
monomials
\be
\boldsymbol{\kappa}^{\boldsymbol n}
=\prod_{s\in\mathcal S}\kappa_s^{n_s}.
\ee
For a fixed multi-index $\boldsymbol n$ and WKB level $K$, the
corresponding spin is
\be
r=1+\sum_{s\in\mathcal S} n_s(s-1)-2K.
\ee
The spin grading then determines whether the contribution belongs to a
prescribed positive-spin source, to the negative-spin tail, or must be
cancelled. In the latter case, the monomial-spin map of
\eqref{eq:monomial-spin-map} selects a unique allowed coefficient of
$P$ or $Q$. Ordering the formal expansion by total coupling degree and
WKB level therefore gives a triangular algebraic recursion: at each step
the new coefficient enters linearly, while its source is determined by
lower-order data. The retained source sectors instead determine the
nonlinear ODE--TBA map.

\section{Conclusions}
\label{sec:conclusions}
We have introduced a scheme for the ODE/IM description of generalised
Gibbs ensembles, and applied it to a chiral sector of the Lee--Yang CFT.
The chemical potentials for the local charges are encoded in a
deformation of the cubic oscillator by terms polynomial in the spectral
parameter, whose coefficients are fixed order by order by the WKB
period: the prescribed GGE sources are retained, every other
positive-spin contribution is either forbidden or cancelled by a
counterterm, and the negative-spin terms yield the deformed eigenvalues
of the conserved charges. The spin-$5$ ensemble provides a non-trivial
example in which this programme can be carried out explicitly. For the
ground state, both the classical deformation and its first quantum
correction are determined in closed form by the periods of the cubic
elliptic curve; the primary of the identity module is described by the
same construction, supplemented by a pole whose position depends on the coupling through a trivial-monodromy condition.

The construction has been tested in several ways. The deformed
eigenvalues of the first local charges obtained from the WKB periods
agree exactly, order by order, with those of an entirely separate
construction of the deformed charges directly in conformal field
theory~\cite{Downing:2024nfb,Karimi:2026kdv}, on both Lee--Yang
primaries. This agreement has a simple structural counterpart: the
natural expansion parameters of the two constructions, although defined
by entirely unrelated procedures, are related by an exact rational
factor once the periods of the elliptic curve are taken into account.
Finally, the formal, divergent series for the deformed ground-state
energy is Borel summable, and its resummation reproduces the numerically
exact solution of the deformed TBA with high accuracy across thirty
decades in the chemical potential, far beyond the perturbative regime
in which the series was derived.

It is important to keep in mind that the asymptotic expansion by itself
says nothing about the exact, finite-coupling solution of the TBA
equation; the resummation provides evidence only along the branch it
reconstructs. It is known from the modular analysis
of~\cite{Downing:2024nfb} that the Lee--Yang GGE TBA admits solutions
beyond this asymptotic series. By analogy with the free-fermion (Ising)
GGE, where such non-asymptotic behaviour has been analysed in
detail~\cite{Downing:2023lnp,Downing:2021mfw,Downing:2023lop}, one expects the physical
spectrum to become complex beyond some critical chemical potential.
Settling this for the Lee--Yang GGE, and thereby the true domain of
validity of the correspondence, is work in progress, combining the
present construction with a direct numerical solution of the TBA
equation with the additional driving terms.

Several directions deserve further investigation. The construction presented here was perturbative in the deformation parameter $\kappa$ and the final expression for $Q$ and $P$ is given as a power series in $\kappa$ and $E$. A non-perturbative expression for the potential may give an exact match to the TBA results without the need for Borel summation.

Also, extending the
construction to descendant states and to massive ODE/IM
systems~\cite{Lukyanov:2010rn,Dorey:2012bx,Ito:2018eon} would bring it
closer to a description of the full GGE partition function; on the ODE
side, descendants require several double poles, with positions coupled
by the trivial-monodromy
conditions~\cite{Bazhanov:2003ni,Fioravanti_2005,Conti:2020zzk},
while higher-rank
ODEs~\cite{Dorey:1999pv,DoreyDunningMasoero:2006,MasoeroRaimondo:2018,ItoZhu:2024}
should provide analogous deformations for more general integrable
theories.
An accessible testing ground for higher-rank ODE descriptions
of GGEs is the free symplectic-fermion theory with central charge $-2$,
whose GGE was studied in~\cite{Karimi:2026tvh}. There, the GGE can
involve commuting charges belonging to a much larger class of integrable
hierarchies, arising from $W$-algebras, which from the ODE/IM perspective
naturally suggests spectral problems associated with higher-order
differential operators \cite{Bazhanov:2001xm, MasoeroRaimondo:2018, Ashok:2024zmw, Ashok:2024ygp,Kudrna:2025bzg}. This theory may thus provide a simple setting in
which to understand how the period-based mechanism for deforming ODEs
extends to more general integrable hierarchies.

It would also be interesting to understand whether the same period-based
techniques can serve as an effective tool for integrable lattice models.
Direct connections between scaling limits of Bethe-ansatz systems and
ordinary differential equations have already emerged in inhomogeneous
XXZ spin chains~\cite{Gehrmann:2024kot,Kotousov:2025qfun}, while higher
conserved charges and their chemical potentials play a natural role in
generalised ensembles~\cite{Pozsgay:2013gqa,Ilievski:2015mza,Sedrakyan:2026chiral}.
This may provide a useful bridge between continuum ODE/IM methods and
generalised ensembles in quantum spin chains.
\section*{Acknowledgements}
We thank  Patrick Dorey, Clare Dunning,  Davide Fioravanti, Katsushi Ito,  Stefano Negro, Tom\'a\v{s} Proch\'azka, Marco Rossi, Ara Sedrakyan, Tigran Sedrakyan, Hongfei Shu and  Gerard Watts for many helpful discussions.
RT is a participant in the project HORIZON-MSCA-2023-SE-01-101182937-HeI.
This work was also partially supported by the INFN RdR ``Statistical Field Theory'' (SFT). FK would like to acknowledge funding from the STFC grant ST/Y509279/1 for this work.

\paragraph{Statement on  AI use:} The work presented in this paper was developed with some assistance from AI tools (ChatGPT and Claude). 

\appendix

\section{Elliptic periods}
\label{app:periods}

This appendix collects the results on the periods of the rescaled
cubic curve
\be
v(\xi)=\xi^3+a\xi-b,\qquad
\cP(a,b)=\oint_\gamma\sqrt{v}\,d\xi,
\ee
already introduced in \eqref{eq:elliptic-period}, that are used
throughout the main text. This includes a vanishing theorem that eliminates one
residue class of periods at every WKB level, a set of identities that
reduce the periods of the regular cubic problem to the two basic ones
$\cA,\cB$, and the hypergeometric representation \eqref{eq:R0R1} of
$\cP$ itself.

\paragraph{The vanishing theorem.}
Exactly one residue class of $n\bmod3$ has a vanishing period at every
WKB level; this fact recurs throughout the paper (starting at
Section~\ref{sec:one-spin-classical}) and is proved once here, for all
levels together. With $\omega=e^{2\pi i/3}$, the curve
$v=\xi^3+a\xi-b$ is invariant under $\xi\to\omega\xi$ together with
$a\to\omega^{-1}a$; expanding any WKB density around $(a,b)=(0,1)$, a
term with $i$ powers of $a$ has an integrand $\xi^{i-2K}f(\xi^3)\,d\xi$,
which acquires the phase $\omega^{i-2K+1}$ under $\xi\to\omega\xi$ and
integrates to zero along a cycle joining two branch points of
$\xi^3=1$ whenever $i-2K+1\equiv0\pmod3$; matching $i$ to the spin
through the weight relations of Section~\ref{sec:family} gives exactly
$r\equiv3\pmod6$. The same statement fixes which of the two periods in
\eqref{eq:R0R1} a given term is proportional to, according to whether
$r\equiv1$ or $5\pmod6$.

\paragraph{Reduction identities}
With $J(m,s):=\oint_\gamma\xi^m(\xi^3-1)^s\,d\xi$, the periods of the
regular cubic problem reduce to
$\mathcal A=J(0,\frac12)$ and $\mathcal B=J(1,-\frac12)$
by three elementary identities. The
algebraic relation $\xi^3=(\xi^3-1)+1$ gives
\be
J(m+3,s)=J(m,s+1)+J(m,s),
\ee
and the exact differential
$d[\xi^m(\xi^3-1)^s]=m\xi^{m-1}(\xi^3-1)^s\,d\xi+3s\,\xi^{m+2}(\xi^3-1)^{s-1}d\xi$,
integrated around the closed cycle, gives
\be
J(m+2,s-1)=-\frac{m}{3s}\,J(m-1,s),
\ee
which, combined with the algebraic relation at $m=0,1$, gives the
same-$m$ recursions
\be
J(0,s)=-\frac{3s}{3s+1}J(0,s-1),\qquad J(1,s)=-\frac{3s}{3s+2}J(1,s-1).
\ee
Together with $J(2,s)=0$ (the vanishing theorem above), these three
identities reduce all periods of the regular cubic problem to
$\cA,\cB$. In the identity-state problem, negative powers of $\xi$
introduce in addition the exceptional $m=-1$ period $\cC$ discussed
in Section~\ref{sec:excited-condition}.

\paragraph{Beta-function representation.}
Let $\gamma$ join the branch points $\omega^{j_1},\omega^{j_2}$ of
$s^3=1$, $\omega=e^{2\pi i/3}$, and deform it into two segments through
the origin. On $0<t<1$, continuity from $s=0$ (where $s^3-1=-1$) fixes
the branch, $(t^3-1)^{1/2-n}=i(-1)^n(1-t^3)^{1/2-n}$, so each segment
evaluates to an elementary Beta function,
\be
\int_0^{\omega^j}\!\!ds\,s^n(s^3-1)^{1/2-n}
=i(-1)^n\omega^{(n+1)j}\,\tfrac13B\!\left(\tfrac{n+1}{3},\tfrac32-n\right).
\ee
The closed cycle is twice the difference of these two segments: for
integer $n$, the monodromy picked up circling either endpoint is
$e^{2\pi i(1/2-n)}=-1$, so traversing the segment once on each of the
integrand's two branches is equivalent to traversing it twice on one
branch, giving
\be
\begin{aligned}
\oint_\gamma s^n(s^3-1)^{1/2-n}ds
&=\mathcal N_{[n]}\,(-1)^n\,\frac13\,B\!\left(\frac{n+1}{3},\frac32-n\right),\\
\quad
\text{with}\quad \mathcal N_{[n]}&=-2i\left(\omega^{(n+1)j_1}-\omega^{(n+1)j_2}\right).
\end{aligned}
\label{eq:In-beta}
\ee
Since $\mathcal
N_{[n]}$ depends on $n$ only through $n\bmod3$ and vanishes for
$n\equiv2\pmod3$ (where $n+1\equiv0\pmod3$ and both powers of $\omega$
equal $1$), this is the vanishing theorem above, restated in this
language.

\paragraph{Closed form for $\mathcal A,\mathcal B$.}
For our cycle, joining $\omega$ and $\omega^2$ ($j_1=1,j_2=2$),
\be
\mathcal N_{[0]}=-2i(\omega-\omega^2)=2\sqrt3~, \quad 
\mathcal N_{[1]}=-2i(\omega^2-\omega^4)=-2\sqrt3~. 
\ee 
Equations~\eqref{eq:In-beta} at $n=0,1$
therefore give $\mathcal A,\mathcal B$ directly:
\be
\mathcal A=2\sqrt3\cdot\frac13B\!\left(\tfrac13,\tfrac32\right)
=\frac{\sqrt3\,\Gamma(\tfrac13)\Gamma(\tfrac16)}{5\sqrt\pi},
\qquad
\mathcal B=2\sqrt3\cdot\frac13B\!\left(\tfrac23,\tfrac12\right)
=\frac{8\pi^{3/2}}{\Gamma(\tfrac13)\Gamma(\tfrac16)}~.
\ee
We have used $\Gamma(\tfrac32)=\tfrac12\sqrt\pi$, $\Gamma(\tfrac12)=\sqrt\pi$,
and the reflection formula $\Gamma(x)\Gamma(1-x)=\pi/\sin(\pi x)$ to
reduce the two Beta functions to the stated closed forms in
$\Gamma(\tfrac13),\Gamma(\tfrac16)$.

\paragraph{Derivation of the hypergeometric functions.}
Recall that 
\be 
\cP(a,b)=\oint_\gamma\sqrt{v}\,d\xi~, \quad \text{with} \quad v=\xi^3+a\xi-b~.
\ee 
Expanding the square root at $b=1$ and splitting the sum by residue
class of $n\bmod3$,
\be
\begin{aligned}
\cP(a,1)=\sum_{n\ge0}c_n\,a^n
&=\sum_{m\ge0}c_{3m}\,a^{3m}+\sum_{m\ge0}c_{3m+1}\,a^{3m+1}
+\sum_{m\ge0}c_{3m+2}\,a^{3m+2}~,\\
\text{with}\quad c_n&=\binom{1/2}{n}\,J\!\left(n,\tfrac12-n\right).
\end{aligned}
\ee
The last sum in the first line vanishes identically, $c_{3m+2}=0$ for every $m$, by the
vanishing theorem above; only the first two survive, and it is these
two sums that resum into hypergeometric functions. Using \eqref{eq:In-beta}\footnote{Near a branch point $s=\omega^{j}$ the integrand $s^n(s^3-1)^{1/2-n}$
behaves as $(s-\omega^{j})^{1/2-n}$, which is integrable only for $n\le1$. For
$n\ge2$ the integral along each of the two segments diverges, so the
segment formula behind \eqref{eq:In-beta} is formal. The integral along the
closed contour $\gamma$, which encircles the branch points, is finite and
equals the analytic continuation in the exponent of the Beta-function expression.
This is how $B(\tfrac{n+1}{3},\tfrac32-n)$, whose second argument is negative for
$n\ge2$, is to be read; the Gamma functions involved have no poles at these
arguments, so the result is finite.},
\be
c_n=\binom{1/2}{n}(-1)^n\,\mathcal N_{[n]}\,\tfrac13\,B\!\left(\tfrac{n+1}{3},\tfrac32-n\right).
\ee
The first argument of the Beta function advances by an integer only
when $n$ advances by three, so $c_{n+3}/c_n$, but not $c_{n+1}/c_n$,
is a rational function of $n$; this is the structural reason the two
surviving classes resum into two independent hypergeometric series
rather than one. For $n=3m$, writing $B(p,q)=\Gamma(p)\Gamma(q)/\Gamma(p+q)$
with $p=m+\frac13$, $q=\frac32-3m$ and using $\Gamma(z+1)=z\Gamma(z)$ on
each of the three Gamma-function ratios this generates, together with
the matching ratio of binomial coefficients, gives
\be
\frac{c_{3m+3}}{c_{3m}}
=-\frac{(m+\frac13)(-\frac16-2m)(\frac56-2m)}{(3m+1)(3m+2)(3m+3)}
=-\frac{4}{27}\,\frac{(m+\frac1{12})(m-\frac5{12})}{(m+\frac23)(m+1)} ,
\label{eq:R0-ratio}
\ee
a rational function of $m$, as required for a hypergeometric series.
Since successive coefficients of ${}_2F_1(\alpha,\beta;\gamma;x)$ have
the ratio $(\alpha+m)(\beta+m)x/[(\gamma+m)(m+1)]$, \eqref{eq:R0-ratio}
identifies
\be
R_0(u)=\sum_m\frac{c_{3m}}{c_0}u^m
={}_2F_1\!\left(-\tfrac5{12},\tfrac1{12};\tfrac23;-\tfrac{4u}{27}\right)
=1+\frac{5}{648}u-\frac{91}{419904}u^2+\dots
\ee
The class $n=3m+1$ goes through identically, with $p=m+\frac23$,
$q=\frac12-3m$ in place of $p,q$ above, giving
\be
\frac{c_{3m+4}}{c_{3m+1}}
=-\frac{4}{27}\,\frac{(m+\frac5{12})(m-\frac1{12})}{(m+1)(m+\frac43)}
\label{eq:R1-ratio}
\ee
and
\be
R_1(u)=\sum_m\frac{c_{3m+1}}{c_1}u^m
={}_2F_1\!\left(-\tfrac1{12},\tfrac5{12};\tfrac43;-\tfrac{4u}{27}\right)
=1+\frac{5}{1296}u-\frac{935}{5878656}u^2+\dots
\ee
The two normalisations are the basic periods, $c_0=\cA$ and
$c_1=\tfrac12\cB$; collecting the two surviving sums gives the closed
form used throughout the main text,
\be
\cP(a,1)=\cA\,R_0(u)+\frac{\cB}{2}\,u^{1/3}R_1(u).
\ee
Restoring $b$ via $\cP(a,b)=b^{5/6}\cP(a/b^{1/3},1)$
reproduces \eqref{eq:R0R1} used in the main text. 

\paragraph{Inverse map.}
Cubing $z=u^{1/3}R_1(u)/R_0(u)$ and inserting the series for $R_0,R_1$
above gives $z^3$ as a series in $u$:
\be
z^3=u-\frac{5}{432}u^2+\frac{401}{1306368}u^3+O(u^4),
\ee
and inverting this series gives $u$ as a
series in $z^3$:
\be
u=z^3+\frac{5}{432}z^6-\frac{17}{435456}z^9+\frac{2017}{564350976}z^{12}+O(z^{15})~.
\ee

\paragraph{Derivatives of $A$ and $B$.} For the first-quantum-level
computation of Section~\ref{sec:hbar2-level}, we have 
\be 
A=b^{5/6}R_0(u)~,\quad B=c_1ab^{1/6}R_1(u)~, \quad\text{with} \quad u=a^3/b^2~.
\ee 
Using $\partial_au=3u/a$, $\partial_bu=-2u/b$, and  denoting $d/du$ with primes, we obtain 
\begin{align}
\partial_aA&=3a^2b^{-7/6}R_0',\qquad
\partial_bA=b^{-1/6}\Big[\tfrac56R_0-2uR_0'\Big],
\nonumber\\
\partial_a\partial_bA&=\frac{3u}{a}\,b^{-1/6}\Big[-\tfrac76R_0'-2uR_0''\Big],
\nonumber\\
\partial_aB&=c_1b^{1/6}\Big[R_1+3uR_1'\Big],\qquad
\partial_bB=c_1ab^{-5/6}\Big[\tfrac16R_1-2uR_1'\Big],
\nonumber\\
\partial_a\partial_bB&=c_1b^{-5/6}\Big[\tfrac16R_1-\tfrac{15}{2}uR_1'-6u^2R_1''\Big],
\label{eq:AB-derivatives}
\end{align}
obtained by direct differentiation of $A,B$ using the chain rule
through $u$.
\section{WKB reductions}
\label{app:wkb}
Let us recall a few facts about the WKB expansion used in Section \ref{sec:family}. Writing $S=\psi'/\psi$, the Schr\"odinger equation
$[-\partial_x^2+V]\psi=0$ becomes the Riccati equation $S'+S^2=V$.
For the equation with a bookkeeping parameter,
$[-\hbar^2\partial_x^2+V]\psi=0$, set $S=\hbar\psi'/\psi$. Then
$\hbar S'+S^2=V$, and expanding
$S=\sum_n\hbar^nS_n$,
\be
S_0^2=V,\qquad S_0'+2S_0S_1=0,\qquad S_1'+2S_0S_2+S_1^2=0,\qquad\dots
\ee
Thus we have $S_0=\sqrt V$, $S_1=-V'/4V$ and, using
$S_1'+S_1^2=-\frac{V''}{4V}+\frac{5(V')^2}{16V^2}$,
\be
S_2=-\frac{S_1'+S_1^2}{2S_0}=\frac{V''}{8V^{3/2}}-\frac{5(V')^2}{32V^{5/2}}.
\ee
The branch of $\sqrt V$ in $S_2$ is that of $S_0$; the recursion fixes
the relative signs of the even-indexed coefficients. The part odd under
exchange of WKB branches controls
the periods, and \eqref{eq:Pi2-reduced} follows from the exact
differential $d(V'/V^{3/2})$. For the cubic potential $V=E(\xi^3+a\xi-b)$
one has moreover $V''=6E^{1/3}\xi$ and
\be
\Pi_2=\frac{1}{48}\oint\frac{6E^{1/3}\xi\;E^{1/3}d\xi}{E^{3/2}(\xi^3+a\xi-b)^{3/2}}
=\frac18E^{-5/6}\oint\frac{\xi\,d\xi}{(\xi^3+a\xi-b)^{3/2}}
=\frac12E^{-5/6}\,\partial_a\partial_b\cP(a,b),
\ee
which is \eqref{eq:Pi2-picard-fuchs}. The same manipulations express
every $\Pi_{2k}$ of the cubic curve as a differential operator in
$(a,b)$ acting on $\cP$: the integrand of $S_{2k}$ is a sum of terms
\be
\frac{V^{(i_1)}\cdots V^{(i_r)}}{V^{(2k-1)/2+r}},\qquad V^{(i)}\in\{V',V'',6\},
\ee
for the cubic, each of which is a derivative of $V^{-1/2}$ with respect
to $a$ and $b$ after integration by parts.

\paragraph{Periods of the WKB coefficients.}
For the computation of Section~\ref{sec:energy-wkb} one needs the
expansion of 
\be 
\pi_{2k}(a,b)=\oint_\gamma s_{2k}[\xi^3+a\xi-b]\,d\xi~,
\ee 
around $(a,b)=(0,1)$. Writing 
\be 
\pi_{2k}=\sum_{i,j}\pi_{2k}^{(i,j)}
a^i(b-1)^j~, 
\ee 
and reducing every $J(m,s)$ to the basic periods
$\mathcal A,\mathcal B$ by the identities of
Appendix~\ref{app:periods}, the
coefficients $\pi_{2k}^{(i,j)}$, $k=0,\dots,5$, $i\le2$, $j\le1$ are tabulated below. The zeros are the class $s\equiv3\pmod 6$; the alternation between
$\mathcal A$ and $\mathcal B$ follows the spin modulo $3$. 
\begin{center}
\renewcommand{\arraystretch}{1.3}
\begin{tabular}{c|cccc}
$k$ & $(0,0)$ & $(0,1)$ & $(1,0)$ & $(1,1)$ \\ \hline
$0$ & $\mathcal A$ & $\frac56\mathcal A$ & $\frac12\mathcal B$ & $\frac1{12}\mathcal B$\\
$1$ & $\frac{1}{24}\mathcal B$ & $-\frac{5}{144}\mathcal B$ & $0$ & $0$\\
$2$ & $0$ & $0$ & $\frac{637}{31104}\mathcal A$ & $-\frac{12103}{186624}\mathcal A$\\
$3$ & $-\frac{21983}{1119744}\mathcal A$ & $\frac{549575}{6718464}\mathcal A$ & $-\frac{55913}{1119744}\mathcal B$ & $\frac{1621477}{6718464}\mathcal B$\\
$4$ & $\frac{26317819}{376233984}\mathcal B$ & $-\frac{131589095}{322486272}\mathcal B$ & $0$ & $0$\\
$5$ & $0$ & $0$ & $\frac{38800584095}{17414258688}\mathcal A$ & $-\frac{1901228620655}{104485552128}\mathcal A$
\end{tabular}
\qquad
\begin{tabular}{c|cc}
$k$ & $(2,0)$ & $(2,1)$ \\ \hline
$0$ & $0$ & $0$\\
$1$ & $-\frac{35}{2592}\mathcal A$ & $\frac{455}{15552}\mathcal A$\\
$2$ & $\frac{1309}{62208}\mathcal B$ & $-\frac{30107}{373248}\mathcal B$\\
$3$ & $0$ & $0$\\
$4$ & $-\frac{1556782955}{2902376448}\mathcal A$ & $\frac{66941667065}{17414258688}\mathcal A$\\
$5$ & $\frac{779901080329}{139314069504}\mathcal B$ & $-\frac{41334757257437}{835884417024}\mathcal B$
\end{tabular}
\end{center}

\section{The WKB expansion for the identity state}
\label{sec:id-wkb}
In this Appendix, we provide details of the WKB analysis of the ODE \eqref{eq:monodromy-condition}.
After the rescaling $x=E^{1/3}\xi$ it takes the form
\be
\left[-\heff^2\frac{d^2}{d\xi^2}+v(\xi)\right]\psi=0,
\ee
with
\be
\begin{aligned}
&v=v_0+\heff^2v_2,\qquad v_0=\xi^3+a\xi-b,\qquad v_2=\frac{2}{(\xi-\zeta_0)^2},\\
&3\zeta_0^2+a=0,\qquad a=\frac{Q}{E^{2/3}},\quad b=\frac PE,\quad \zeta_0=\frac{z_0}{E^{1/3}},\quad \heff^2=E^{-5/3}.
\end{aligned}
\ee
For any $v$, with the formal parameter set to one (the odd densities are total derivatives), the
even WKB densities are $s_0=\sqrt v$,
\be
s_2=\frac{v''}{8v^{3/2}}-\frac{5v'^2}{32v^{5/2}},\qquad
s_4=\frac{v''''}{32v^{5/2}}-\frac{7v'v'''}{32v^{7/2}}-\frac{19v''^2}{128v^{7/2}}
+\frac{221v'^2v''}{256v^{9/2}}-\frac{1105v'^4}{2048v^{11/2}},
\ee
and $s_6$ follows from the same recursion \eqref{eq:S0S1S2}.

\paragraph{The periods and their integrands.} 
Since $v_2$ is multiplied by $\heff^2$, expanding $s_{2K}[v_0+\heff^2v_2]$ and collecting powers of
$\heff^2$ gives $\Pi=E^{5/6}\sum_K\heff^{2K}\pi_{2K}$ with
\be
\pi_{2K}=\sum_{m=0}^{K}\frac1{m!}\oint_\gamma\delta^{m}s_{2(K-m)}\,d\xi,\qquad
\delta^m s:=\partial_\eta^{\,m}\,s[v_0+\eta v_2]\Big|_{\eta=0}.
\ee
Explicitly,
\begin{align}
\pi_0&=\oint_\gamma\sqrt{v_0}\,d\xi,\\
\pi_2&=\oint_\gamma\Big[\frac{v_0''}{8v_0^{3/2}}-\frac{5v_0'^2}{32v_0^{5/2}}
+\frac{1}{(\xi-\zeta_0)^2\sqrt{v_0}}\Big]d\xi,\\
\pi_4&=\oint_\gamma\Big[s_4[v_0]+\delta s_2-\frac{v_2^2}{8v_0^{3/2}}\Big]d\xi,\qquad
\delta s_2=\frac{v_2''}{8v_0^{3/2}}-\frac{3v_2v_0''}{16v_0^{5/2}}-\frac{5v_0'v_2'}{16v_0^{5/2}}
+\frac{25v_0'^2v_2}{64v_0^{7/2}},
\end{align}
and similarly, $\pi_6=\oint[s_6[v_0]+\delta s_4+\tfrac12\delta^2s_2+\tfrac16\delta^3s_0]\,d\xi$. In $\pi_2$ the
first two terms integrate to the ground-state contribution $\tfrac18\oint\xi\,v_0^{-3/2}d\xi$ and the third is the
pole term. In each $\pi_{2K}$ one can thus distinguish the ground-type term ($m=0$) and the terms of
increasing order in $v_2$.

Every term is of the form
\be
\xi^m(\xi-\zeta_0)^{-k}\,v_0^{-p-1/2},
\ee
where $k\ge0$ counts the powers of $(\xi-\zeta_0)^{-1}$ coming from $v_2$ and its derivatives ($k=0$ for the
ground-type terms, $k=2$ for the first pole term). Expanding
\be
\begin{aligned}
(\xi-\zeta_0)^{-k}&=\sum_{n\ge0}\binom{n+k-1}{n}\,\zeta_0^{\,n}\,\xi^{-k-n},\\
v_0^{-p-1/2}&=\sum_{j\ge0}\frac{(-1)^j(p+\tfrac12)_j}{j!}\,(a\xi-\delta)^j\,(\xi^3-1)^{-p-1/2-j},
\end{aligned}
\ee
with $b=1+\delta$ and $(x)_j$ denoting the rising factorial, reduces each term to some $J(m,s)$, which is
reduced to $\cA,\cB,\cC$ by the identity \eqref{eq:excited-J}:
\be
J(m,s)=\frac{m+3s+4}{m+1}\,J(m+3,s),\qquad m\le-2,
\ee
the identities of Appendix~\ref{app:periods}, and $J(-1,s+1)=-J(-1,s)$. For example, at $b=1$,
\be
\oint_\gamma\frac{d\xi}{(\xi-\zeta_0)^2\sqrt{v_0}}
=\sum_{n,j\ge0}(n+1)\,\frac{(-1)^j\binom{2j}{j}}{4^{j}}\,\zeta_0^{\,n}a^{\,j}\,
J\!\left(j-n-2,\,-\tfrac12-j\right).
\ee
Applying this to each density, the periods $\pi_{2K}$ become power series in $\zeta_0$ and $a$ whose
coefficients are rational multiples of $\cA$, $\cB$ and $\cC$. We count $\zeta_0$ as weight $1$ and $a$ as
weight $2$, since $\zeta_0\sim z^{1/2}$ and $a\sim z$. The classical solution is $a=z$, $b=1$ up to
corrections of weight $6$ and higher (the first is $b_0=-z^3/108$), which do not matter at the weights
considered here, and the pole sits at $\zeta_0^2=-a/3$. Setting $b=1$ and $a=-3\zeta_0^2$, the results for the
period integrals, before any counterterm is included, are
\be 
\begin{aligned}
\pi_0&=\cA-\tfrac32\cB\,\zeta_0^2+O(\zeta_0^6),\\
\pi_2&=-\tfrac{11}{24}\cB-\tfrac56\cA\,\zeta_0+\tfrac54\cB\,\zeta_0^3+\tfrac{385}{288}\cA\,\zeta_0^4+O(\zeta_0^6),\\
\pi_4&=-\tfrac{55}{144}\cB\,\zeta_0-\tfrac{11557}{10368}\cA\,\zeta_0^2+\tfrac{23749}{6912}\cB\,\zeta_0^4
+\tfrac{8645}{1728}\cA\,\zeta_0^5+O(\zeta_0^6),\\
\pi_6&=-\tfrac{84227}{1119744}\cA-\tfrac{98923}{373248}\cB\,\zeta_0^2+O(\zeta_0^3).
\end{aligned}
\ee
The terms proportional to $\cC$ cancel at every weight shown: in $\pi_2$ they carry the factor
$a+3\zeta_0^2=v_0'(\zeta_0)$, and in $\pi_4$ they cancel between the terms linear and quadratic in $v_2$.
\subsection{The asymptotic expansion}
A monomial $\zeta_0^{w}$ in $\pi_{2K}$ contributes $\zeta_0^w\heff^{2K}E^{5/6}$ to $\Pi$. Using
$\heff^{2K}=E^{-5K/3}$ and $\zeta_0^2=-z/3$, $z=\kappa E^{10/3}$, this is
\be
\rho^{\,w}\,E^{5r/6},\qquad r=1+2w-2K,\qquad \rho:=\sqrt{-\kappa/3},
\ee
where $\rho$ is the part of $\zeta_0$ independent of $E$, $\zeta_0=\rho\,E^{5/3}$. Thus $E$ enters
through the prefactor, through $\heff^{2K}$ and through $\zeta_0^w$; the spin $r$ determines which of
$D,D_5,\widetilde I_s$ the monomial belongs to, and the power of $\rho$ gives its $\kappa$-dependence:
even powers of $\rho$ are integer powers of $\kappa$ ($\rho^2=-\kappa/3$), odd powers are half-integer
powers. Collecting the periods above by spin,
\begin{align}
\Pi\Big|_{\rm no\ c.t.}={}&E^{35/6}\Big[\tfrac{385}{288}\cA\rho^4+\tfrac{8645}{1728}\cA\rho^5+\dots\Big]\nonumber\\
&+E^{25/6}\Big[-\tfrac32\cB\rho^2+\tfrac54\cB\rho^3+\tfrac{23749}{6912}\cB\rho^4+\dots\Big]\nonumber\\
&+E^{5/6}\Big[\cA-\tfrac56\cA\rho-\tfrac{11557}{10368}\cA\rho^2+\dots\Big]\nonumber\\
&+E^{-5/6}\Big[-\tfrac{11}{24}\cB-\tfrac{55}{144}\cB\rho-\tfrac{98923}{373248}\cB\rho^2+\dots\Big]\nonumber\\
&+E^{-25/6}\Big[-\tfrac{84227}{1119744}\cA+\dots\Big],
\end{align}
with no terms of spin $\pm3$. Two things differ from the form required by the defining condition.
First, it contains a spin-$7$ term ($E^{35/6}$), which the defining condition does not allow. Second,
odd powers of $\rho$, i.e.\ half-integer powers of $\kappa$, appear in the slots of spin $7$, $5$, $1$
and $-1$. These must cancel, since $D$, $D_5$ and $\widetilde I_s$ are power series in $\kappa$
(equivalently, for real $\kappa>0$, $\rho$ is imaginary, so odd terms would be imaginary contributions
to a real quantity). In the slots of spin $1$ and $5$ the cancellation is achieved by the counterterms
of the next subsection; in the negative-spin slots there is no freedom, so this is a consistency check.
\subsection{Counterterms and the recalculated periods}
\label{sec:id-counterterms}
As for the ground state, the potential is corrected order by order in $\heff^2$,
$a=z+\heff^2a_1+\heff^4a_2+\dots$ and $b=1+\heff^2b_1+\heff^4b_2+\dots$, and a monomial
$\zeta_0^w$ in $a_{K'}$ or $b_{K'}$ contributes to $\Pi$ with spin $r=1+2w-2K'$. By
\eqref{eq:monomial-spin-map} $b$ carries spins $r\equiv1$ and $a$ spins $r\equiv5\pmod 6$, so up to
weight $5$ the available slots are
\begin{center}
\begin{tabular}{c|cc|cc}
 & \multicolumn{2}{c|}{$b$} & \multicolumn{2}{c}{$a$}\\\hline
$K'=1$ & $w=1\ (r{=}1)$ & $w=4\ (r{=}7)$ & $w=3\ (r{=}5)$ & \\
$K'=2$ & $w=2\ (r{=}1)$ & $w=5\ (r{=}7)$ & $w=4\ (r{=}5)$ &
\end{tabular}
\end{center}
Counterterms in $a$ also affect the next order: a monomial of weight $w_c$ in $a$ feeds, through the
motion of the pole, into the next order at weight $w_c-1$. To compute $\pi_4^{\rm eff}$ through weight $5$
one therefore needs $a_1$ through $\zeta_0^6$ (spin $11$), although the table stops at weight $5$.
We write the corresponding counterterms as
\be
\begin{aligned}
a_1&=\alpha\,\zeta_0^3, & b_1&=\beta\,\zeta_0+b_{14}\,\zeta_0^4,\\
a_2&=\alpha_2\,\zeta_0^4, & b_2&=\gamma\,\zeta_0^2+g_5\,\zeta_0^5 .
\end{aligned}
\label{eq:id-counterterms}
\ee
Recall that $\zeta_0$ stands for $\sqrt{-z/3}$, so that $\zeta_0^w=(-z/3)^{w/2}$ and the counterterms are series
in $z^{1/2}$: the even powers of $\zeta_0$ are the integer powers of $z$ familiar from the ground state, and
the odd powers are new for the identity state.
The slots of spin $1$ and $5$ are not fixed by the defining condition, since they renormalise $D$
and $D_5$. As for the ground state, we set the integer-power ones ($\gamma,\alpha_2$) to
zero.\footnote{These are free slots: $\gamma$ gives $P\ni-\tfrac\gamma3\kappa E$, to leading order a
shift $E\to E(1+c\,\kappa)$ that only changes $D$, and $\alpha_2$ gives
$Q\ni\tfrac{\alpha_2}9\kappa^2E^4$, a shift $\kappa\to\kappa(1+c\,\kappa)$ that only changes $D_5$ and
the relation to $\kappa_{\rm TBA}$. Setting them to zero is the same convention as for the ground
state.} The half-integer ones
($\beta,\alpha$) we fix so that $D$ and $D_5$ are analytic in $\kappa$; any other value merely
reparametrises $E$ and $\kappa$, and drops out of physical quantities. 

The corrected $a,b$ are inserted into $\pi_{2K}(a,b)$ and expanded around the classical point, the
pole following $a$ through $\zeta_0^2=-a/3$ (so $\partial_a=-\frac1{6\zeta_0}\partial_{\zeta_0}$ on
the identity-state periods). At $\heff^2$ and $\heff^4$,
\begin{align}
\pi_2^{\rm eff}&=\pi_2+\partial_b\pi_0\,b_1+\partial_a\pi_0\,a_1,\\
\pi_4^{\rm eff}&=\pi_4+\partial_b\pi_2\,b_1+\partial_a\pi_2\,a_1+\tfrac12\partial_b^2\pi_0\,b_1^2
+\partial_a\partial_b\pi_0\,a_1b_1+\partial_b\pi_0\,b_2+\partial_a\pi_0\,a_2+\dots
\end{align}
The ingredients are
\begin{align}
&\partial_b\pi_0=\tfrac56\cA-\tfrac14\cB\zeta_0^2+\dots,~~ \partial_a\pi_0=\tfrac12\cB+\dots,~~
\tfrac12\partial_b^2\pi_0=-\tfrac5{72}\cA+\dots,~~ \partial_a\partial_b\pi_0=\tfrac1{12}\cB+\dots,\\
&\partial_b\pi_2=\tfrac{55}{144}\cB+\tfrac{35}{36}\cA\zeta_0-\tfrac{55}{24}\cB\zeta_0^3-\tfrac{5005}{1728}\cA\zeta_0^4+\dots,\\
&\partial_a\pi_2=\tfrac{5}{36}\frac{\cA}{\zeta_0}-\tfrac58\cB\zeta_0-\tfrac{385}{432}\cA\zeta_0^2+\dots
\end{align}

\paragraph{Order $\heff^2$:}
With the counterterms,
\be
\pi_2^{\rm eff}=-\tfrac{11}{24}\cB+\tfrac56\cA(\beta-1)\zeta_0
+\tfrac14\cB(2\alpha-\beta+5)\zeta_0^3+\tfrac5{288}\cA(48b_{14}+77)\zeta_0^4+O(\zeta_0^6).
\ee
We make the following observations:
\begin{itemize}
    \item The odd term that could contribute to $D$ (spin $1$) is removed by $\beta=1$. 
    \item Using this, we see that the odd term that could contribute to $D_5$ (spin $5$) vanishes for $\alpha=-2$. 
    \item The spin-$7$ term is cancelled by $b_{14}=-\tfrac{77}{48}$, i.e.\
$b_1=\zeta_0-\tfrac{77}{432}z^2+\dots$, to be compared with $\tfrac7{432}z^2$ for the ground state.
\end{itemize}
The slots of spin $3$ and $9$ ($w=2,5$) vanish without any counterterm. The same logic applied to the terms of weight $6$ and $7$ fixes the higher terms of $a_1$ and $b_1$ (spins
$11$ and $13$). Together with the classical $b_0=\tfrac14\zeta_0^6=-\tfrac{z^3}{108}$ of the main text, which
first enters at weight $6$, one finds $a_1\ni\tfrac{77}{24}\zeta_0^6$; the $\zeta_0^7$ term of $b_1$ only
matters at weight $7$. Thus we have fixed the first quantum corrections $a_1$ and $b_1$.

\paragraph{Order $\heff^4$:}
Inserting $\beta=1$, $\alpha=-2$, $b_{14}=-\tfrac{77}{48}$ into $\pi_4^{\rm eff}$, and keeping the
parameters $\gamma$, $\alpha_2$, $g_5$ of $b_2$ and $a_2$ undetermined, we find
\be
\pi_4^{\rm eff}=\Big(\tfrac56\gamma-\tfrac{5077}{10368}\Big)\cA\,\zeta_0^2
+\Big(\tfrac12\alpha_2-\tfrac14\gamma+\tfrac{5941}{3456}\Big)\cB\,\zeta_0^4
+\tfrac56\Big(g_5+\tfrac{259}{72}\Big)\cA\,\zeta_0^5+O(\zeta_0^6).
\ee
We make the following observations:
\begin{itemize}
\item The terms of weight $0$, $1$ and $3$ vanish. There is no spin $-3$ term. The odd term in
$\widetilde I_1$ (spin $-1$, $w=1$) cancels thanks to the same $\beta=1$ that removed the odd term in 
$D$ at order $\heff^2$, which is a consistency check, since no freedom remains in this slot.
The spin-$3$ term vanishes without any counterterm.
\item The terms of weight $2$ and $4$ have spins $1$ and $5$ and integer powers of $\kappa$. They only
renormalise $D$ and $D_5$, so no condition is imposed on $\gamma$ and $\alpha_2$: they remain free, and
we set them to zero.
\item The term of weight $5$ has spin $7$ and a half-integer power of $\kappa$, so it must vanish. This
requires $g_5=-\tfrac{259}{72}$, where $\tfrac{259}{72}=\tfrac{49}{16}+\tfrac16\cdot\tfrac{77}{24}$: the
second piece is the contribution of the $\tfrac{77}{24}\zeta_0^6$ term of $a_1$ through the motion of the pole.
\end{itemize}
Thus $b_2=\gamma\,\zeta_0^2-\tfrac{259}{72}\zeta_0^5$ and $a_2=\alpha_2\,\zeta_0^4$, with $\gamma$ and
$\alpha_2$ free.

\paragraph{The recalculated periods.}
With $(\beta,\alpha,b_{14},\gamma,g_5,\alpha_2)=(1,-2,-\tfrac{77}{48},0,-\tfrac{259}{72},0)$ the
coefficients of $\zeta_0^w$ in $\pi_{2K}^{\rm eff}$ are
\begin{center}
\begin{tabular}{c|cccccc}
 & $w=0$ & $1$ & $2$ & $3$ & $4$ & $5$\\\hline
$K=0$ & $\cA$ & $0$ & $-\tfrac32\cB$ & $0$ & $0$ & $0$\\
$K=1$ & $-\tfrac{11}{24}\cB$ & $0$ & $0$ & $0$ & $0$ & $0$\\
$K=2$ & $0$ & $0$ & $-\tfrac{5077}{10368}\cA$ & $0$ & $\tfrac{5941}{3456}\cB$ & $0$\\
$K=3$ & $-\tfrac{84227}{1119744}\cA$ & $0$ & $\tfrac{126797}{373248}\cB$ & & &
\end{tabular}
\end{center}
Collecting the nonzero entries of the table by spin, with $\rho^2=-\kappa/3$, the large-$E$ expansion
is now of the required form, with no spin $7$ and no odd powers of $\rho$:
\begin{align}
\Pi={}&E^{5/6}\Big[\cA-\tfrac{5077}{10368}\cA\,\rho^2+\dots\Big]
+E^{25/6}\Big[-\tfrac32\cB\,\rho^2+\tfrac{5941}{3456}\cB\,\rho^4+\dots\Big]\nonumber\\
&+E^{-5/6}\Big[-\tfrac{11}{24}\cB+\tfrac{126797}{373248}\cB\,\rho^2+\dots\Big]
+E^{-25/6}\Big[-\tfrac{84227}{1119744}\cA+\dots\Big]+\dots\nonumber\\
={}&E^{5/6}\,\cA\Big[1+\tfrac{5077}{31104}\kappa+\dots\Big]
+E^{25/6}\,\tfrac12\cB\kappa\Big[1+\tfrac{5941}{15552}\kappa+\dots\Big]\nonumber\\
&-E^{-5/6}\,\tfrac{11}{24}\cB\Big[1+\tfrac{11527}{46656}\kappa+\dots\Big]
-E^{-25/6}\,\tfrac{84227}{1119744}\cA\,\big[1+O(\kappa)\big]+\dots
\end{align}
In the first two lines each term is the entry $(K,w)$ of the table, multiplying $\rho^wE^{5r/6}$ with
$r=1+2w-2K$: the brackets contain the entries $(0,0),(2,2)$, then $(0,2),(2,4)$, then $(1,0),(3,2)$, and
$(3,0)$, respectively.  

\paragraph{The dictionary and the eigenvalues of $I_1$ and $I_5$.}
Let $x:=\cA^5\kappa_{\rm TBA}/\cB$. From $\kappa_{\rm TBA}=D_5/D^5$ and the brackets above,
\be
x=\frac{\kappa}{2}\Big[1-\frac{4501}{10368}\,\kappa+O(\kappa^2)\Big],\qquad
\kappa=2x\Big[1+\frac{4501}{5184}\,x+O(x^2)\Big],
\ee
$\widetilde{I}_1(\kappa)$ and $\widetilde{I}_5(\kappa)$ can be read off from the coefficients of $E^{-5/6}$ and $E^{-25/6}$ respectively. 
Using the relation \eqref{eq:I1I5-relation},
$N_sI_s=D^s\widetilde I_s$, and expanding as a power series expansion in $x$ then gives the linear-in-$x$ correction to the eigenvalues $I_s$ as a series in $x$. Continuing in the same way to include the contributions from the higher-order WKB periods (this can be automated), we find
\begin{align}
I_1^{(\id)}
&=\frac{11}{60}+\frac{84227}{559872}\,x-\frac{42166065305}{2902376448}\,x^2+O(x^3),\\
I_5^{(\id)}&=\frac{341}{756000}-\frac{4877509}{111974400}\,x+\frac{4410941121587}{69657034752}\,x^2+O(x^3)
\end{align}
which, with $\gamma=\tfrac{216125}{648}\,x$, reproduces the expansion \eqref{eq:I1-cft-identity} of the CFT
eigenvalue through second order. By comparing with the results of \cite{Karimi:2026kdv}, we also confirm that we match the higher-order terms of both eigenvalues.

\bibliographystyle{JHEP}
\bibliography{ref}

@article{Martins:1991zy,
    author = "Martins, Marcio Jose",
    title = "{Complex excitations in the thermodynamic Bethe ansatz approach}",
    reportNumber = "UCSBTH-91-14",
    doi = "10.1103/PhysRevLett.67.419",
    journal = "Phys. Rev. Lett.",
    volume = "67",
    pages = "419--421",
    year = "1991"
}

@article{Fendley:1999gb,
    author = "Fendley, Paul",
    title = "{Airy functions in the thermodynamic Bethe ansatz}",
    doi = "10.1023/A:1007658622109",
    journal = "Lett. Math. Phys.",
    volume = "49",
    pages = "229--233",
    year = "1999",
    eprint = "hep-th/9906114",
    archivePrefix = "arXiv"
}

@article{Fendley:1991bj,
    author = "Fendley, Paul",
    title = "{Excited state thermodynamics}",
    reportNumber = "BUHEP-91-16",
    doi = "10.1016/0550-3213(92)90404-Y",
    journal = "Nucl. Phys. B",
    volume = "374",
    pages = "667--691",
    year = "1992",
    eprint = "hep-th/9109021",
    archivePrefix = "arXiv"
}

@unpublished{DAST-TBA,
    author = "Ashok, Sujay K. and Downing, Max and Sudhakar, Adarsh and Tateo, Roberto",
    title  = "{Excited states and complex spectra in the Lee--Yang generalised Gibbs ensemble}",
    note   = "in preparation",
    year   = "2026"
}

@article{Aniceto:2018bis,
  author  = {Aniceto, In{\^e}s and Ba{\c{s}}ar, G{\"o}khan and Schiappa, Ricardo},
  title   = {{A Primer on Resurgent Transseries and Their Asymptotics}},
  journal = {Phys. Rept.},
  volume  = {809},
  pages   = {1--135},
  year    = {2019},
  doi     = {10.1016/j.physrep.2019.02.003},
  eprint  = {1802.10441},
  archivePrefix = {arXiv},
  primaryClass  = {hep-th}
}

@article{Dorigoni:2014hea,
  author  = {Dorigoni, Daniele},
  title   = {{An Introduction to Resurgence, Trans-Series and Alien Calculus}},
  journal = {Annals Phys.},
  volume  = {409},
  pages   = {167914},
  year    = {2019},
  doi     = {10.1016/j.aop.2019.167914},
  eprint  = {1411.3585},
  archivePrefix = {arXiv},
  primaryClass  = {hep-th}
}

@book{BenderOrszag:1978,
    author    = "Bender, Carl M. and Orszag, Steven A.",
    title     = "{Advanced Mathematical Methods for Scientists and Engineers}",
    publisher = "McGraw-Hill",
    address   = "New York",
    year      = "1978"
}

@article{Arutyunov:2009ur,
    author = "Arutyunov, Gleb and Frolov, Sergey",
    title = "{Thermodynamic Bethe Ansatz for the AdS(5) x S(5) Mirror Model}",
    eprint = "0903.0141",
    archivePrefix = "arXiv",
    primaryClass = "hep-th",
    reportNumber = "ITP-UU-09-09, SPIN-09-09, TCDMATH-09-09, HMI-09-05",
    doi = "10.1088/1126-6708/2009/05/068",
    journal = "JHEP",
    volume = "05",
    pages = "068",
    year = "2009"
}

@article{Arutyunov:2007tc,
    author = "Arutyunov, Gleb and Frolov, Sergey",
    title = "{On String S-matrix, Bound States and TBA}",
    eprint = "0710.1568",
    archivePrefix = "arXiv",
    primaryClass = "hep-th",
    reportNumber = "ITP-UU-07-50, SPIN-07-37, TCDMATH-07-15",
    doi = "10.1088/1126-6708/2007/12/024",
    journal = "JHEP",
    volume = "12",
    pages = "024",
    year = "2007"
}

@article{Fioravanti_2005,
   title={Geometrical loci and CFTs via the Virasoro symmetry of the mKdV-SG hierarchy: an excursus},
   volume={609},
   ISSN={0370-2693},
   url={http://dx.doi.org/10.1016/j.physletb.2005.01.037},
   DOI={10.1016/j.physletb.2005.01.037},
   number={1-2},
   journal={Physics Letters B},
   publisher={Elsevier BV},
   author={Fioravanti, Davide},
   year={2005},
   month=Mar, pages={173–179} }

@article{Downing:2025huv,
    author = "Downing, Max and Karimi, Faisal and Sengupta, Tanmoy and Sudhakar, Adarsh and Watts, G{\'e}rard M. T.",
    title = "{Modular properties of $ {\mathcal{W}}_3 $ generalised Gibbs ensembles}",
    eprint = "2508.16258",
    archivePrefix = "arXiv",
    primaryClass = "hep-th",
    doi = "10.1007/JHEP06(2026)114",
    journal = "JHEP",
    volume = "06",
    pages = "114",
    year = "2026"
}

@article{Ashok:2026cyi,
    author = "Ashok, Sujay K. and Sengupta, Tanmoy and Sudhakar, Adarsh and Watts, G{\'e}rard M. T.",
    title = "{Universal modular properties of generalized Gibbs ensembles and chiral deformations}",
    eprint = "2603.28244",
    archivePrefix = "arXiv",
    primaryClass = "hep-th",
    doi = "10.1007/JHEP08(2026)168",
    journal = "JHEP",
    volume = "08",
    pages = "168",
    year = "2026"
}

@article{Sibuya:1984fe,
    author = "Sibuya, Yasutaka",
    title = "{On the Functional Equation $f(\lambda)+f(\omega\lambda)f(\omega^{-1}\lambda)=1$, $(\omega^5=1)$}",
    journal = "Les rencontres physiciens-mathematiciens de Strasbourg - RCP25",
    volume = "34",
    pages = "91--103",
    year = "1984",
    publisher = "Institut de Recherche Mathematique Avancee - Universite Louis Pasteur"
}

@article{Kotousov:2025qfun,
    author = "Kotousov, Gleb A. and Lukyanov, Sergei L. and Shabetnik, Daria A.",
    title = "{Scaling limit of $Q$-functions for the ${\cal Z}_r$ invariant inhomogeneous XXZ spin-$\frac{1}{2}$ chain near free fermion point}",
    eprint = "2511.20530",
    archivePrefix = "arXiv",
    primaryClass = "hep-th",
    doi = "10.1016/j.nuclphysb.2026.117315",
    journal = "Nucl. Phys. B",
    volume = "1024",
    pages = "117315",
    year = "2026"
}

@article{Gehrmann:2024kot,
    author = "Gehrmann, Sascha and Kotousov, Gleb A. and Lukyanov, Sergei L.",
    title = "{Scaling limit of the ground state Bethe roots for the inhomogeneous XXZ spin-$\frac{1}{2}$ chain}",
    eprint = "2406.12102",
    archivePrefix = "arXiv",
    primaryClass = "hep-th",
    doi = "10.1016/j.nuclphysb.2024.116624",
    journal = "Nucl. Phys. B",
    volume = "1006",
    pages = "116624",
    year = "2024"
}

@article{CastroAlvaredo:2022ys,
    author = "Castro-Alvaredo, Olalla A.",
    title = "{Y-systems for generalised Gibbs ensembles in integrable quantum field theory}",
    eprint = "2204.11749",
    archivePrefix = "arXiv",
    primaryClass = "hep-th",
    doi = "10.1088/1751-8121/ac9162",
    journal = "J. Phys. A",
    volume = "55",
    number = "40",
    pages = "405402",
    year = "2022"
}

@book{ItoShu:2025,
    author = "Ito, Katsushi and Shu, Hongfei",
    title = "{ODE/IM Correspondence and Quantum Periods}",
    series = "SpringerBriefs in Mathematical Physics",
    volume = "51",
    publisher = "Springer",
    address = "Singapore",
    doi = "10.1007/978-981-96-0499-9",
    isbn = "978-981-96-0498-2",
    year = "2025"
}

@article{Kuniba:2010ir,
    author = "Kuniba, Atsuo and Nakanishi, Tomoki and Suzuki, Junji",
    title = "{T-systems and Y-systems in integrable systems}",
    eprint = "1010.1344",
    archivePrefix = "arXiv",
    primaryClass = "hep-th",
    doi = "10.1088/1751-8113/44/10/103001",
    journal = "J. Phys. A",
    volume = "44",
    pages = "103001",
    year = "2011"
}

@article{Ravanini:1992fi,
    author = "Ravanini, Francesco and Tateo, Roberto and Valleriani, Alessandro",
    title = "{Dynkin TBAs}",
    eprint = "hep-th/9207040",
    archivePrefix = "arXiv",
    reportNumber = "DFUB-92-08",
    doi = "10.1142/S0217751X93000709",
    journal = "Int. J. Mod. Phys. A",
    volume = "8",
    pages = "1707--1728",
    year = "1993"
}

@article{Klassen:1990dx,
    author = "Klassen, Timothy R. and Melzer, Ezer",
    title = "{The Thermodynamics of purely elastic scattering theories and conformal perturbation theory}",
    doi = "10.1016/0550-3213(91)90159-U",
    journal = "Nucl. Phys. B",
    volume = "350",
    pages = "635--689",
    year = "1991"
}

@article{Smirnov:1989fa,
    author = "Smirnov, F. A.",
    title = "{The Perturbed $c<1$ Conformal Field Theories as Reductions of Sine-Gordon Model}",
    doi = "10.1142/S0217751X89001714",
    journal = "Int. J. Mod. Phys. A",
    volume = "4",
    pages = "4213--4220",
    year = "1989"
}

@article{Essler:2014mya,
    author = "Essler, Fabian H. L. and Mussardo, Giuseppe and Panfil, Miłosz",
    title = "{Generalized Gibbs Ensembles for Quantum Field Theories}",
    eprint = "1411.5352",
    archivePrefix = "arXiv",
    primaryClass = "cond-mat.quant-gas",
    doi = "10.1103/PhysRevA.91.051602",
    journal = "Phys. Rev. A",
    volume = "91",
    number = "5",
    pages = "051602",
    year = "2015"
}

@article{Rigol:2006py,
    author = "Rigol, Marcos and Dunjko, Vanja and Yurovsky, Vladimir and Olshanii, Maxim",
    title = "{Relaxation in a Completely Integrable Many-Body Quantum System: An Ab Initio Study of the Dynamics of the Highly Excited States of 1D Lattice Hard-Core Bosons}",
    eprint = "cond-mat/0604476",
    archivePrefix = "arXiv",
    doi = "10.1103/PhysRevLett.98.050405",
    journal = "Phys. Rev. Lett.",
    volume = "98",
    pages = "050405",
    year = "2007"
}

@article{ItoZhu:2024,
    author = "Ito, Katsushi and Zhu, Mingshuo",
    title = "{Integrals of motion in conformal field theory with W-symmetry and the ODE/IM correspondence}",
    eprint = "2408.12917",
    archivePrefix = "arXiv",
    primaryClass = "hep-th",
    doi = "10.1016/j.nuclphysb.2024.116756",
    journal = "Nucl. Phys. B",
    volume = "1010",
    pages = "116756",
    year = "2025"
}

@incollection{FeiginFrenkel:2007,
    author = "Feigin, Boris and Frenkel, Edward",
    title = "{Quantization of soliton systems and Langlands duality}",
    eprint = "0705.2486",
    archivePrefix = "arXiv",
    primaryClass = "math.QA",
    booktitle = "{Exploring New Structures and Natural Constructions in Mathematical Physics}",
    series = "Adv. Stud. Pure Math.",
    volume = "61",
    pages = "185--274",
    year = "2011",
    doi = "10.2969/aspm/06110185"
}

@article{MasoeroRaimondo:2018,
    author = "Masoero, Davide and Raimondo, Andrea",
    title = "{Opers for higher states of quantum KdV models}",
    eprint = "1812.00228",
    archivePrefix = "arXiv",
    primaryClass = "math-ph",
    doi = "10.1007/s00220-020-03792-3",
    journal = "Commun. Math. Phys.",
    volume = "378",
    number = "1",
    pages = "1--74",
    year = "2020"
}

@article{DoreyDunningMasoero:2006,
    author = "Dorey, Patrick and Dunning, Clare and Masoero, Davide and Suzuki, Junji and Tateo, Roberto",
    title = "{Pseudo-differential equations, and the Bethe ansatz for the classical Lie algebras}",
    eprint = "hep-th/0612298",
    archivePrefix = "arXiv",
    doi = "10.1016/j.nuclphysb.2007.02.029",
    journal = "Nucl. Phys. B",
    volume = "772",
    pages = "249--289",
    year = "2007"
}

@article{ItoMarinoShu:2018,
    author = "Ito, Katsushi and Mari{\~n}o, Marcos and Shu, Hongfei",
    title = "{TBA equations and resurgent Quantum Mechanics}",
    eprint = "1811.04812",
    archivePrefix = "arXiv",
    primaryClass = "hep-th",
    doi = "10.1007/JHEP01(2019)228",
    journal = "JHEP",
    volume = "01",
    pages = "228",
    year = "2019"
}

@article{Ito:2021hdy,
    author = "Ito, Katsushi and Kondo, Takayasu and Kuroda, Kohei and Shu, Hongfei",
    title = "{WKB periods for higher order ODE and TBA equations}",
    eprint = "2104.13680",
    archivePrefix = "arXiv",
    primaryClass = "hep-th",
    doi = "10.1007/JHEP10(2021)167",
    journal = "JHEP",
    volume = "10",
    pages = "167",
    year = "2021"
}

@article{Aramini:2022wxi,
    author = "Aramini, Fabrizio and Brizio, Nicolò and Negro, Stefano and Tateo, Roberto",
    title = "{Deforming the ODE/IM correspondence with $T\bar T$}",
    eprint = "2212.13957",
    archivePrefix = "arXiv",
    primaryClass = "hep-th",
    doi = "10.1007/JHEP03(2023)084",
    journal = "JHEP",
    volume = "03",
    pages = "084",
    year = "2023"
}

@article{Fioravanti:2021gbd,
    author = "Fioravanti, Davide and Rossi, Marco",
    title = "{On the origin of the correspondence between classical and quantum integrable theories}",
    eprint = "2106.07600",
    archivePrefix = "arXiv",
    primaryClass = "hep-th",
    doi = "10.1016/j.physletb.2023.137706",
    journal = "Phys. Lett. B",
    volume = "838",
    pages = "137706",
    year = "2023"
}

@article{Ito:2018eon,
    author = "Ito, Katsushi and Shu, Hongfei",
    title = "{Massive ODE/IM Correspondence and Non-linear Integral Equations for $A_r^{(1)}$-type modified Affine Toda Field Equations}",
    eprint = "1805.08062",
    archivePrefix = "arXiv",
    primaryClass = "hep-th",
    doi = "10.1088/1751-8121/aad63f",
    journal = "J. Phys. A",
    volume = "51",
    number = "38",
    pages = "385401",
    year = "2018"
}

@article{Gaiotto:2009hg,
    author = "Gaiotto, Davide and Moore, Gregory W. and Neitzke, Andrew",
    title = "{Wall-crossing, Hitchin Systems, and the WKB Approximation}",
    eprint = "0907.3987",
    archivePrefix = "arXiv",
    primaryClass = "hep-th",
    doi = "10.1016/j.aim.2012.09.027",
    journal = "Adv. Math.",
    volume = "234",
    pages = "239--403",
    year = "2013"
}

@article{Lencses:2014cba,
    author = "Lencsés, M. and Takács, G.",
    title = "{Excited state TBA and renormalized TCSA in the scaling Potts model}",
    eprint = "1405.3157",
    archivePrefix = "arXiv",
    primaryClass = "hep-th",
    doi = "10.1007/JHEP09(2014)052",
    journal = "JHEP",
    volume = "09",
    pages = "052",
    year = "2014"
}

@article{Fioravanti:1996rz,
    author = "Fioravanti, Davide and Mariottini, A. and Quattrini, E. and Ravanini, F.",
    title = "{Excited state Destri-De Vega equation for Sine-Gordon and restricted Sine-Gordon models}",
    eprint = "hep-th/9608091",
    archivePrefix = "arXiv",
    doi = "10.1016/S0370-2693(96)01409-8",
    journal = "Phys. Lett. B",
    volume = "390",
    pages = "243--251",
    year = "1997"
}

@article{Bazhanov:1996aq,
    author = "Bazhanov, Vladimir V. and Lukyanov, Sergei L. and Zamolodchikov, Alexander B.",
    title = "{Integrable Quantum Field Theories in Finite Volume: Excited State Energies}",
    eprint = "hep-th/9607099",
    archivePrefix = "arXiv",
    doi = "10.1016/S0550-3213(97)00022-9",
    journal = "Nucl. Phys. B",
    volume = "489",
    pages = "487--531",
    year = "1997"
}

@article{Dymarsky:2018lhf,
    author = "Dymarsky, Anatoly and Pavlenko, Kirill",
    title = "{Generalized Gibbs Ensemble of 2d CFTs at large central charge in the thermodynamic limit}",
    eprint = "1810.11025",
    archivePrefix = "arXiv",
    primaryClass = "hep-th",
    doi = "10.1007/JHEP01(2019)098",
    journal = "JHEP",
    volume = "01",
    pages = "098",
    year = "2019"
}

@article{Maloney:2018yrz,
    author = "Maloney, Alexander and Ng, Gim Seng and Ross, Simon F. and Tsiares, Ioannis",
    title = "{Generalized Gibbs Ensemble and the Statistics of KdV Charges in 2D CFT}",
    eprint = "1810.11054",
    archivePrefix = "arXiv",
    primaryClass = "hep-th",
    doi = "10.1007/JHEP03(2019)075",
    journal = "JHEP",
    volume = "03",
    pages = "075",
    year = "2019"
}

@article{Downing:2024nfb,
    author = "Downing, Max and Karimi, Faisal",
    title = "{Modular Properties of Generalised Gibbs Ensembles}",
    eprint = "2410.06288",
    archivePrefix = "arXiv",
    primaryClass = "hep-th",
    doi = "10.21468/SciPostPhys.18.3.085",
    journal = "SciPost Phys.",
    volume = "18",
    pages = "085",
    year = "2025"
}

@article{Maloney:2018hdg,
    author = "Maloney, Alexander and Ng, Gim Seng and Ross, Simon F. and Tsiares, Ioannis",
    title = "{Thermal Correlation Functions of KdV Charges in 2D CFT}",
    eprint = "1810.11053",
    archivePrefix = "arXiv",
    primaryClass = "hep-th",
    doi = "10.1007/JHEP02(2019)044",
    journal = "JHEP",
    volume = "02",
    pages = "044",
    year = "2019"
}

@article{Downing:2021mfw,
    author = "Downing, Max and Watts, Gerard M. T.",
    title = "{Free fermions, KdV charges, generalised Gibbs ensembles and modular transforms}",
    eprint = "2111.13950",
    archivePrefix = "arXiv",
    primaryClass = "hep-th",
    reportNumber = "KCL-MTH-21-02",
    doi = "10.1007/JHEP06(2022)036",
    journal = "JHEP",
    volume = "06",
    pages = "036",
    year = "2022"
}

@article{Downing:2023lop,
    author = "Downing, Max",
    title = "{Modular transform of free fermion generalised Gibbs ensembles and generalised power partitions}",
    eprint = "2310.07601",
    archivePrefix = "arXiv",
    primaryClass = "hep-th",
    month = "10",
    year = "2023"
}

@article{Downing:2023lnp,
    author = "Downing, Max and Watts, G\'erard M. T.",
    title = "{Free fermions, KdV charges, generalised Gibbs ensembles, modular transforms and line defects}",
    eprint = "2311.04564",
    archivePrefix = "arXiv",
    primaryClass = "hep-th",
    doi = "10.1007/JHEP01(2024)041",
    journal = "JHEP",
    volume = "01",
    pages = "041",
    year = "2024"
}

@article{Zamolodchikov:1989cf,
    author = "Zamolodchikov, A. B.",
    title = "{Thermodynamic Bethe Ansatz in Relativistic Models. Scaling Three State Potts and Lee-yang Models}",
    reportNumber = "ITEP-89-144",
    doi = "10.1016/0550-3213(90)90333-9",
    journal = "Nucl. Phys. B",
    volume = "342",
    pages = "695--720",
    year = "1990"
}

@article{Dorey:1996re,
    author = "Dorey, Patrick and Tateo, Roberto",
    title = "{Excited states by analytic continuation of TBA equations}",
    eprint = "hep-th/9607167",
    archivePrefix = "arXiv",
    reportNumber = "DTP-96-29",
    doi = "10.1016/S0550-3213(96)00516-0",
    journal = "Nucl. Phys. B",
    volume = "482",
    pages = "639--659",
    year = "1996"
}

@article{Dorey:1997rb,
    author = "Dorey, Patrick and Tateo, Roberto",
    title = "{Excited states in some simple perturbed conformal field theories}",
    eprint = "hep-th/9706140",
    archivePrefix = "arXiv",
    reportNumber = "DTP-97-29",
    doi = "10.1016/S0550-3213(97)00838-9",
    journal = "Nucl. Phys. B",
    volume = "515",
    pages = "575--623",
    year = "1998"
}

@article{Ashok:2024ygp,
    author = "Ashok, Sujay K. and Parihar, Sanhita and Sengupta, Tanmoy and Sudhakar, Adarsh and Tateo, Roberto",
    title = "{Thermal Correlators and Currents of the $\mathcal{W}_3$ Algebra}",
    eprint = "2410.11748",
    archivePrefix = "arXiv",
    primaryClass = "hep-th",
    month = "10",
    year = "2024"
}

@article{Ashok:2024zmw,
    author = "Ashok, Sujay K. and Parihar, Sanhita and Sengupta, Tanmoy and Sudhakar, Adarsh and Tateo, Roberto",
    title = "{Integrable structure of higher spin CFT and the ODE/IM correspondence}",
    eprint = "2405.12636",
    archivePrefix = "arXiv",
    primaryClass = "hep-th",
    doi = "10.1007/JHEP07(2024)179",
    journal = "JHEP",
    volume = "07",
    pages = "179",
    year = "2024"
}

@article{Bajnok:2014fca,
    author = "Bajnok, Zoltan and el Deeb, Omar and Pearce, Paul A.",
    title = "{Finite-Volume Spectra of the Lee-Yang Model}",
    eprint = "1412.8494",
    archivePrefix = "arXiv",
    primaryClass = "hep-th",
    doi = "10.1007/JHEP04(2015)073",
    journal = "JHEP",
    volume = "04",
    pages = "073",
    year = "2015"
}

@article{Bazhanov:1994ft,
    author = "Bazhanov, Vladimir V. and Lukyanov, Sergei L. and Zamolodchikov, Alexander B.",
    title = "{Integrable structure of conformal field theory, quantum KdV theory and thermodynamic Bethe ansatz}",
    eprint = "hep-th/9412229",
    archivePrefix = "arXiv",
    reportNumber = "CLNS-94-1316, RU-94-98",
    doi = "10.1007/BF02101898",
    journal = "Commun. Math. Phys.",
    volume = "177",
    pages = "381--398",
    year = "1996"
}

@article{Cardy:1989fw,
    author = "Cardy, John L. and Mussardo, G.",
    title = "{S Matrix of the Yang-Lee Edge Singularity in Two-Dimensions}",
    reportNumber = "UCSB-TH-89-11",
    doi = "10.1016/0370-2693(89)90818-6",
    journal = "Phys. Lett. B",
    volume = "225",
    pages = "275--278",
    year = "1989"
}

@article{Dorey:1998pt,
  author        = "Dorey, Patrick and Tateo, Roberto",
  title         = "{Anharmonic oscillators, the thermodynamic Bethe ansatz and nonlinear integral equations}",
  journal       = "J. Phys. A",
  volume        = "32",
  pages         = "L419--L425",
  year          = "1999",
  doi           = "10.1088/0305-4470/32/38/102",
  eprint        = "hep-th/9812211",
  archivePrefix = "arXiv"
}

@article{Bazhanov:1998wj,
  author        = "Bazhanov, Vladimir V. and Lukyanov, Sergei L. and Zamolodchikov, Alexander B.",
  title         = "{Spectral determinants for Schr\"odinger equation and Q-operators of conformal field theory}",
  journal       = "J. Stat. Phys.",
  volume        = "102",
  pages         = "567--576",
  year          = "2001",
  doi           = "10.1023/A:1004838616921",
  eprint        = "hep-th/9812247",
  archivePrefix = "arXiv"
}

@article{Dorey:1999uk,
  author        = "Dorey, Patrick and Tateo, Roberto",
  title         = "{On the relation between Stokes multipliers and the T-Q systems of conformal field theory}",
  journal       = "Nucl. Phys. B",
  volume        = "563",
  pages         = "573--602",
  year          = "1999",
  doi           = "10.1016/S0550-3213(99)00609-4",
  eprint        = "hep-th/9906219",
  archivePrefix = "arXiv",
  note          = "[Erratum: Nucl. Phys. B 603 (2001) 581]"
}

@article{Dorey:2007zx,
  author        = "Dorey, Patrick and Dunning, Clare and Tateo, Roberto",
  title         = "{The ODE/IM correspondence}",
  journal       = "J. Phys. A",
  volume        = "40",
  pages         = "R205",
  year          = "2007",
  doi           = "10.1088/1751-8113/40/32/R01",
  eprint        = "hep-th/0703066",
  archivePrefix = "arXiv"
}

@article{Zamolodchikov:1991et,
  author        = "Zamolodchikov, Al. B.",
  title         = "{On the thermodynamic Bethe ansatz equations for reflectionless ADE scattering theories}",
  journal       = "Phys. Lett. B",
  volume        = "253",
  pages         = "391--394",
  year          = "1991",
  doi           = "10.1016/0370-2693(91)91737-D"
}

@article{Dorey:1999pv,
  author  = "Dorey, Patrick and Tateo, Roberto",
  title   = "{Differential equations and integrable models: the SU(3) case}",
  journal = "Nucl. Phys. B",
  volume  = "571",
  pages   = "583--606",
  year    = "2000",
  eprint  = "hep-th/9910102",
  archivePrefix = "arXiv",
  note    = "[Erratum: Nucl. Phys. B 603 (2001) 582]. CHECK"
}

@article{Bazhanov:2001xm,
  author  = "Bazhanov, Vladimir V. and Hibberd, Anthony N. and Khoroshkin, Sergey M.",
  title   = "{Integrable structure of W(3) conformal field theory, quantum Boussinesq theory and boundary affine Toda theory}",
  journal = "Nucl. Phys. B",
  volume  = "622",
  pages   = "475--547",
  year    = "2002",
  eprint  = "hep-th/0105177",
  archivePrefix = "arXiv",
  note    = "CHECK"
}

@unpublished{Karimi:2026kdv,
    author = "Karimi, Faisal",
    title = "{\it Modular Properties of Generalised Gibbs Ensembles: The Deformed KdV Hierarchy, In preparation}",
    note = "In preparation" 
}

@article{Pozsgay:2013gqa,
    author = "Pozsgay, Balazs",
    title = "{The generalized Gibbs ensemble for Heisenberg spin chains}",
    eprint = "1304.5374",
    archivePrefix = "arXiv",
    primaryClass = "cond-mat.stat-mech",
    doi = "10.1088/1742-5468/2013/07/P07003",
    journal = "J. Stat. Mech.",
    volume = "2013",
    pages = "P07003",
    year = "2013"
}

@article{Ilievski:2015mza,
    author = "Ilievski, Enej and De Nardis, Jacopo and Wouters, Bram and Caux, Jean-Sebastien and Essler, Fabian H. L. and Prosen, Tomaz",
    title = "{Complete Generalized Gibbs Ensembles in an Interacting Theory}",
    eprint = "1507.02993",
    archivePrefix = "arXiv",
    primaryClass = "quant-ph",
    doi = "10.1103/PhysRevLett.115.157201",
    journal = "Phys. Rev. Lett.",
    volume = "115",
    number = "15",
    pages = "157201",
    year = "2015"
}

@article{Bazhanov:2003ni,
    author = "Bazhanov, Vladimir V. and Lukyanov, Sergei L. and Zamolodchikov, Alexander B.",
    title = "{Higher-level eigenvalues of Q-operators and Schroedinger equation}",
    eprint = "hep-th/0307108",
    archivePrefix = "arXiv",
    doi = "10.4310/ATMP.2003.v7.n4.a4",
    journal = "Adv. Theor. Math. Phys.",
    volume = "7",
    number = "4",
    pages = "711--725",
    year = "2003"
}

@article{Conti:2020zzk,
    author = "Conti, Riccardo and Masoero, Davide",
    title = "{Counting monster potentials}",
    eprint = "2009.14638",
    archivePrefix = "arXiv",
    primaryClass = "math-ph",
    doi = "10.1007/JHEP02(2021)059",
    journal = "JHEP",
    volume = "02",
    pages = "059",
    year = "2021"
}

@article{Lukyanov:2010rn,
  author  = "Lukyanov, Sergei L. and Zamolodchikov, Alexander B.",
  title   = "{Quantum sine(h)-Gordon model and classical integrable equations}",
  journal = "JHEP",
  volume  = "07",
  pages   = "008",
  year    = "2010",
  eprint  = "1003.5333",
  archivePrefix = "arXiv",
  note    = "CHECK"
}

@article{Dorey:2012bx,
  author  = "Dorey, Patrick and Faldella, Simone and Negro, Stefano and Tateo, Roberto",
  title   = "{The Bethe ansatz and the Tzitz\'eica-Bullough-Dodd equation}",
  journal = "Phil. Trans. Roy. Soc. A",
  volume  = "371",
  pages   = "20120052",
  year    = "2013",
  eprint  = "1209.5517",
  archivePrefix = "arXiv",
  note    = "CHECK"
}

@article{Sedrakyan:2026chiral,
    author = {Sedrakyan, Tigran A. and Pang, Jun-Jun and
              Wei, Chenan and Wang, Baigeng},
    title = {Spontaneous symmetry breaking in chiral
             current-carrying high-energy eigenstates},
    journal = {Phys. Rev. B},
    volume = {114},
    number = {5},
    pages = {055127},
    year = {2026},
    doi = {10.1103/ld5r-jkh7},
    eprint = {2512.09107},
    archivePrefix = {arXiv},
    primaryClass = {quant-ph}
}

@article{Karimi:2026tvh,
    author = "Karimi, Faisal and Watts, G{\'e}rard M. T.",
    title = "{Modular Properties of Symplectic Fermion Generalised Gibbs Ensemble}",
    eprint = "2603.19383",
    archivePrefix = "arXiv",
    primaryClass = "hep-th",
    month = "3",
    year = "2026"
}

@incollection{Dorey:2025odeim,
  author    = {Dorey, Patrick and Dunning, Clare and Tateo, Roberto},
  title     = {The {ODE/IM} Correspondence},
  booktitle = {Encyclopedia of Mathematical Physics (Second Edition)},
  editor    = {Szabo, Richard and Bojowald, Martin},
  publisher = {Academic Press},
  address   = {Oxford},
  year      = {2025},
  pages     = {145--161},
  isbn      = {978-0-323-95706-9},
  doi       = {10.1016/B978-0-323-95703-8.00065-3}
}

@article{Kudrna:2025bzg,
    author = "Kudrna, Mat{\v{e}}j and Proch{\'a}zka, Tom{\'a}{\v{s}}",
    title = "{On W-algebras and ODE/IM correspondence}",
    eprint = "2508.20793",
    archivePrefix = "arXiv",
    primaryClass = "hep-th",
    month = "8",
    year = "2025"
}

\end{document}